\documentclass[a4paper,prb,floatfix,twocolumn,showpacs]{revtex4-1}
\usepackage{amsmath}
\usepackage{amssymb}
\usepackage{graphicx}
\usepackage{esint}

\makeatletter

\renewcommand{\vec}[1]{\boldsymbol{#1}}
\renewcommand{\Re}{\operatorname{Re}}
\renewcommand{\Im}{\operatorname{Im}}
\makeatother

\begin{document}

\title{On the nature of the phase transitions in two-dimensional  type
II superconductors}

\author{Niels R. Walet}
\email{Niels.Walet@manchester.ac.uk}
\author{M. A. Moore}
\email{m.a.moore@manchester.ac.uk}

\affiliation{School of Physics and Astronomy, University of Manchester, 
Manchester M13 9PL, United Kingdom}
\begin{abstract}
We have simulated the thermodynamics of vortices in a thin film of a
type-II superconductor. We make the lowest Landau level approximation,
and use quasi-periodic boundary conditions.  Our work is consistent with the
results of previous simulations where evidence was found for an
apparent first order transition between the vortex liquid state and
the vortex crystal state. We show, however, that these results are just an
artifact of studying systems which are too small.  There are
substantial difficulties in simulating larger systems using
traditional approaches.  By means of the optimal energy diffusion
algorithm we have been able to study systems containing up to about
one thousand vortices, and for these larger systems the evidence for a
first order transition disappears.  By studying both crystalline and
hexatic order, we show that the KTHNY scenario seems to apply, where
melting from the crystal is first to the hexatic liquid state and next
to the normal vortex liquid, in both cases via a continuous
transition.
\end{abstract}

\maketitle

\section{Introduction}

It was Abrikosov\cite{abrikosov1957onthe}  who first studied the phase
transition from the normal fluid of vortices to the vortex crystal. In
a layer  of superconducting film of  thickness $d$ such  that than the
effective  penetration  depth\cite{pearl_current_1964},  $(2 \lambda^2/d)$,  is
greater  than  the  linear extent  $L_x$  or  $L_y$  of the  film  the
Ginzburg-Landau expansion  of the free  energy can be  very accurately
approximated as \cite{rosenstein2010ginzburglandau},
\begin{equation}
\frac{F_{\text{GL}}}{k_BT^{\text{MF}}_c}=d\int d^{2}r\,\left[\alpha(T)|\Psi|^{2}+\frac{\beta_{\kappa}}{2}|\Psi|^{4}+\frac{1}{2m}|\vec{D}\Psi|^{2}\right],
\label{eq:Fcl}
\end{equation}
where $\vec{D} =-i\hbar\vec{\nabla}-2e\vec{A}$, and the vector
potential $\vec{A}$ is that appropriate for a field $B$ normal to the
film, $\vec{A} =(0,Bx,0)$.  The temperature-dependent coefficient in
this expression, $\alpha(T)$, behaves near the mean-field critical
point in zero magnetic field, $T^{\text{MF}}_c$, as
\begin{equation}
\alpha(T)=\alpha_{0}\left(T-T^{\text{MF}}_c\right).
\end{equation}

Abrikosov found the mean-field solution for the functional in
Eq.~(\ref{eq:Fcl}) in the LLL (Lowest-Landau-Level) approximation, by
minimizing $F_{\text{GL}}$.  This yields the solution $\Psi=0$ when $T
>T^{\text{MF}}_c(B)$ and a solution corresponding to a lattice crystal
of vortices when $T<T^{\text{MF}}_c(B)$.  In his work Abrikosov
assumed a crystal with square symmetry, but it was later shown
\cite{kleiner_bulk_1964} that a triangular lattice has a slightly
lower free energy.  The transition at the magnetic-field dependent
temperature $T^{\text{MF}}_c(B)$, which is normally called the
$H_{c2}$ line, is a second order transition at mean-field level, and
is a transition from a vortex liquid to the vortex crystal phase.

For conventional superconductors the mean-field solution is an
excellent first approximation, but fluctuations around the mean-field
can never be entirely neglected.  The effect of fluctuations on the
transition in various dimensions has been studied by renormalization
group (RG) methods.  An expansion about the upper critical dimension
$6$ in $\epsilon$ (when the dimensionality is $6-\epsilon$) was
carried out by Br\'{e}zin, Nelson and Thiaville
\cite{brezin_fluctuation_1985} within the LLL approximation.  They
could not find a stable fixed point and concluded that as a
consequence the transition to the crystalline state from the vortex
liquid state might be a first order transition.  Later measurements of
the specific heat of high-temperature superconductors in a field found
strong evidence for a first order
transition\cite{schilling_calorimetric_1996} in three dimensional
system. However, for the two-dimensional thin film system studied in
experiments by Urbach \emph{et al.}\cite{urbach_specific_1992} there
was no sign of a first order transition. On the other hand, a number
of Monte Carlo simulations of thin films have indicated that there
might be a first order phase transition after all
\cite{kato1993firstorder,kato1993montecarlo,
  hu1994correlations,vsavsik1994calculation,vsavsik1995phasecoherence,
  li2003fluxlattice}.  There have been many other theoretical
approaches to vortex lattice melting; these have been extensively
reviewed by Rosenstein and Li\cite{rosenstein2010ginzburglandau}.

In this paper we revisit the problem of simulating two-dimensional
superconducting films with quasi-periodic boundary conditions using a
novel method \cite{trebst2004optimizing,bauer2010optimized}.  As a
consequence we are able to equilibrate larger systems than were
studied previously, and have been able to investigate the behavior in
more detail by using the microcanonical (constant energy) ensemble.
We find that the evidence for a first order transition goes away as
the number $N$ of vortices in the simulation is increased. We
therefore attribute the apparent evidence for a first order transition
in two dimensional superconducting films with quasi-periodic boundary
conditions (which geometrically can be thought of as ``the flat
torus'') to finite size effects.

One of us (MAM) has been arguing for some years that there might be no
freezing transition of the vortices in two and three dimensions, and
that the correlation length of crystalline short-range order just grew
as the temperature $T$ was reduced, reaching infinity only at $T=0$.
This argument was partly based on approximate analytical calculations
\cite{yeo1996parquetgraph,yeo1996nonperturbative} and general scaling
arguments \cite{moore1997counter}, but also on Monte Carlo simulations
where the vortices moved on the surface of a sphere rather than a flat
torus
\cite{dodgson_vortices_1997,lee1994montecarlo,oneill_monte_1992}.  In
these simulations on the sphere there was no sign of a first order
transition. Since the choice of boundary conditions should not affect
thermodynamics in the limit of large $N$, this result is consistent
with the finding in this paper that the previously reported first
order transition with quasi-periodic boundary conditions is just a
finite-size artifact.  In the simulations which we are reporting in
this paper, the correlation length of crystalline order obtained from
the density-density correlation function grows similarly to that
reported earlier for the sphere, in that in most of the liquid region,
the correlation length seems to be diverging only as $T \to 0$, except
that over a narrow temperature interval we find evidence that the
hexatic correlation length is diverging to infinity at a finite
temperature.  This divergence was not observed in the earlier
simulations as only the correlation length of crystalline order in the
liquid was studied although later some limited evidence was found for
a rapidly growing hexatic correlation length \cite{Kim_Stephenson}.

The divergence at finite temperature of the hexatic correlation length
suggests           that          the           KTHNY          scenario
\cite{halperin_theory_1978,nelson_dislocation-mediated_1979,young1979melting}
might be  relevant.  In  this scenario the  vortex crystal melts  at a
continuous transition  involving the  unbinding of dislocations  as in
the  Kosterlitz-Thouless  picture  \cite{kosterlitz1973ordering} to  a
hexatic liquid, which  in turn changes to a normal  liquid at a higher
temperature  when  the  disclinations  unbind. We  believe  that  this
scenario describes our simulations best, although the evidence for the
transition from  crystal to the hexatic  state is not as  clear as one
might   have  hoped  for,   because  of   finite  size   effects  (see
Sec.~\ref{Results}).     Recently   a    cut-down   model    for   the
crystal-hexatic        transition        has       been        studied
\cite{iaconis2010continuous}  which  gives   results  which  are  also
consistent with the KTHNY scenario.

Our interest in this whole  topic was reawakened by the investigations
of  Bowell   \emph  {et  al.}    on  the  specific  heat   of  niobium
\cite{bowell_absence_2010}.   The  niobium  studied  had a  very  high
degree of purity.  No evidence for a first order transition was found.
Bossen  and  Schilling\cite{bossen_estimates_2012}  have reviewed  the
literature on vortex lattice melting  of both low and high temperature
superconductors and  concluded that the first  order transition should
have  been detectable  for  niobium if  the  conventional approach  to
vortex lattice  melting applies to  it.  Our simulations being  in two
dimensions alas  cast no  direct light on  this mystery.   However, in
three  dimensions the  KTHNY theory  of the  unbinding  of topological
defects  does not  apply and  a hexatic  state is  not  expected.  The
apparent absence  of a  first order transition  in very  clean niobium
could indicate  that the old  approach of Moore\cite{moore1997counter}
and     the     numerical      studies     in     three     dimensions
\cite{kienappel_numerical_1997} still have utility.  The present study
in two dimensions appears consistent with  the  growth predicted
in  Ref.~\onlinecite{moore1997counter}  of  the   correlation  length  of
crystalline order in the vortex liquid phase,  that is, it appears to diverge
as $T \to 0$, and it only fails to work
very  close  to  the  temperature  at which  the  topological  defects
bind.  In three  dimensions there  seems to  be no  analogue  of KTHNY
theory  and  it  is   possible  therefore  that  the  expectations  of
Ref.~\onlinecite{moore1997counter} that  there  is no  true transition  in
three dimensions  remains valid. But we shall  leave such speculations
to be settled by future work  and just concentrate for now on the case
of two dimensions.

The three important approximations applied in this paper and commonly
used in the large literature on the subject are: the description of a
thin film as just a two-dimensional system, the restriction to the
Lowest-Landau level and the assumption of a uniform $B$ field.  An
early paper discussing the first approximation in a theoretical
framework is that of Ruggieri and Thouless
\cite{ruggeri_perturbation_1976}, who argued that if the correlation
length in the perpendicular direction, $\xi_z$, is larger than the
layer thickness $d$, we get a dimensional reduction from three to two.
Since this length is divergent at the zero-magnetic field critical
temperature, there is always a regime where this approximation
applies.  The lowest-Landau level approximation has been discussed in
detail in the work of Rosenstein and Li
\cite{rosenstein2010ginzburglandau}.  Finally, the success of the
analysis of the experimental data in papers such as that by Urbach
\emph{et al.}  \cite{urbach_specific_1992}, where it is shown that
experimental data at different magnetic fields $B$ and temperatures
$T$ collapse onto a single curve as a function of the parameter
$\alpha_B(T)$, (as in our Eq.~(\ref{eq:FGLalpha})), shows that all
three approximations are applicable to their arrangement of thin
films.  They used many stacked thin layers with a large separation
between them. This means that the magnetic field $B$ is likely to be
fairly uniform across the system, so that edge effects, such as
fringing fields, can be ignored to a good approximation.

The plan of the paper is as follows. In Sec.~\ref{basis} we set up the
basis we use to simulate the superconducting film with quasi-periodic
boundary conditions. This work is described in some detail (see also
Appendix A) because we found it hard to obtain a consistent set of the
detailed results needed for doing the simulations from the old
literature.  In Sec.~\ref{vortexpositions} we explain how one can
extract from the chosen basis set the actual positions of the
vortices.  This is necessary when we have to calculate the hexatic
order parameter and its correlations.  It is also needed to construct
the Delaunay diagrams showing the topological defects present in the
system at various temperatures. In the same section we also discuss
the translational symmetry which quasi-periodic boundary conditions
allows. We have noticed that for some values of $\sqrt{N}$ there are
several distinct orientations possible for the lattice which have
exactly the same ground state energy. This feature is studied as it
plays a role in the form of the density-density correlation at
non-zero temperatures. In Sec.~\ref{Correlations} and Appendix B we
define the correlation functions which we study in the simulation:
they include a modified form of the density-density correlation
function and also the hexatic correlation function.  We also give
details of the formalism used to calculate the shear modulus of the
system. In Sec.~\ref{MonteCarlo} and Appendices C and D are given
details of the Monte Carlo methods which were used and why we found it
necessary to use the optimal diffusion algorithm when studying large
numbers of vortices. Results are finally discussed in
Sec.~\ref{Results}, which is the heart of the paper. We conclude with
a discussion of some of the key results in Sec.~\ref{discussion}.


\section{Expansion in a basis}
\label{basis}
In this paper we choose to model the infinite system by applying quasi-periodic
boundary conditions, i.e., by working on the flat torus. Here we shall
follow the convention of Yoshioka \emph{et al.} \cite{yoshioka1983groundstate}
and work on a domain of size $[0,L_{x}]\otimes[0,L_{y}]$ with quasi-periodic
boundary conditions. Functions satisfying such quasi-periodic boundary
conditions have also been considered in mathematics in the theory
of $\theta$ functions, and this particular area of research goes
by the name ``theta-representation of the Heisenberg group'' (see
the book by Mumford \cite{mumford2006tatalectures}); indeed we shall
see many of the results rely heavily on the properties of $\theta$
functions. 

The most general quasi-periodic boundary conditions are of the form
\begin{align}
\Psi(x+L_{x},y) & =e^{i\theta_{x}}e^{i2\pi Ny/L_{y}}\Psi(x,y),\label{eq:BC1}\\
\Psi(x,y+L_{y}) & =e^{i\theta_{y}}\Psi(x,y),\label{eq:BC2}
\end{align}
where $N$ is an integer, so that 
\begin{equation}
\Psi(x+L_{x},y+L_{y})=e^{i(\theta_{x}+\theta_{y})}\Psi(x,y).
\end{equation}
We can give a physical interpretation of the boundary conditions 
in terms of the  quantization of the magnetic flux. 
Since the area for each flux quantum can be expressed in terms of the magnetic length $l_{B}$ as
\begin{align}
2\pi l_{B}^{2} & =h/(2eB),
\end{align}
the periodicity of the wave function requires that
\begin{align}
L_{x}L_{y} & =N 2\pi l_{B}^{2}. \label{eq:lquant}
\end{align}
The integer $N$ thus specifies the number of vortices on the torus.
With the specific choice of quasi-periodic boundary conditions made here,
the center of mass of the exactly $N$ zeroes 
$z^{(0)}_i$ of $\Psi$ on the torus, which should be interpreted
as the positions of the core of the vortices, is quantized to be  \cite{mumford2006tatalectures,haldane1985periodic}
\begin{equation}
\sum_{i=1}^N z^{(0)}_i = \left[(n_x) N_y +\frac{\theta_y}{2\pi}\right] L_x
+i \left[(n_y) N_x +\frac{\theta_x}{2\pi}\right] L_y.
\label{eq:CoMQuant}
\end{equation}

We expand the order parameter field $\Psi$ in terms of quasi-periodic
basis functions,
\begin{equation}
\Psi(x,y)=\sum_{j=1}^{N}c_{j}\phi_{j}(x,y;\theta_{x},\theta_{y}),
\end{equation}
with \cite{yoshioka1983groundstate,chung1997symmetries} 
\begin{align}
\phi_{j}(x,y;\theta_{x},\theta_{y}) & =\sum_{s=-\infty}^{\infty}f_{j,s}(x)g_{j,s}(y)\label{eq:phi2},
\end{align}
where
\begin{align}
f_{j,s}(x) & =\frac{1}{\sqrt{l_{B}\sqrt{\pi}}}\exp\left(-\left(x-X_{js}\right)^{2}/\left(2l_{B}^{2}\right)\right),\label{eq:fx}\\
g_{j,s}(y) & =\frac{1}{\sqrt{L_{y}}}\exp\left(i\left(y+\frac{\theta_{x}}{2\pi N}L_{y}\right)X_{js}/l_{B}^{2}\right),\label{eq:gy}
\end{align}
and
\begin{align}
X_{js} & =L_{x}\left[\left(j-\frac{\theta_{y}}{2\pi}\right)\frac{1}{N}+s\right].\label{eq:X}
\end{align}

This function can be re-expressed in terms of $\theta_{3}$, the 
Jacobi theta function, also denoted as $\vartheta$ or $\vartheta_{00}$  \cite{mumford2006tatalectures},
\begin{widetext}
\begin{align}
\phi_{j}(x,y;\theta_{x},\theta_{y}) & =i\frac{1}{\sqrt{l_{B}L_{x}N\sqrt{\pi}}}\exp\left[\frac{-x^{2}}{2l_{B}^{2}}\right]\exp\left(\frac{\left(i\frac{\theta_{x}}{2\pi N}L_{y}+z\right)^{2}}{2l_{B}^{2}}\right)\theta_{3}\left(\pi\left[\frac{1}{L_{x}}\left(-z+\Theta\right)+\frac{j}{N}\right]\middle|\tau\right)\label{eq:thetasum},
\end{align}
\end{widetext}
where
\begin{align}
z & =x+iy,\\
\Theta & =\frac{1}{2\pi N}\left[L_{x}\theta_{y}-iL_{y}\theta_{x}\right],\\
\tau & =\frac{i}{N}\frac{L_{y}}{L_{x}}.
\end{align}
These basis functions satisfy the boundary conditions
\begin{align}
\phi_{j}(x+L_{x},y;\theta_{x},\theta_{y}) & =e^{i\left(\theta_{x}+2\pi N\frac{y}{L_{y}}\right)}\phi_{j}(x,y;\theta_{x},\theta_{y}),\\
\phi_{j}(x,y+L_{y};\theta_{x},\theta_{y}) & =e^{i\theta_{y}}\phi_{j}(x,y;\theta_{x},\theta_{y}),
\end{align}
provided that the quantization condition Eq.~(\ref{eq:lquant}) is
satisfied.  It is most easy to interpret the basis functions $\phi_j$
by looking at Eqs.~(\ref{eq:phi2},\ref{eq:fx},\ref{eq:gy}).  We see
that $f$ describes a Gaussian centered at the point $X_{js}$, and $g$
is a plane wave in $y$, with wave vector directly linked to $X_{js}$.
Finally, the parameters $\theta_{x}$ and $\theta_{y}$ describe
micro-shifts for the vortex positions by $z^{(0)}_i\rightarrow z^{(0)}_i+\Theta$.

Since the Hamiltonian, as derived below, can easily be show to be
independent of the angles $\vec{\theta}$\footnote{That is true if the
  flux contained in the area is a constant. For slowly varying fields,
  these angles become time dependent, and describe electronic
  transport.}, we shall from now on take these angles to equal 0. We
shall also work on a rectangle shaped to fit a triangular lattice,
where
\begin{equation}
L_{y}=\frac{2\sqrt{\pi}l_{B}N_{y}}{\sqrt[4]{3}},\quad L_{x}=\sqrt[4]{3}\sqrt{\pi}l_{B}N_{x}, \label{eq:LxLytri}
\end{equation}
and the integers $N_{x}$ and $N_{y}$ multiply to give the number
of vortices, 
\begin{equation}
N_{x}N_{y}=N.
\end{equation}
If we decompose the order parameter field in the basis functions (\ref{eq:phi2})
\begin{equation}
\Psi=\mathcal{N}\sum_{j=0}^{N-1}c_{j}\phi_{j},\label{eq:thetasum2}
\end{equation}
where  $\mathcal{N}$ is a normalization constant,  we find that the quadratic term reduces  (using orthonormality
of the $\phi$'s) to
\begin{equation}
d \sum_{j=0}^{N-1}|c_{j}|^{2}\left[\alpha_{0}\left(T-T^{\text{MF}}_c\right)+\frac{e\hbar B}{m}\right]\mathcal{N}^{2}.
\end{equation}
This shows that it is useful to introduce a new coupling constant that
depends on the magnetic field. 

With the choice 
\begin{equation}
\mathcal N=\pi l_B/\sqrt{\beta_\kappa},
\end{equation}
we can absorb all of the dimensioned factors in a single dimensionless
temperature-dependent coupling constant
\begin{equation}
\alpha_{B}(T)=\sqrt{2 \pi d/\beta_{\kappa}}\, l_B\left(\alpha_{0}\left(T-T^{\text{MF}}_c\right)+\frac{e\hbar B}{m}\right),\label{eq:alphaT}
\end{equation}
where $\beta_{\kappa}$ is the strength of the effective quartic term
in the GL functional.  With a little work (see appendix
\ref{app:quartic}) we can now express the Ginzburg-Landau free energy
in terms of the $N$ complex variables $c_{i}$,
\begin{widetext}
\begin{equation}
F_{\text{GL}}(\{c\})/\left(k_{B}T^{\text{MF}}_c\right)=
\frac{\pi}{2}\left(|\alpha_{B}(T)|^{2}\left(\text{sgn}\left(\alpha_{B}(T)\right)\sum_{n=0}^{N-1}|c_{n}|^{2}+
\frac{\pi}{2^{3/2}3^{1/4}N_{y}}\sum_{n_{s}=0}^{2N-1}\left|Q_{n_{s}}\right|^{2}\right)\right),\label{eq:FGLalpha}
\end{equation}
where
\begin{equation}
Q_{n_{s}}=\sum_{n_{p}=0}^{2N-1}\delta_{n_{s}+n_{p},\text{even}}\left[\sum_{s_{p}=-\infty}^{\infty}e^{-\frac{\pi(n_{p}+2Ns_{p})^{2}}{\sqrt{3}N_{y}^{2}}}\right]c_{\left\lceil (n_{p}+n_{s})/2\right\rceil }c_{\left\lceil (n_{p}-n_{s})/2\right\rceil }\label{eq:Qns}
\end{equation}
\end{widetext}
is the periodic analogue of a Gaussian weighted convolution between
pairs of coefficients on a circle of circumference $N$.

We have introduced a short-hand notation for the periodic continuation
of indices,
\[
c_{\left\lceil k\right\rceil }\equiv c_{k ~\text{mod}N}.
\]
In summary, we have found that
\begin{equation}
\frac{F_{\text{GL}}(\{c\})}{kT^{\text{MF}}_c\,\alpha_{B}(T)^{2}}=E(\{c\})= E_{2}(\{c\})+E_{4}(\{c\}),\label{eq:E}
\end{equation}
with the temperature independent scaled free energy $E$ split into a
quadratic and quartic part,
\begin{align}
E_{2}(\{c\}) & =\frac{\pi}{2}\text{sgn}\left(\alpha_{B}(T)\right)\sum_{n=0}^{N-1}|c_{n}|^{2},\\
E_{4}(\{c\}) & =\pi^2 2^{-5/2}3^{-1/4}N_{y}^{-1}\sum_{n_{s}=0}^{2N-1}\left|Q_{n_{s}}\right|^{2},\label{eq:E2E4}
\end{align}
where $Q_{n_{s}}$ is given in Eq.~(\ref{eq:Qns}). The energy
expression effectively describes the LLL vortex problem as $N$ complex
($2N$ real) coupled quartic oscillators on a circle of radius $N$,
with an intermediate range interaction between the oscillators that
acts over a distance $\propto N_{y}\sim\sqrt{N}$ along the circle.

We want to simulate the partition function for this model for $N$ vortices (or flux lines)
\begin{equation}
Z(T)=\int e^{-F_{\text{GL}}(\{c\})/kT^{\text{MF}}_c}d\{c\},
\end{equation}
which  can be written as 
\begin{equation}
Z(\alpha_{B}(T))=\int e^{-\alpha_{B}^{2}(T)\left(E_{2}(\{c\})+E_{4}(\{c\})\right)}d\{c\}.
\end{equation}
Here we use the notation 
\begin{equation}d\{c\}=\prod_{i=1}^Nd^2c_i.
\end{equation}

\section{Vortex positions}
\label{vortexpositions}
The wave function $\Psi$ is only an indirect measure of the vortex
nature of the system; the square of its absolute value is the vortex
density. As we shall see in the next section, we can evaluate
density-density correlations directly from $\Psi$, but other important
data on the behavior of the system, such as the hexatic order
parameter requires that we know the position of the vortices, which
are given by the zeroes of $\Psi$. We thus need to find an efficient
way to relate a representation in terms of the order parameter wave
function to one in terms of its zeroes, and vice-versa.

\subsection{$\theta$ functions}

It is known, see e.g. Ref.~\onlinecite{mumford2006tatalectures}, that the
number of zeroes in the fundamental domain of any quasi-periodic
function satisfying the boundary conditions
(\ref{eq:BC1},\ref{eq:BC2}) is exactly $N$. It is not a trivial task
to find their positions from the decomposition in terms of a sum of
Jacobi theta functions $\theta_{3}$, Eqs.~(\ref{eq:thetasum},
\ref{eq:thetasum2}).  If we wish to formulate $\Psi$ in terms of its
zeroes, it is better to use an alternative expression of $\Psi$ as the
product of periodic functions with a single zero in the fundamental
domain, i.e., Jacobi functions of the first kind, $\theta_{1}$.

Following Haldane and Rezayi\cite{haldane1985periodic} we thus write
the order parameter wave function as a product of $N$ terms each with a zero at the position $z^{(0)}_i$,
\begin{align}
\psi_{\text{prod}}(x,y) & =\exp\left(-x^{2}/2l_{B}^{2}\right)
\nonumber\\ & \quad
\exp(ikz)\prod_{i=1}^{N}\theta_{1}\left(\pi(z-z^{(0)}_{i})/L_{x}\middle|\tau'\right) 
\nonumber \\ &
 \equiv\exp(-x^{2}/2l_{B}^{2})\theta_{\text{prod}}(z|\{z^{(0)}\}),\label{eq:thetaprod}
\end{align}
where 
\begin{equation}
\tau'=i\frac{L_{y}}{L_{x}}
\end{equation}
This leads to restrictions for $k$ and for the center of mass (see Ref.~\onlinecite{haldane1985periodic})
{[}remember that we have chosen the parameters $\theta_{i}$ in Eq.
(\ref{eq:thetasum}), which correspond to the $\phi_{i}$'s in 
Ref.~\onlinecite{haldane1985periodic}, equal to zero{]}
\begin{equation}
k=N(m+1)\pi/L_{x},\quad\frac{1}{N}\sum_{i=1}^{N}z^{(0)}_{i}=\frac{n}{N_{x}}L_{x}+i\frac{m}{N_{y}}L_{y}.\label{eq:CoMrestrict}
\end{equation}
The equivalence between Eqs.~(\ref{eq:thetasum2}) and (\ref{eq:thetaprod}),
up to a normalization constant, will allow useful transformations to be performed.

\subsection{Vortex positions from $c$'s}
  If we wish to determine the vortex positions give the $c$'s, we need
  to find the zeroes of the product form (\ref{eq:thetaprod}), given
  the coefficients in the linear ``summation'' form. To facilitate
  this calculation, we write, comparing Eq.~(\ref{eq:thetaprod}) with
  Eq.~(\ref{eq:thetasum}),
\begin{align}
\psi_{\text{sum}}(x,y) & =\sum_{j=0}^{N-1}c_{j}\frac{i}{\sqrt{l_{B}L_{x}N\sqrt{\pi}}}\exp\left[\frac{-x^{2}}{2l_{B}^{2}}\right]
\nonumber\\
&\quad
\exp\left(\frac{z^{2}}{2l_{B}^{2}}\right)\theta_{3}\left(\pi\left[-\frac{1}{L_{x}}z+\frac{j}{N}\right]\middle|\tau\right)\nonumber \\
 & =\exp(-x^{2}/2l_{B}^{2})\theta_{\text{sum }}(z|\{c\}).\label{eq:thetasum2a}
\end{align}
We must now solve the equations 
\begin{equation}
\theta_{\text{prod}}(z|\{z^{(0)}\})=\theta_{\text{sum}}(z|\{c\})
\end{equation}
for the set $\{z^{(0)}\}$. This is a complex non-linear problem. We
approach it by first using a contour-finding technique to determine
all closed contours of $|\theta_{\text{sum}}|$ for a chosen value near
zero.  We then determine the center of gravity of each contour, which
is usually already a very good approximation to the position of a
zero. The position of each $z^{(0)}_{i}$ is then improved by using a root
finding algorithm in the complex variable $z$, which normally
converges quickly. The remaining normalization of
$\theta_{\text{prod}}$ can be found by fixing the value at any point
that does not coincide with a root. The most likely scenario for
failure of the method is the occurrence of multiple or very close
roots, and making small mistakes in the calculation of a vortex that
is close to the boundary, thus misidentifying it as inside or outside
the fundamental domain.

\subsection{Determining $c$'s from vortex positions}
 Clearly, given only the vortex positions, we can only determine
$\theta_{\text{prod}}$ up to a normalization. A change of
normalization in the sum-representation (\ref{eq:thetasum2})
corresponds to a simultaneous scaling of all the $c$'s. In other
words, the vortex positions alone define a one-dimensional manifold of
fields. That means that we have just found that the \emph{linear}
equations stating that there are vortices at all the $z^{(0)}_{i}$'s, i.e.,
the set of equations ($1\leq j\leq N$)
\begin{equation}
 \theta_{\text{sum}}(z^{(0)}_{j}|\{c\})=0,
\end{equation}
must have a one-dimensional family of solutions $\{c\}$. This
condition can thus be written as the determinantal condition $\det\left(M\right)=0,$ where
the entries in the $N\times N$ matrix $M$ take the simple form 
\begin{widetext}
\begin{align}
M_{ij} & =\sum_{s=-\infty}^{s=\infty}\exp\left[-\frac{1}{2}\left[2\pi\left(j+Ns\right)\right]^{2}\frac{L_x}{L_y N}+2\pi\left(j+Ns\right)z^{(0)}_{i}/L_{y}\right].
\end{align}
\end{widetext}
The coefficients $c_{j}$ are obviously given, up to the unknown normalization
constant, by the eigenvector for eigenvalue $0$ (the singular value)
of the matrix $M$.%
\footnote{For values of $z^{(0)}_{i}$ satisfying the center-of-mass quantization condition,
we have always found at least one singular value, and it is highly
likely an existence proof can be given. In a few special cases there
are even multiple solutions.}

\subsection{Symmetry breaking}
Since we are studying a 2D system with an interaction of intermediate
range, we expect the Mermin-Wagner theorem to hold, so that no spontaneous
breaking of continuous symmetries can exist. In other words, the one-body densities
satisfy these symmetries.
Unfortunately, the flat torus -- periodic boundary conditions -- inherently
breaks both rotational and translational symmetries, and we thus need
to consider what spurious effects this may have.

\subsubsection{Translations}
Since the center-of-mass coordinate can only take a finite number $N$
of values, see Eq.~(\ref{eq:CoMrestrict}), translational symmetry is
broken to a discrete subgroup. The number of ground-states has
$N$-fold degeneracy, and we could alternatively have used these as a
basis for the calculations. This corresponds to the Eilenberger basis,
a well-known alternative to the basis choice employed in this
work\cite{eilenberger_thermodynamic_1967}.

\subsubsection{Rotations}
\begin{figure}
\begin{centering}
\includegraphics[width=6cm]{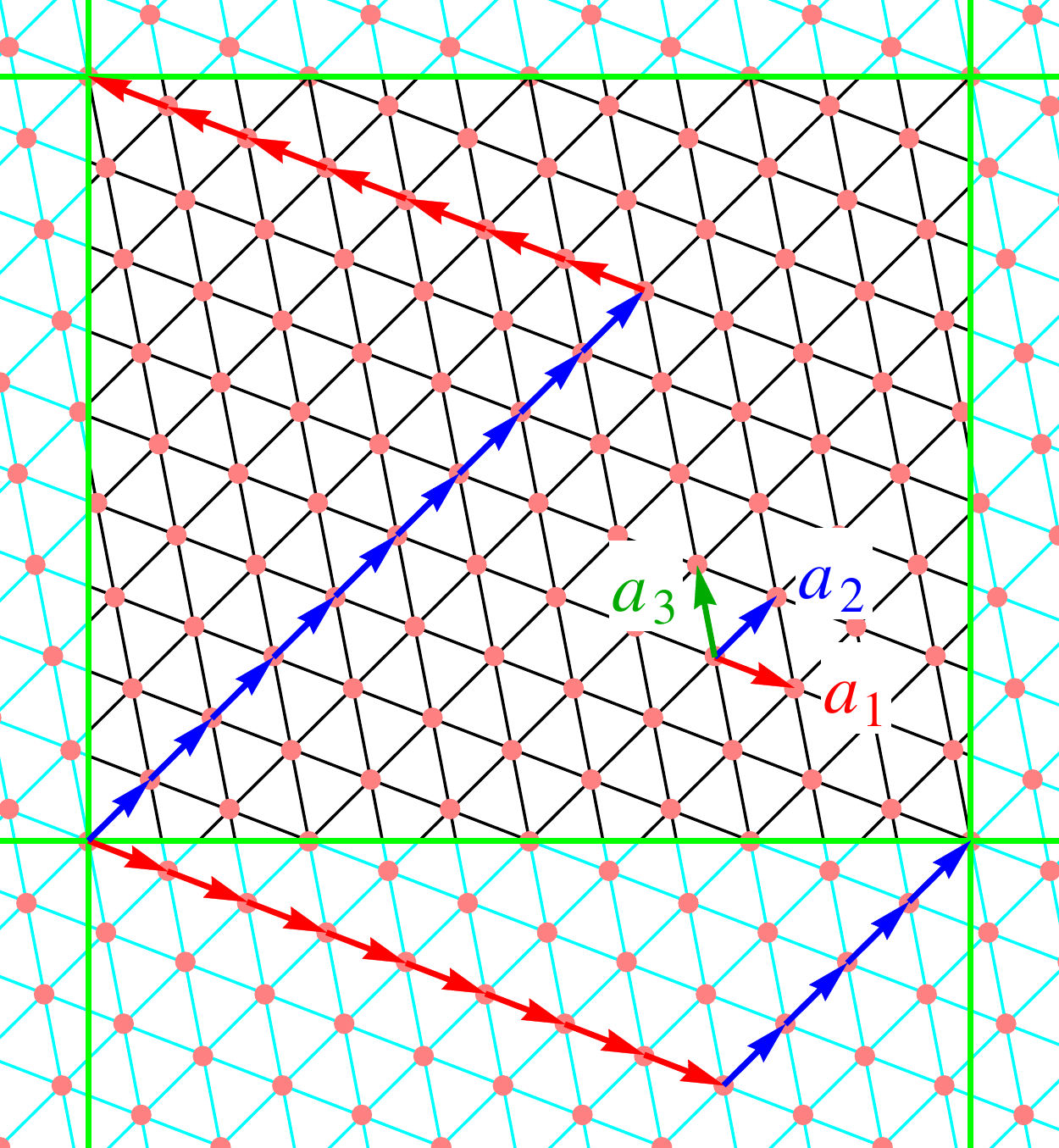}
\end{centering}
\caption{(Color online) A picture of the labeling of a periodic
  triangular lattice on the flat torus. The fundamental domain is
  inside the green rectangle, and the lattice connections are shown in
  black.  Each such can be labeled uniquely by two vectors that wrap
  along the $x$ and $y$-axes, respectively, decomposed in the basis
  $\vec{a}_{1}$, $\vec{a}_{2}$.  In the case shown here this label
  would be
  $\left(\left(8,4\right),\left(-7,9\right)\right)$.\label{fig:periodic}}
\end{figure}

There is only limited room for restoring rotational symmetry on a
torus. Whereas we have designed the shape of the fundamental domain to
fit one triangular lattice, there are a few special cases where we can
fit an the same equilateral-triangular lattice to a flat torus in a
different way.  \cite{webber1997monohedral,giomi2008elastic}.
However, we can easily find many ways to fit a general triangular
lattice to a torus, some of which are very close in energy to the
ground state. Enumerating the winding around the torus in the standard
way for carbon nanotubes, but now in both $x$ and $y$ directions, we
find that we can label such a lattice with two integer vectors
$\vec{m}=(m_{1},m_{2})$ and
$\mbox{\ensuremath{\vec{n}}}=(n_{1},n_{2})$, describing the periodic
vectors along the $x$ and $y$ axes in terms of the lattice basis, see
Fig.~\ref{fig:periodic}.

Since the number of vortices in the fundamental domain is fixed, we
find that the vectors satisfy the Diophantine equation
\begin{equation}
m_{1}n_{2}-m_{2}n_{1}=N_{x}N_{y}=N,\label{eq:diophantine}
\end{equation}
where we have added the (simplifying, but not strictly necessary)
requirement that the determinant on the left-hand-side is positive.
The case of negative determinant can be reached by a simple inversion
of one of the the two vectors $\vec{m}$ or $\vec{n}$, i.e., a reflection
in the $y$ or $x$ axis. There is an issue with the integer labels
not being unique; we can choose any two out of the three vectors $\vec{a}_{i}$
as a basis, and multiply any of these vectors by an overall sign and
get a different result for $\vec{n}$ and $\vec{m}$, without changing
the crystal. The ordering principle we apply is that we always choose
the two shortest vectors $\vec{a}_{i}$, and require that the largest
components of each vector $\vec{a}_{1}$, $\vec{a}_{2}$ are positive.

An expression for the energy for a crystal of this type can be found
in the literature, e.g., Eq.~(77) in Ref.~\onlinecite{rosenstein2010ginzburglandau},
and can be written as
\begin{equation}
E/N=-\frac{1}{2\beta},
\end{equation}
where the parameter $\beta$ is given by the lattice sum
\begin{equation}
\beta=\sum_{k,l=-\infty}^{\infty}\exp\left(-\frac{\pi}{A}\left|k\vec{a}_{1}+l\vec{a}_{2}\right|^{2}\right),
\end{equation}
and $A$ is the area of the basic triangular unit cell of the crystal.
There is an equivalent, but slightly less practical, expression in
terms of the elliptic constant $\tau$ \cite{aftalion2006lowestlandau}.
The triangular ground state is obtained for the well-known value 
\begin{equation}
\beta=\beta_{A}=1.159595,
\label{eq:betaA}
\end{equation} the ``Abrikosov parameter''.

Since we shall always apply this on an $(N_{x},\frac{1}{2}\sqrt{3}N_{y})$
basic lattice (using $\sqrt[4]{3}\sqrt{\pi}l_{B}$ as the unit of length, see Eq.~(\ref{eq:LxLytri})), we find that
\begin{equation}
\left(\begin{array}{cc}
m_{1} & m_{2}\\
n_{1} & n_{2}
\end{array}\right)\left(\begin{array}{c}
\vec{a}_{1}\\
\vec{a}_{2}
\end{array}\right)=\left(\begin{array}{c}
(N_{x},0)\\
(0,N_{y}\sqrt{3}/2)
\end{array}\right),
\end{equation}
or 
\begin{align}
\left(\begin{array}{c}
\vec{a}_{1}\\
\vec{a}_{2}
\end{array}\right) & =\frac{1}{N_{x}N_{y}}\left(\begin{array}{cc}
n_{2} & -m_{2}\\
-n_{1} & m_{1}
\end{array}\right)\left(\begin{array}{c}
(N_{x},0)\\
(0,N_{y}\sqrt{3}/2)
\end{array}\right)\nonumber \\
 & =\left(\begin{array}{c}
(n_{2}/N_{y},-m_{2}/N_{x}\sqrt{3}/2)\\
(-n_{1}/N_{y},m_{1}/N_{x}\sqrt{3}/2)
\end{array}\right).
\end{align}
The third vector $\vec{a}_{3}$ is the shortest of 
\begin{equation}
((n_{2}\mp n_{1})/N_{y},(-m_{2}\pm m_{1})/N_{x}\sqrt{3}/2).
\end{equation}
The lengths squared are thus 
\begin{multline}
n_{2}^{2}/N_{y}^{2}+\frac{3}{2}m_{2}^{2}/N_{x}^{2},n_{1}^{2}/N_{y}^{2}+\frac{3}{2}m_{1}^{2}/N_{x}^{2}\nonumber \\
\text{ and }(n_{1}\pm n_{2})^{2}/N_{y}^{2}+\frac{3}{2}(m_{1}\mp m_{2})^{2}/N_{x}^{2}.
\end{multline}

There are two classes of solutions we shall be interested in: First
of all states that are close to the crystal aligned with the boundary
conditions, and secondly low energy states that are very differently
orientated from the default crystal -- in the most extreme scenario
this will be another ground state with different orientation. In Sec.~\ref{Results} when discussing non-universal behavior we shall 
analyze the state of almost default orientation, and the solutions
to the Diophantine equation (\ref{eq:diophantine}) for small deformations
of the ground state.

\section{Correlations }
\label{Correlations}
The best way to study the structure of the phase diagram is to look at
the correlations that are present. Most of the past Monte Carlo
work concentrated on the crystalline correlations; there is some
work on the shear modulus, which is linked to the 
hexatic-crystal phase transition. We shall look at these,
but also include  hexatic order correlations in our arsenal of analysis tools.

\subsection{Density-density correlations: the function $\Delta$\label{sec:Delta}}
The most studied correlation function for the vortex problem is  a modification 
of the density-density correlation. We start from 
\begin{align}
g(\vec{q)} & \equiv\int d^{2}r\, d^{2}r'e^{i\vec{q}.(\vec{r}-\vec{r}')}\nonumber \\
 & \times\left[\left\langle \left|\Psi(r)\right|^{2}\left|\Psi(r')\right|^{2}\right\rangle -\left\langle \left|\Psi(r)\right|^{2}\right\rangle \left\langle \left|\Psi(r')\right|^{2}\right\rangle \right]/\mathcal{N}^{4}\nonumber \\
 & =\left\langle \left|\rho(\vec{q})\right|^{2}\right\rangle -\left|\left\langle \rho(\vec{q})\right\rangle \right|^{2},
\end{align}
with 
\begin{align}
\rho(\vec{q}) & =\int d^{2}r|\Psi(r)|^{2}e^{i\vec{q}\cdot\vec{r}}/\mathcal{N}^{2}.
\end{align}
We choose 
\begin{equation}
q_{y}=k_{y}/l_{B}=m_{y}2\pi/L_{y}=\frac{2\sqrt{\pi}m_{y}}{\sqrt[4]{3}l_{B}N_{y}},
\end{equation}
and follow the approach set out in detail in Appendix \ref{app:dens-dens}.

It is convenient to multiply $g(q)$ by the  Gaussian factor
  $\exp(k^{2}l_{B}^{2}/2)$, 
to obtain the correlator
\begin{equation}
\Delta(\vec{k})=\left\langle \left|\delta(\vec{k})\right|^{2}\right\rangle -\left|\left\langle \delta(\vec{k)}\right\rangle \right|^{2},
\end{equation}
with 
\begin{equation}
\delta(\vec{k})=\sum_{j'=0}^{N-1}\left(c_{\left\lceil j'\right\rceil }\exp\left(\frac{2i\pi m_{x}j'}{N}\right)\right)c_{\left\lceil j'+m_{y}\right\rceil }^{*},\label{eq:delta}
\end{equation}
 This is often normalized to
\begin{equation}
\tilde{\Delta}(\vec{k})=\Delta(\vec{k})/\left|\left\langle \delta(0)\right\rangle \right|^{2},
\end{equation}
so that the results no longer depend on the normalization of the wave function.
The periodicity of the expression (\ref{eq:delta}) shows that
\emph{both} $m_{x}$ and $m_{y}$ take integer values from  $0$ to
$N-1$. Thus
\begin{align}
0 & \le k_{x}\le \frac{\sqrt[4]{3}\sqrt{\pi}\left(N-1\right)}{N_{x}},\\
0 & \le k_{y}\le \frac{2\sqrt{\pi}(N-1)}{\sqrt[4]{3}N_{y}}.
\end{align}

\subsubsection{Extraction of the correlation length}
We would like to use the density-density correlation function to
extract a correlation length. The difficulty, as one can see later,
e.g., in Fig.~\ref{fig:DDcorrelations}, is that there is no peak at
$k=0$, i.e., there is a correlation hole. The first peaks when we have
crystal, or the first ring when we have a liquid, occur at a finite
$k$ value. There are various ways to extract a correlation length from these
features.

In a liquid, we can take the circular average of the density-density
correlation function, and either fit a Lorentzian to the first peak,
or look at the curvature of the peak near the top, assuming it is
Lorentzian. We find that both of these approaches lead to a very similar
value for a correlation length. 

In a state where we can see discrete Bragg peaks, the situation is
much more complex. Even though the angular average gives the best
statistics, it appears to be a less sensible approach. Several of the
discrete Bragg peaks correspond to states that are not aligned with
the natural crystal axes, and therefore must have some defects or
vacancies and some deformation of the remaining crystal structure to
fit the simulation box. Thus both the direction and magnitude of the
$k$-vector changes.  This tends to broaden the apparent peaks, and
leads to an underestimate of the correlation length when mixing 
discrete points at different angles.

Since at least one of the crystalline ground states aligns with the
$x$-axis, we instead look at the width near the maximum along this
axis in $k$ space, as well as the two lines making an angle $\pi/3$
with this axis. We then determine the correlation length by the
curvature at the top of the peak.  This measure is sensible both  in
liquid and crystal-like states, as long as we take account of the fact
that the position of the peak in $k$ space varies a little with temperature.

\subsection{Hexatic order parameter}

The standard model of 2D melting is the KTHNY scenario, where there is
an intermediate phase with hexatic ordering between crystal and liquid.
In order to see whether this plays a role, it would be useful to
measure the hexatic order parameter $\Psi_{6}$ and its correlation
function. If we know the vortex positions, we just need to determine
the positions of the nearest neighbors of each vortex, which can be
obtained from a Delaunay triangulation, which will give us access to
the local coordination number of each vortex of the lattice, as well
as a list of $N_{\text{nn}}(\vec{r}_i)$ nearest neighbors.  We then
define an order parameter
\begin{equation}
\Psi_{6}(\vec{r}_{i})=\frac{1}{N_{\text{nn}}(\vec{r}_{i})}\sum_{k=1}^{N_{\text{nn}}(\vec{r}_{i})}\exp(i6\theta_{k}),
\end{equation}
where $\theta_{k}$ is the angle between the vector pointing from
the vortex at $\vec{r}_{i}$ to its $k$-th nearest neighbor and
the $x$-axis. Notwithstanding the obvious bias in its definition,
this quantity is commonly used in the analysis of colloidal systems.
We define the hexatic correlation function as 
\[
\chi_{6}(\vec{q})=\left\langle\sum_{ij}\Psi_{6}^{*}(\vec{r}_{i})e^{i\vec{q}\cdot(\vec{r}_{j}-\vec{r}_{i})}\Psi_{6}(\vec{r}_{j})\right\rangle.
\]
A hexatic length parameter is found through the standard Ballesteros-type
analysis \cite{ballesteros2000critical}, and define the correlation
length by the width of $\chi_{6}$ as determined by the curvature at the top of the curve,
\begin{equation}
\xi_6 \equiv \sqrt{(\chi_6(0)/\chi_6(\vec k)-1)}/(2\sin(k/2)).\label{eq:ballesteros}
\end{equation}
The only difference with the standard approach, is that we take the average over
 $\vec k$  as the first allowed reciprocal lattice vector in the
$x$ and $y$ directions.

\subsection{Shear Modulus}

If we shear the crystal by an angle $\theta$, the shear modulus 
can be evaluated to be expressed as \cite{vsavsik1994calculation},
\begin{align}
\mu&=-(k_BT/N)\left[\partial_{\theta,\theta}\ln Z(a_{B}(T),\theta)\right]_{\theta=0} 
\nonumber\\
& =
\frac{k_BT}{N}
\biggl[
  \alpha_B(T)^{2}
  \left\langle \partial_{\theta,\theta}E(\{c\},\theta)\right\rangle_{\theta=0}-
\nonumber\\ &\qquad 
  \alpha_B(T)^{4}
  \left\langle
  \left(\partial_{\theta}E(\{c\},\theta )\right)^{2}
\right\rangle_{\theta=0}
\biggr].
\end{align}
We shall refer to $\left\langle\left(\partial_{\theta}E(\{c\},\theta )\right)^{2}\right\rangle_{\theta=0}$
as the shear susceptibility, and
$\left\langle \partial_{\theta,\theta}E(\{c\},\theta)\right\rangle_{\theta=0}$
as the shear stiffness, since these labels give a sensible
interpretation of these two quantities. We shall see below that it is
the shear susceptibility that gives us information about the
phases of the system; the shear stiffness is almost temperature independent.

With the choice of boundary conditions as in
Eqs.~(\ref{eq:BC1},\ref{eq:BC2}) the easiest way to implement shearing
is in the $y$ direction. As discussed by {\v S}{\'a}{\v s}ik \emph{et
  al} \cite{vsavsik1994calculation,vsavsik1995phasecoherence}, such a
sheared system can be described by the transformation
$\Psi(x,y)\rightarrow\Psi(x,y+x\sin\theta)$.  The transformation can
be absorbed in a change of the basis functions
\begin{widetext}
\begin{align}
\phi^\theta_{j}(x,y) &=
\phi_{j}(x,y+x\sin\theta) \nonumber\\
& =\frac{1}{\sqrt{L_{y}}}\frac{1}{\sqrt{l_{B}\sqrt{\pi}}}\exp\left(-\left(x-X_{js}\right)^{2}/\left(2l_{B}^{2}\right)\right)\exp\left(i\left(y+x\sin\theta\right)X_{js}/l_{B}^{2}\right)\nonumber \\
 & =\frac{1}{\sqrt{L_{y}}}\frac{1}{\sqrt{l_{B}\sqrt{\pi}}}
\exp\left(-\left[{X_{js}}^2-{X^\theta_{js}}^2\right]/\left(2l_{B}^{2}\right)\right)
\exp\left(-\left(x-X_{js}^{\theta}\right)^{2}/\left(2l_{B}^{2}\right)\right)\exp\left(iyX_{js}/l_{B}^{2}\right),
\end{align}
where
\begin{equation}
X_{js}^{\theta}=X_{js}(1-i\sin\theta).\label{eq:Xjstheta}
\end{equation}
If we wrap around the torus in the $x$ direction, we obtain an additional
phase in the wave function 
\begin{align}
\phi^\theta_{j}(x+L_{x},y) & =\sum_{s=-\infty}^{s=\infty}f_{j,s}(x)\exp\left(i\sin\theta L_{x}X_{js}/l_{B}^{2}\right)\frac{1}{\sqrt{L_{y}}}\exp\left(iy\, X_{js}/l_{B}^{2}\right).
\label{eq:phitheta}
\end{align}
It is rather straightforward to obtain the $\theta$ dependence of
the scaled free energy
\begin{align}
E(\{c\},\tan\theta ) & =\frac{\pi}{2}\sum_{n=0}^{N-1}\left|c_{n}\right|^{2}\text{sgn}\left(\alpha_{B}(T)\right)+
\pi^22^{-5/2}3^{-1/4}N_{y}^{-1}\sum_{n_{s}=0}^{2N-1}\left|Q_{n_{s}}(\tan\theta)\right|^{2},
\label{eq:Etheta}\\
Q_{n_{s}}(\tan\theta) & =\sum_{n_{p}=0}^{2N-1}\delta_{n_{s}+n_{p},\text{even}}\left[\sum_{s_{p}=-\infty}^{\infty}e^{-\frac{\left(\pi-i\tan(\theta)/2N\right)(n_{p}+Ns_{p})^{2}}{\sqrt{3}N_{y}^{2}}}\right]c_{\left\lceil (n_{p}+n_{s})/2\right\rceil }c_{\left\lceil (n_{p}-n_{s})/2\right\rceil }.\label{eq:QNstheta}
\end{align}
This allows us to calculate the first and second derivatives at $\theta=0$
by
\begin{align}
\left.\partial_{\tan\theta }E(\{c\},\tan\theta \right|_{\theta=0} & = \pi^2 2^{-5/2}3^{-1/4}N_{y}^{-1}\sum_{n_{s}=0}^{2N-1}(-2)\Im\left(Q_{n_{s}}^{\prime}Q_{n_{s}}^{*}\right),\nonumber \\
\left.\partial_{\tan\theta}^{2}E(\{c\},\tan\theta )\right|_{\theta=0} & =\pi^2 2^{-5/2}3^{-1/4}N_{y}^{-1}\sum_{n_{s}=0}^{2N-1}\left|Q_{n_{s}}^{\prime}\right|^{2}-2\Re\left(Q_{n_{s}}^{\prime\prime}Q_{n_{s}}^{*}\right),
\end{align}
with 
\begin{equation}
Q_{n_{s}}^{\prime}=\left[\partial_{\theta}Q_{n_{s}}(\tan\theta)\right]_{\theta=0},\quad Q_{n_{s}}^{\prime\prime}=\left[\partial_{\theta}^{2}Q_{n_{s}}(\tan\theta)\right]_{\theta=0}.\quad
\end{equation}
\end{widetext}

\begin{figure}
\begin{centering}
\includegraphics[width=8cm]{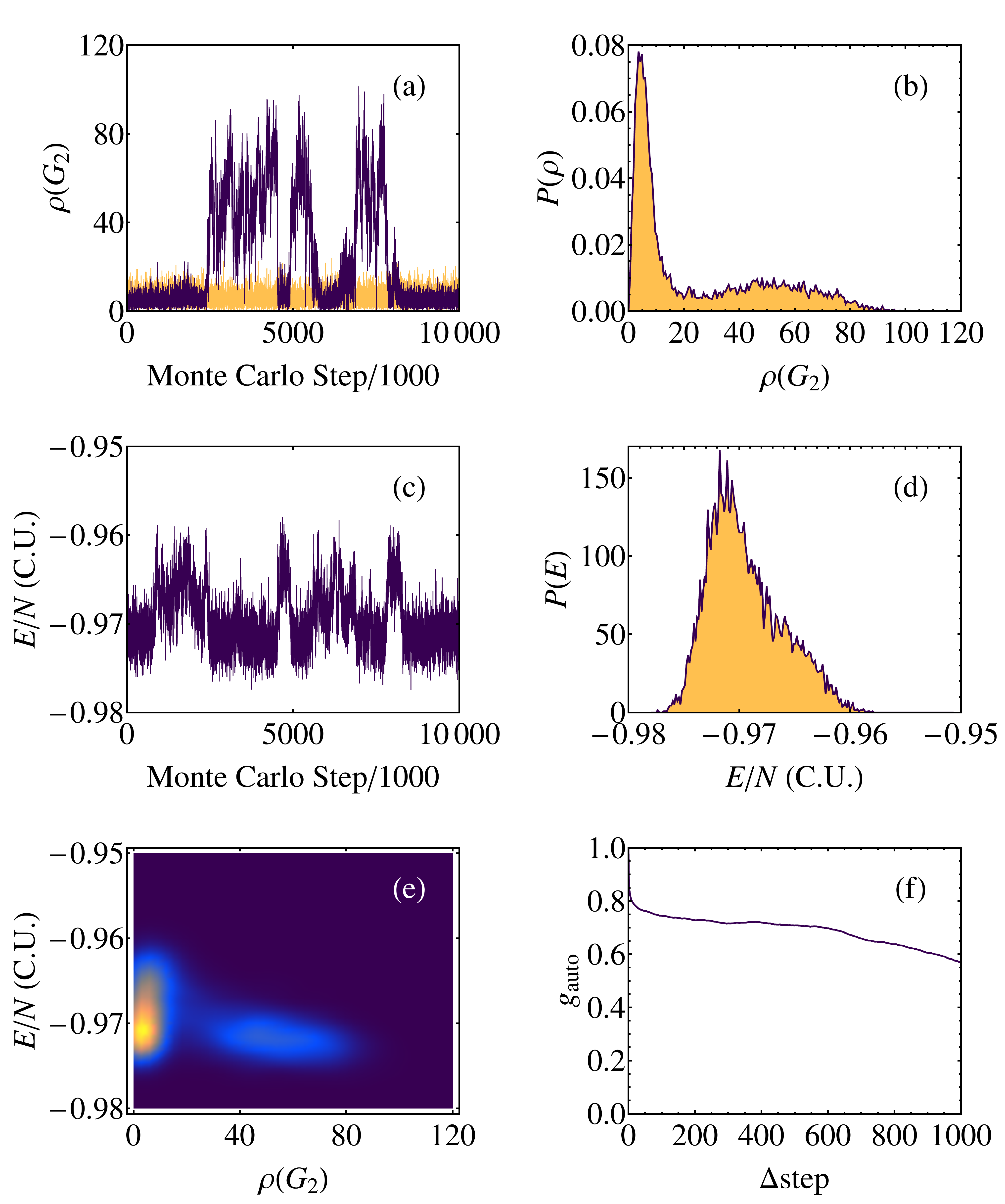}
\end{centering}
\caption{(Color online) An example of a standard Monte Carlo
  simulation for $\alpha_{B}(T)^{2}=91$, for a system of $16\times16$
  vortices. In (a) we show both a measure of the density-correlation
  (purple) and average density (yellow) at the point $G_{2}$ -- the
  second smallest $k$ vector in the reciprocal crystal lattice -- as a
  function of the Monte Carlo time. To the right of this (b) we show
  the probability distribution for the density-density correlation at
  this point.  On the second row we have similar plots for the energy
  (c) and (d).  The final row shows a density plot of $E$
  vs. $\rho(G_{2})$ (e), and to the right of this, (f), we show the
  energy auto-correlation function as a function of the number of
  Monte Carlo steps.\label{fig:MC_16_16_91}}
\end{figure}

\section{Monte Carlo Techniques}
\label{MonteCarlo}
As discussed in great detail in Appendix \ref{app:Monte-Carlo-techniques},
\ref{app:Monte-Carlo-techniques}, we have studied the application of
three different Monte Carlo techniques: The standard canonical
Metropolis algorithm, and two microcanonical ones (sometimes referred
to as ``broad-histogram techniques''), the Wang-Landau and the optimal
energy diffusion algorithm.

The reason we have not persisted with the use of the standard Metropolis
sampling of the Boltzmann factor, is that, as shown below, there appear
to be serious difficulties with obtaining reproducible results. This
points to a long time scale associated with diffusion through the
abstract configuration space, especially at certain temperatures. Such
behavior is commonly associated with a phase transition.

To really study the detailed behavior in the interesting temperature
range, one might use parallel tempered Monte Carlo
\cite{neuhaus_efficient_2007}. In such simulations one often reverts
to ``reweighting'' the results to a single temperature to get the
temperature dependent probability distribution $P(E,T)$. We follow a
different route that gives us more direct access to the underlying
density of state.

As we shall show below, most of the relevant information is actually contained
in the microcanonical density of states $g(E)$, and especially the  derivative
of its logarithm,
\begin{equation}
  S_\text{mc}(E)=\frac{d}{dE} \ln g(E).
\end{equation}
There is a very powerful algorithm that directly determines $g(E)$,
the Wang-Landau algorithm \cite{wang2001determining}.  As discussed in  Appendix~\ref{app:Monte-Carlo-techniques}, it is very
different from a standard Monte Carlo algorithm, since the simulation
weights are being updated as the simulation progresses, finally
converging to the density of states. 

As argued cogently in Ref.~\onlinecite{wu2005overcoming}, this algorithm can
run into difficulties if there are barriers to the Monte Carlo process,
modeled as a random walk
as a function of energy.  In the problem we are considering these
barriers appear to be substantial -- which is closely linked to the
apparent first order nature of the phase transition for smaller system
sizes. Also, it is not very simple to write parallel versions of the
Wang-Landau algorithm.

The approach from Ref.~\onlinecite{trebst2004optimizing}, where we optimize
the current in the random walk between the extremes in energy, seems
to give the best of both worlds. It is microcanonical, and the
diffusion becomes optimal as the simulation progresses. As a
by-product we obtain an understanding where barriers to energy
diffusion are -- and there are clear links between the barriers and the
associated phase structure.

As will be shown below, the optimal energy diffusion algorithm proved
the most illuminating for the problem studied here.

\section{Results}
\label{Results}

In the results reported in this section, 
we use a unit of energy where the crystalline state for $N$ vortices
has a natural value $E=-N$; this means that we scale the energy as
defined in Eqs.~(\ref{eq:E},\ref{eq:E2E4}) by twice the Abrikosov
parameter (\ref{eq:betaA}), i.e., we multiply these expressions by 
$2\beta_A$.  Expressed in these units, the crystalline state thus has
energy $E(\{c_{\text{crystal}}\})/N=-1$.  Such units will be denoted
as ``C.U.''  below.

Our first set of simulations uses the classical Metropolis Monte Carlo
algorithm without optimizations.  We obtain results of similar quality
to those which can be found in the literature for lattice sizes of up to about 14 by 14 --
we find that around that point results have rather limited
reproducibility with a sensible length of simulation.  A typical
example of a set of simulations just beyond that size is shown in
Fig.~\ref{fig:MC_16_16_91}. Here we analyze a system with $256$
vortices (boundary conditions are chosen such that the lowest energy
state is a triangular lattice of 16 rows of 16 vortices each).  In
that figure we show both energy and the density fluctuations at a
suitably chosen point of the triangular lattice, for
$\alpha_{B}(T)^{2}=91$, which is at a temperature where we see phase-coexistence. We can clearly see the density fluctuations
suggestive of a two-phase system; the energy seems to behave
similarly, even though the correlation between energy and density on
the final row shows a more complicated picture. Nevertheless, it is
entirely possible to pick out two populations.  Finally the energy
auto-correlation of the Monte Carlo process shows fast decay for a few
steps, and then little or no decay.  We were very concerned when we
first saw this behavior, since it seems to invalidate the
simulations. A simple statistical model with $P(E)$ the sum of two
disjoint Gaussian probability distributions shows exactly this
behavior, so it may be that energy autocorrelation is not a very good
quality measure in an area of phase coexistence.  The density plot of
energy versus $\rho(G_{2})$ lends support to such a model.

\begin{figure}
\begin{centering}
\includegraphics[width=7cm]{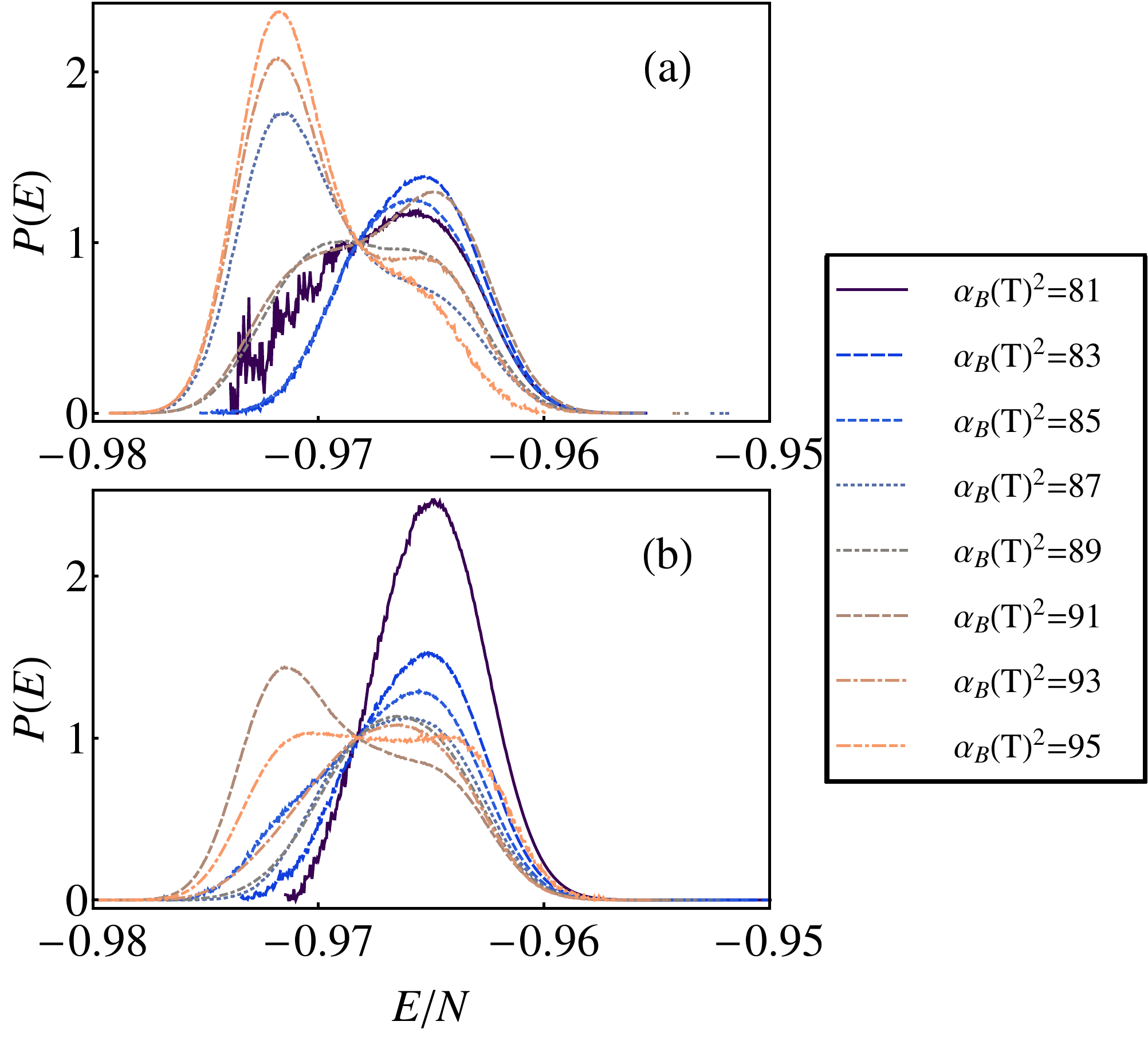}
\end{centering}
\caption{(Color online) The $P(E)$ curves for the $16\times16$ system,
  rescaled to a single value of $\alpha_{B}(T)^{2}=90.1$. The curves are
  normalized to cross at a single point, $E/N=-0.968\ \text{C.U.}$,
  which is roughly in the middle of the two peaks. Panel (a) shows results
  for simulations with decreasing $\alpha_{B}(T)^{2}$; panel (b) for
  increasing values of this parameter.  \label{fig:rescaled}}
\end{figure}

The fact that we have two distinct populations means that we will find
it difficult to simulate reliably near this temperature, as is shown
in Fig.~\ref{fig:rescaled}. There we show results from two groups of
simulations. In each series of simulation we use the final
configuration from a simulations for a previous value of
$\alpha_{B}(T)$ to start the next simulation for a different coupling
constant. In one group we increase the values of $\alpha_{B}(T)^{2}$
(starting from an initial random configuration), in the other we
decrease $\alpha_{B}(T)^{2}$ (starting from an initial crystalline
configuration). We extract an energy histogram from the Metropolis
simulation, and reweigh these histograms to correspond to a single
coupling constant,
\begin{equation}
n_{\text{RW}}(E)=\mathcal{N} \exp\left(-\left(\alpha_{B}(T_{0})^{2}-\alpha_{B}(T)^{2}\right)E\right)\, n(E).
\end{equation}
We choose the normalization constant $\mathcal{N}$ so that the curves
have the same height at one energy. We see a clear indication of two
peaks, but the peak-heights are rather different -- thus showing no real
agreement between the simulations. Even though we can run the
simulations for longer, the very long time scales from the
autocorrelation function make convergence rather doubtful. This calls
for an improved approach.

\begin{figure}
\begin{centering}
\includegraphics[width=7cm]{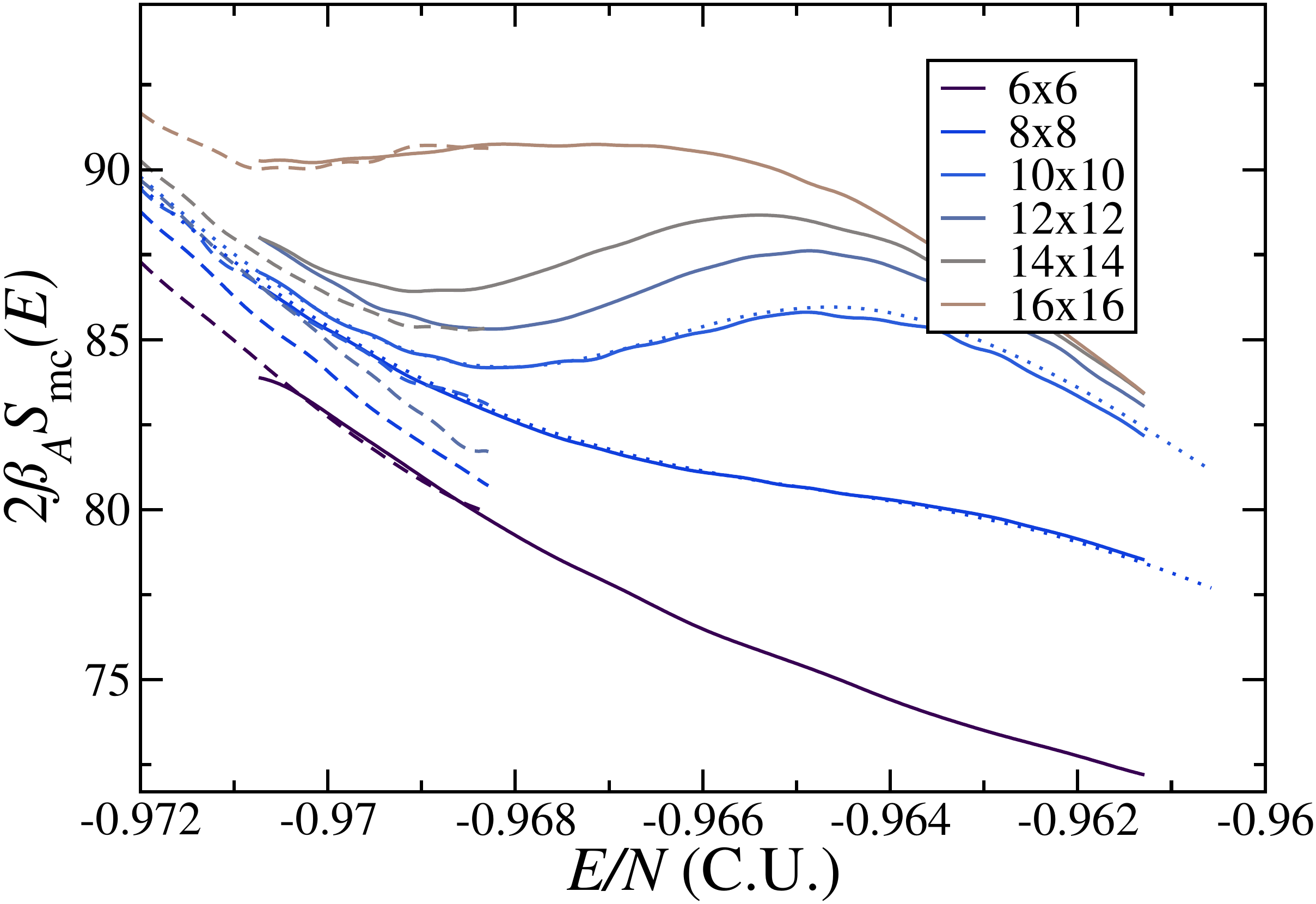}
\end{centering}
\caption{(Color online) The micro-canonical entropy as as a function
  of the energy in the system, obtained using the Wang-Landau
  method. The dashed, and solid curves show simulations in two different energy intervals; The dotted curves are simulations over a much larger energy interval.
\label{fig:The-coupling-strengthWL}}
\end{figure}

We thus investigate the use broad histogram techniques; we first apply
the one most widely reported in the literature, the Wang-Landau
technique \cite{wang2001determining}, which we have tested in quite a few
variants.  We find that the best way to represent the results is by
using the micro-canonical entropy $d\ln g/dE$ versus $E$. From
\begin{equation}
\frac{d}{dE}P_{\beta}(E)=\frac{d}{dE}\left(g(E)e^{-\beta E}\right)=0,
\end{equation}
we see that for an energy satisfying
\begin{equation}
2\beta_{A}d\ln g(E)/dE=\alpha_{B}(T)^{2}
\end{equation}
we have an extremum of the probability density. When there is one
solution this is the maximum; where there are multiple (as well shall see below,
that usually means three) solutions we may have
a first-order phase transition. Numerically the quality of the calculations can be
slightly questionable, since we need to take a numerical derivative of
the density of states. This latter quantity is obtained through a
Monte Carlo technique, and thus has associated statistical errors,
which are considerably amplified when taking derivatives, see the
discussion in the Appendix.  
This means that we must impose a very high threshold to the convergence of
the algorithm. As a consequence, we find questionable
quality of the Wang-Landau results for system sizes beyond about
16 by 16, almost independent of the variant of the procedure we have
applied, see Fig.~\ref{fig:The-coupling-strengthWL}. In that figure we
see that we get good (but not perfect) overlap between the results in
the various energy regions we have split the simulation in; we also
see that $16\times16$ is much flatter than all the other simulations,
but needs better data to draw reliable conclusions.

The Wang-Landau procedure is different from a normal Monte Carlo procedure,
since the acceptance criteria (i.e., the weights) change during a
simulation. This means that it is difficult to run such simulations
in parallel. Even if we could, the apparent long time scales associated
with equilibration and phase coexistence suggest that we would rather
use a method that only occasionally changes the Monte Carlo weights,
such as the optimal energy diffusion (OED) algorithm we discuss
in the previous section and in Appendix \ref{app:Monte-Carlo-techniques}.

\paragraph{Optimal Energy Diffusion results}

\begin{figure}
\begin{center}
\includegraphics[width=4cm]{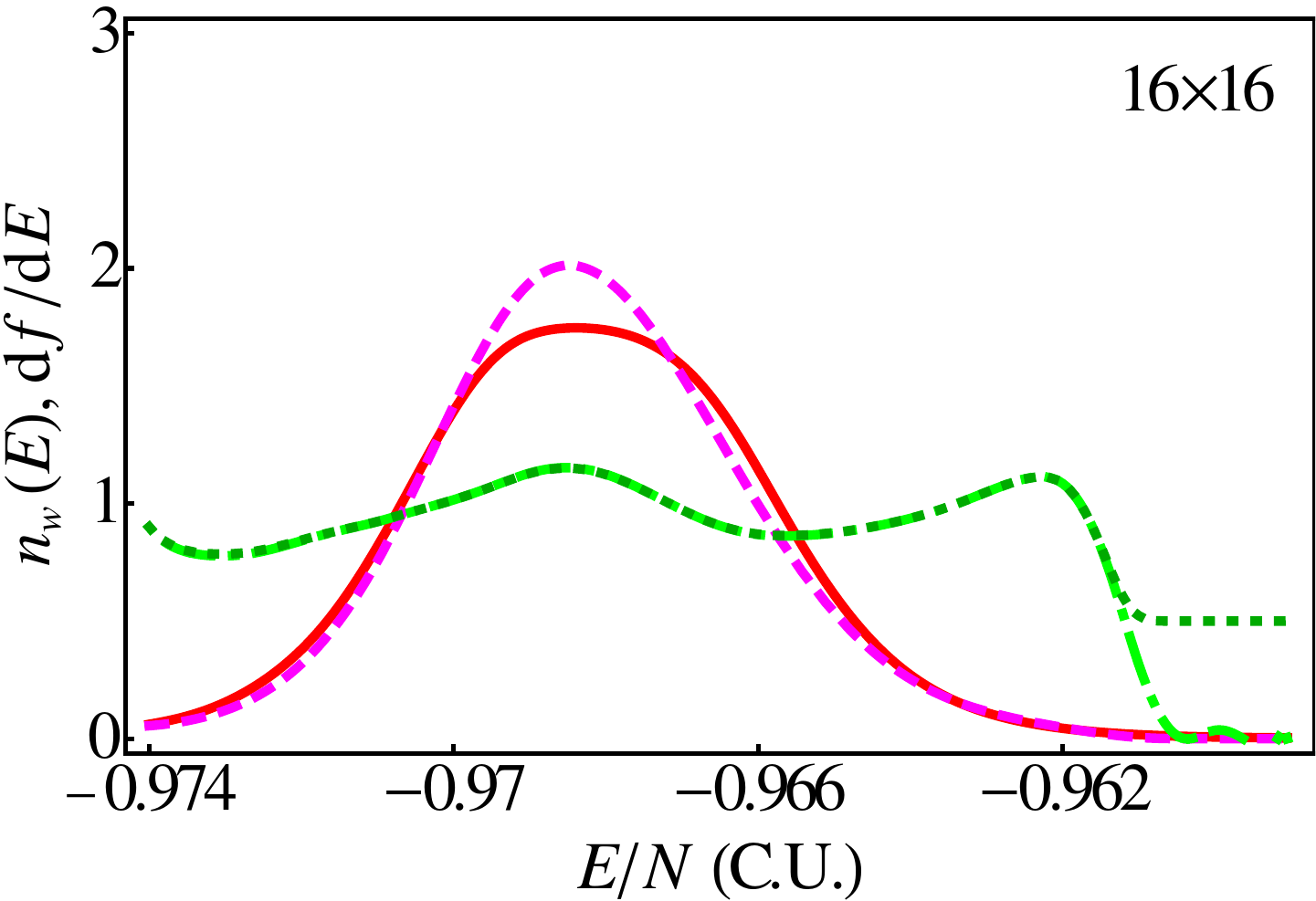}\includegraphics[width=4cm]{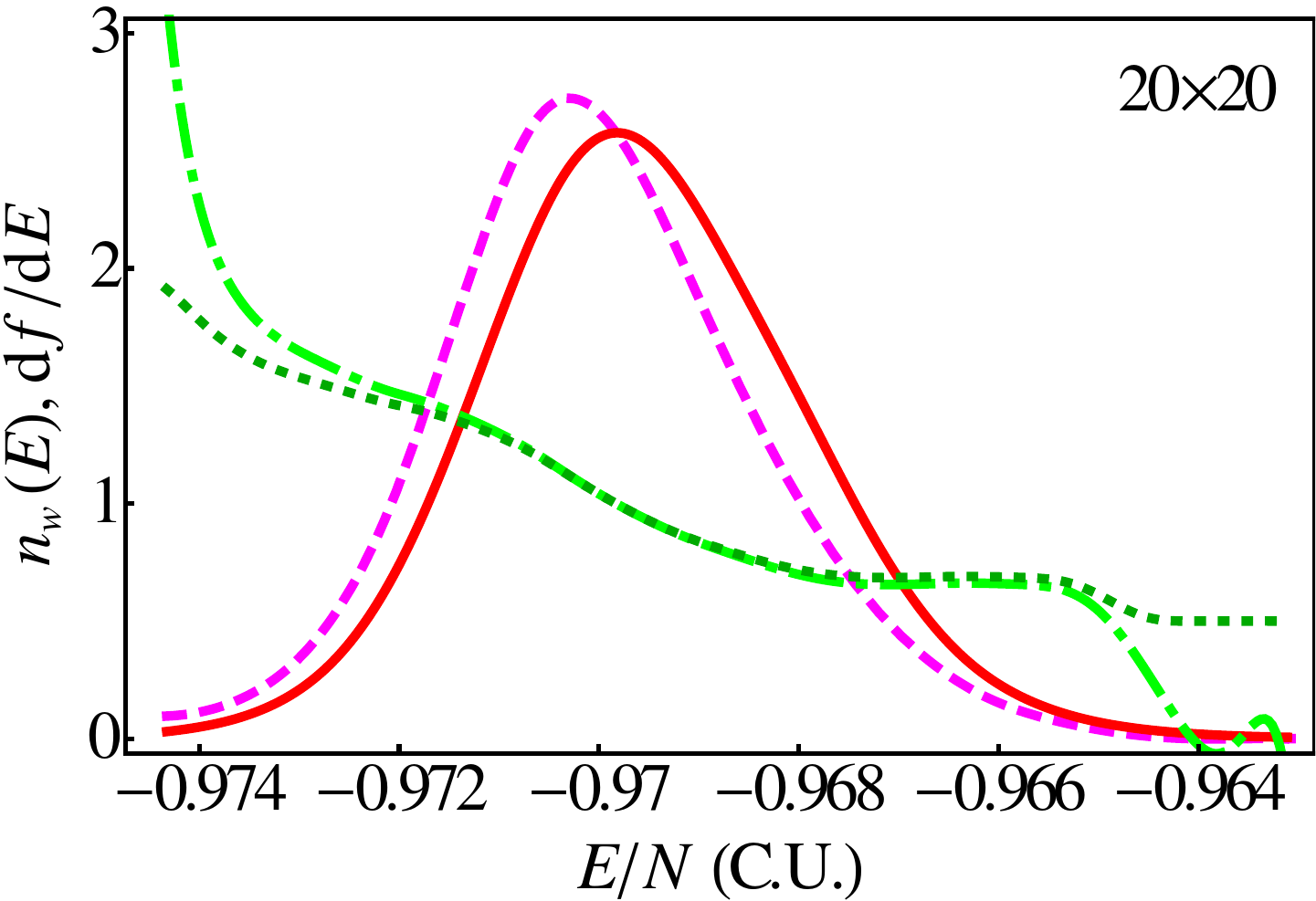}\\
\includegraphics[width=4cm]{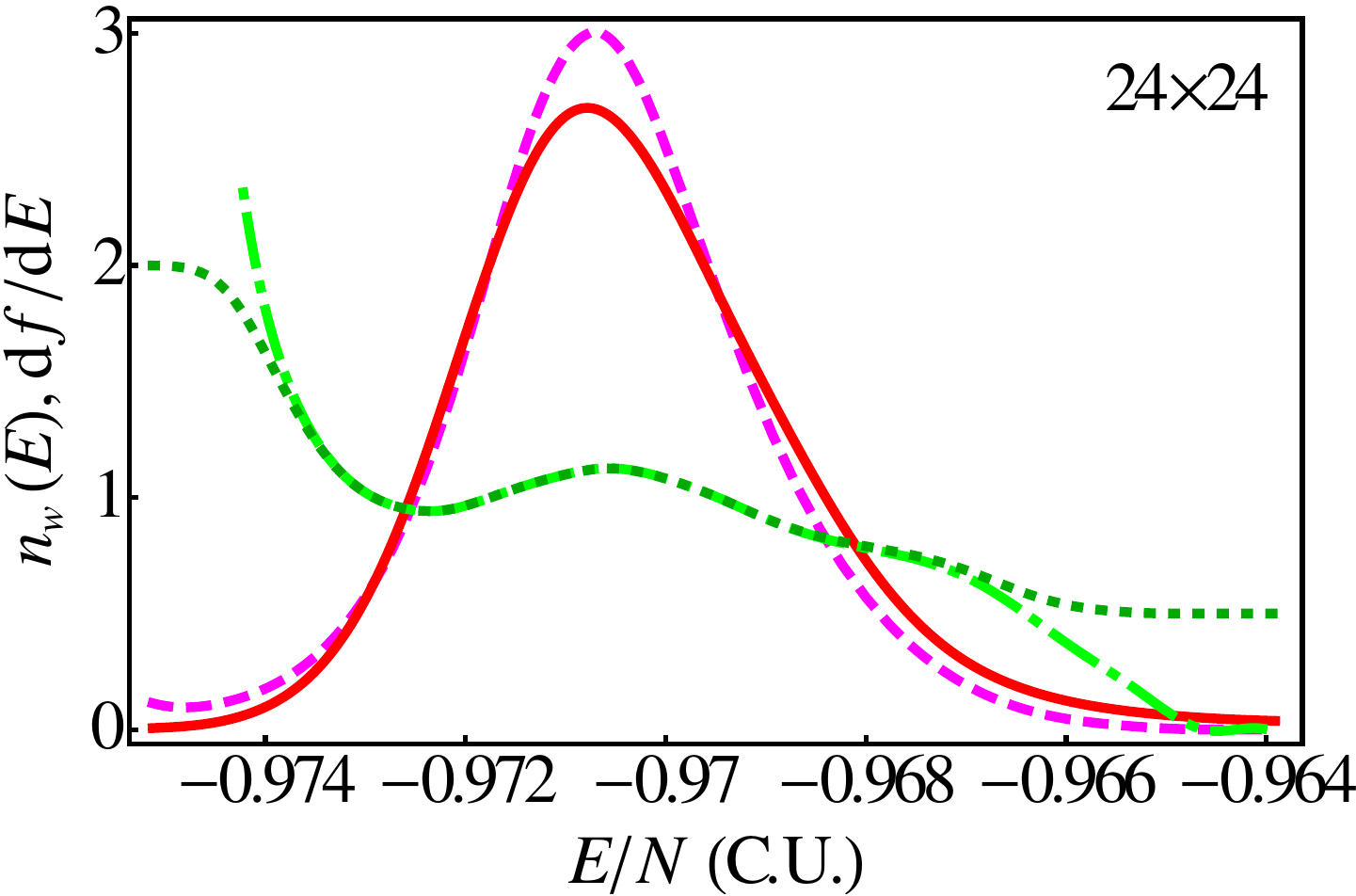}\includegraphics[width=4cm]{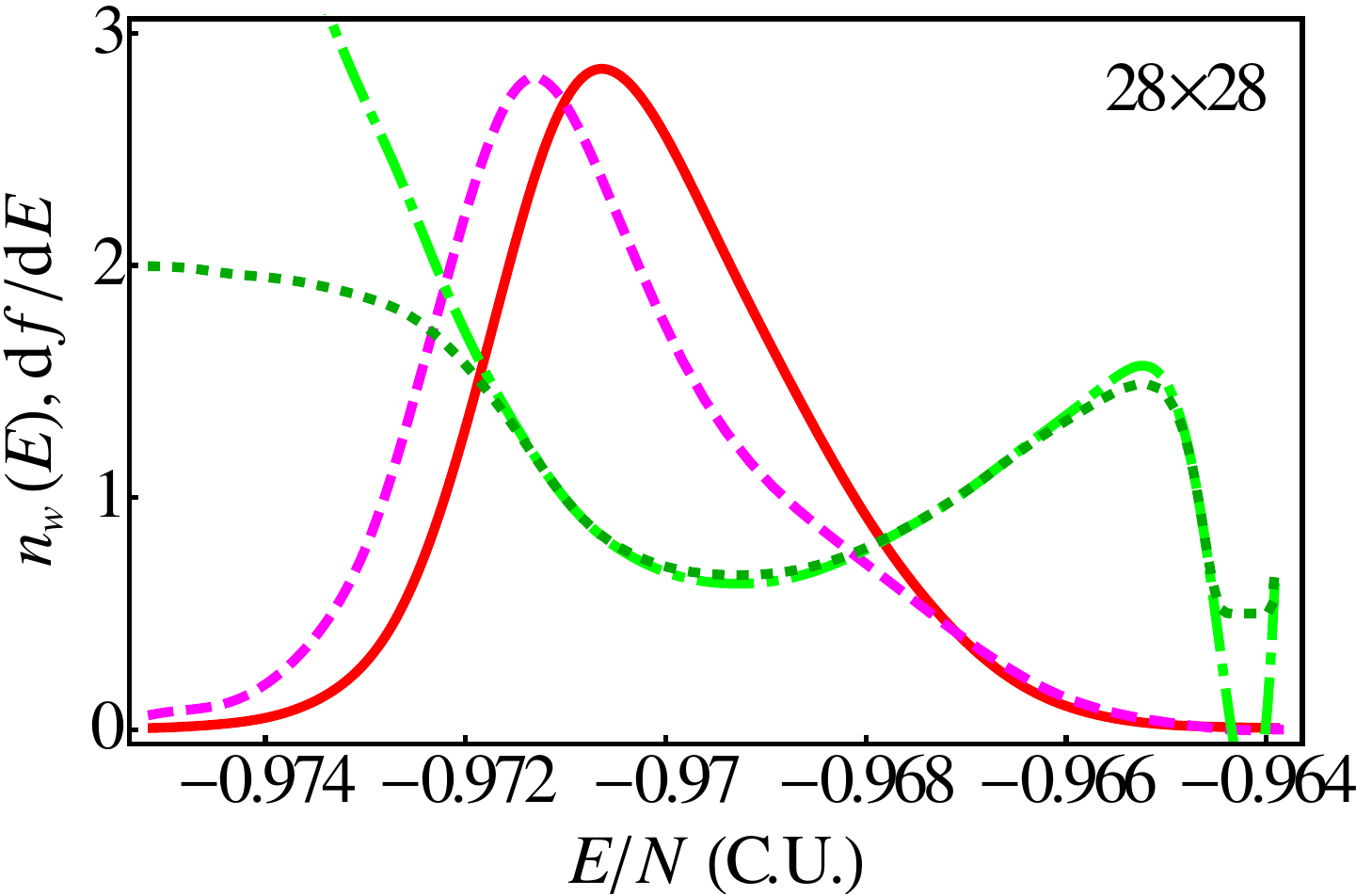}\\
\includegraphics[width=4cm]{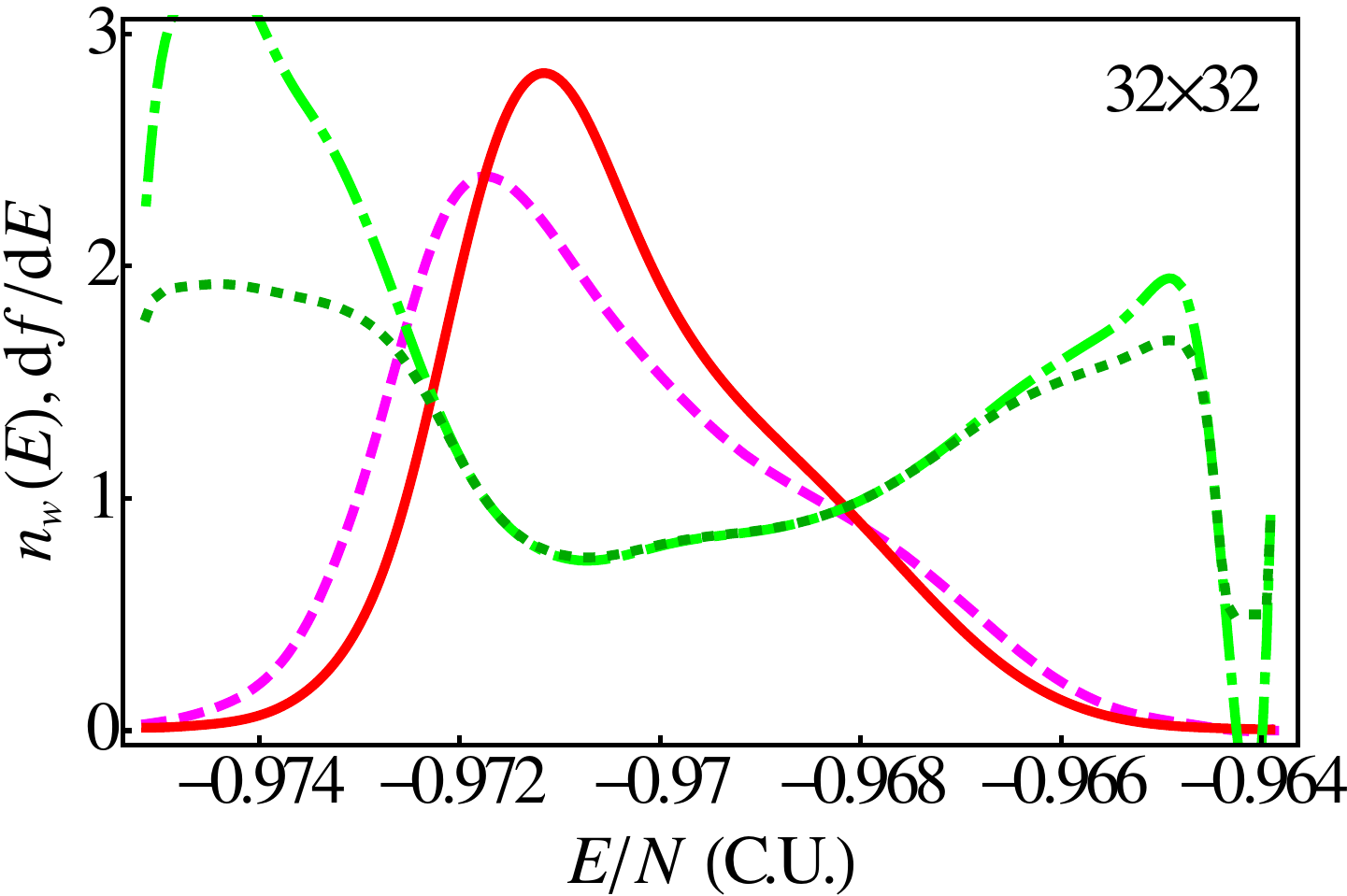}\includegraphics[width=4cm]{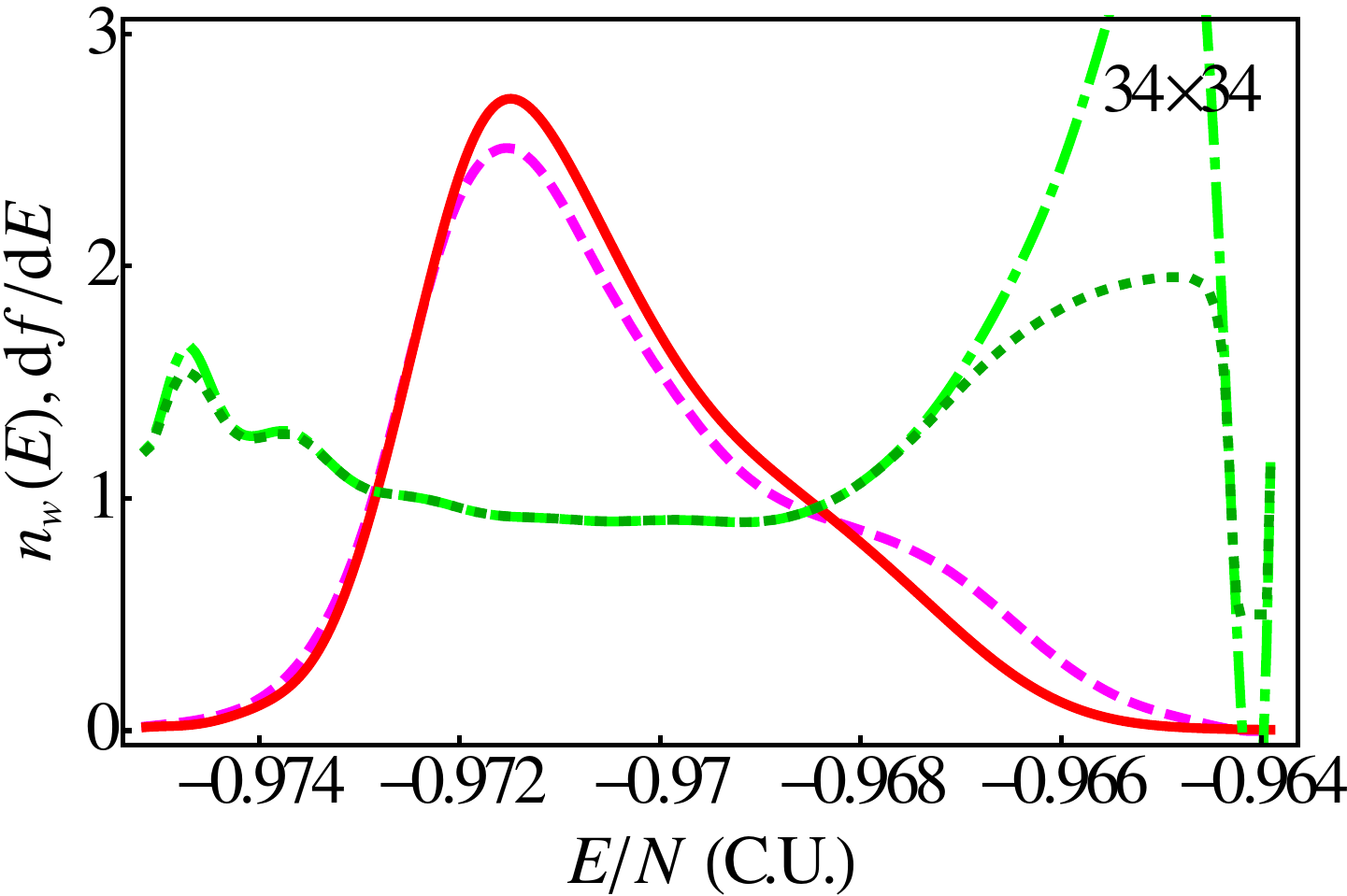}
\end{center}
\caption{(Color online) Some typical results from optimal energy
  diffusion simulations.  The plots differ in the value of
  $N=N_x\times N_y$ displayed in the top right-hand corner.  In each
  plot, the solid red curve is a normalized probability histogram for
  walkers ($n_{w}(E)$), the dashed purple curve is (a smoothed
  version) of the derivative of positive direction walker density,
  $df/dE$.  The light green curve shows the ratio between the red and
  purple curves, which should be one on convergence; the dark green
  curve gives a smoothed version of this ratio as described in the text,
  see Appendix \ref{app:Monte-Carlo-techniques} for the details. The
  difference between largest and smallest values of the red and purple
  curves is about 200 for the largest system size
  shown. \label{fig:TTtypical}}
\end{figure}

We show a typical result from one iteration of the OED algorithm in our
implementation in Fig.~\ref{fig:TTtypical}. We see that we obtain
reasonable convergence for the weights, and seem to have optimized
the current well, since the two relevant curves ($n_{w}(E)$ and $df/dE$,
see the discussion around Eq.~(\ref{eq:nw}) for details)
almost coincide. The peak in these curves shows a pronounced minimum
in energy diffusivity (typically the difference between largest and
smallest values grows with system size, and is about 200 for the largest
system size shown). As we go to larger system sizes, there seems to
be a trend that is highly suggestive of the development of two peaks
in the inverse diffusivity. If we link such peaks to features in the
energy landscape that give rise to phase transitions, this might be
the first indication of the pair of transitions predicted in the KTHNY
picture.

\begin{figure}
\begin{center}
\includegraphics[width=7cm]{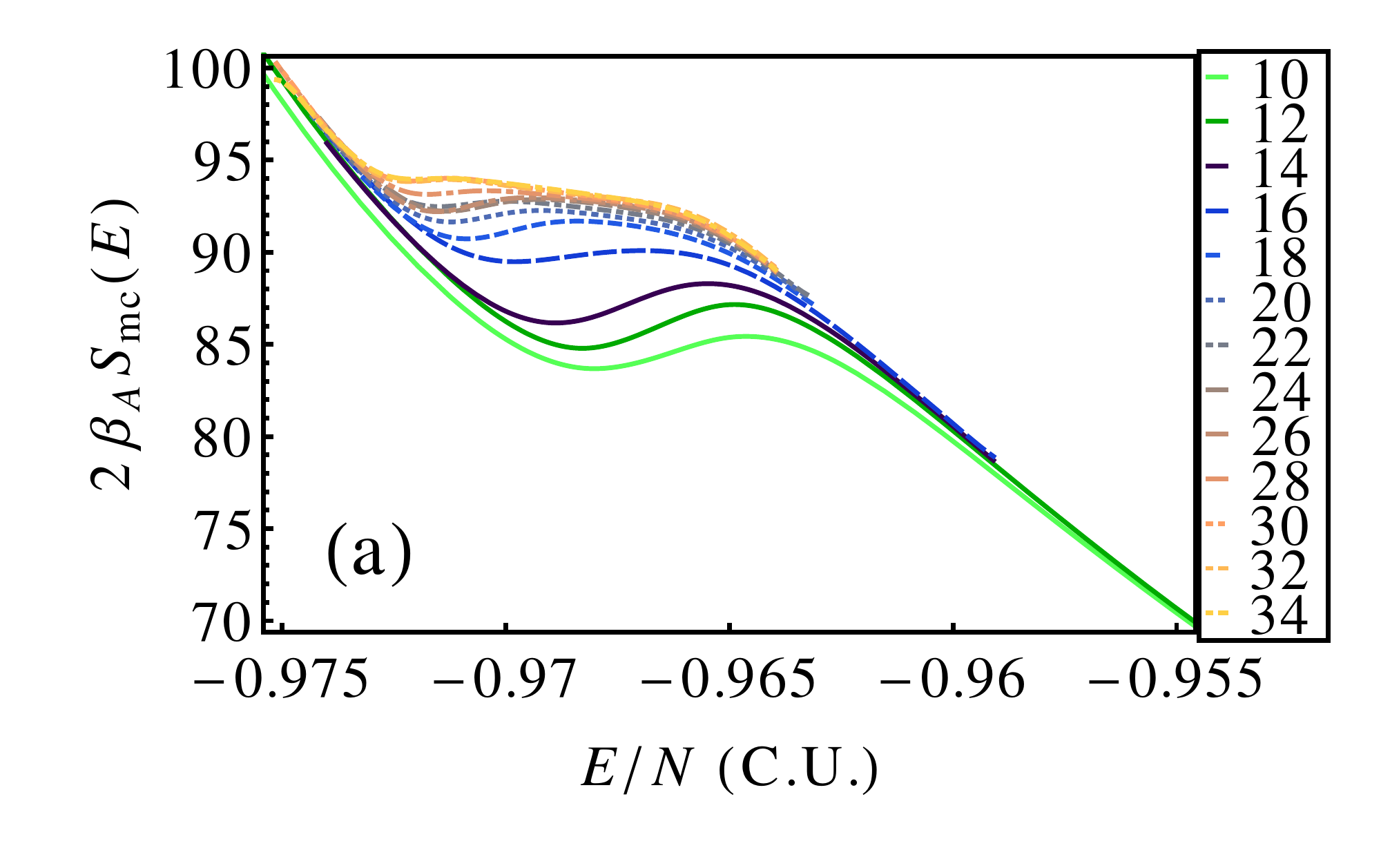}\\
\includegraphics[width=7cm]{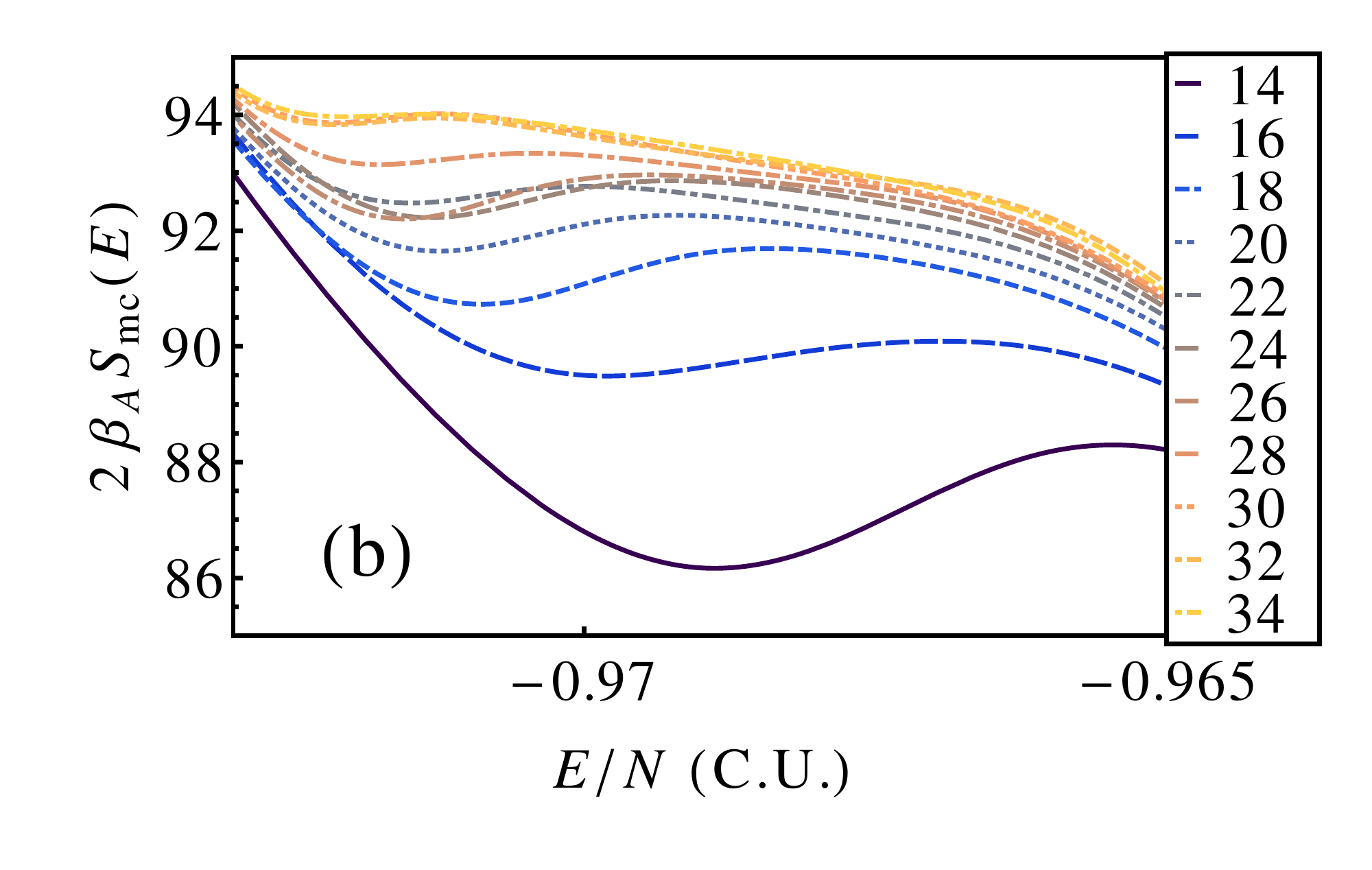}
\end{center}
\caption{(Color online) The microcanonical entropy (the derivative of
  the logarithm of the density of states), as a function of the energy
  in the system. The values on the vertical axis have been scaled by
  $2\beta_{A}$ so that their values equal the value of
  $\alpha_{B}(T)^{2}$ that makes the probability density peak at the
  energy for each point. The label of each curve is $\sqrt{N}$. 
  Figures (a) and (b) show the same data on different scales
  \label{fig:The-microcanonical-entropy}}
\end{figure}

We show the microcanonical entropy $S_{\text{mc}}(E)\equiv\frac{d}{dE}\ln g(E)$
for all simulations we have performed in Figs.~\ref{fig:The-microcanonical-entropy}.
The advantage of that quantity, when expressed in natural energy units by
multiplication of the energy by  $2\beta_{A}$, is that it allows us to read off
the value of $\alpha_{B}(T)^{2}$ that gives us the peak (or, in a
few cases peaks) in the canonical probability at that energy directly
from the graph. Having one intersection corresponds to  a single peak of $P_{\beta}(E)$, and three
intersections corresponds to  two peaks and a minimum in between. As we can
see the separation and depth of the peaks moves rapidly together -- actually
at the largest values of $\sqrt{N}$ it is debatable whether we even have
three intersections (see also Fig.~\ref{fig:Maxwell} below).

\begin{figure}
\begin{centering}
\includegraphics[width=6cm]{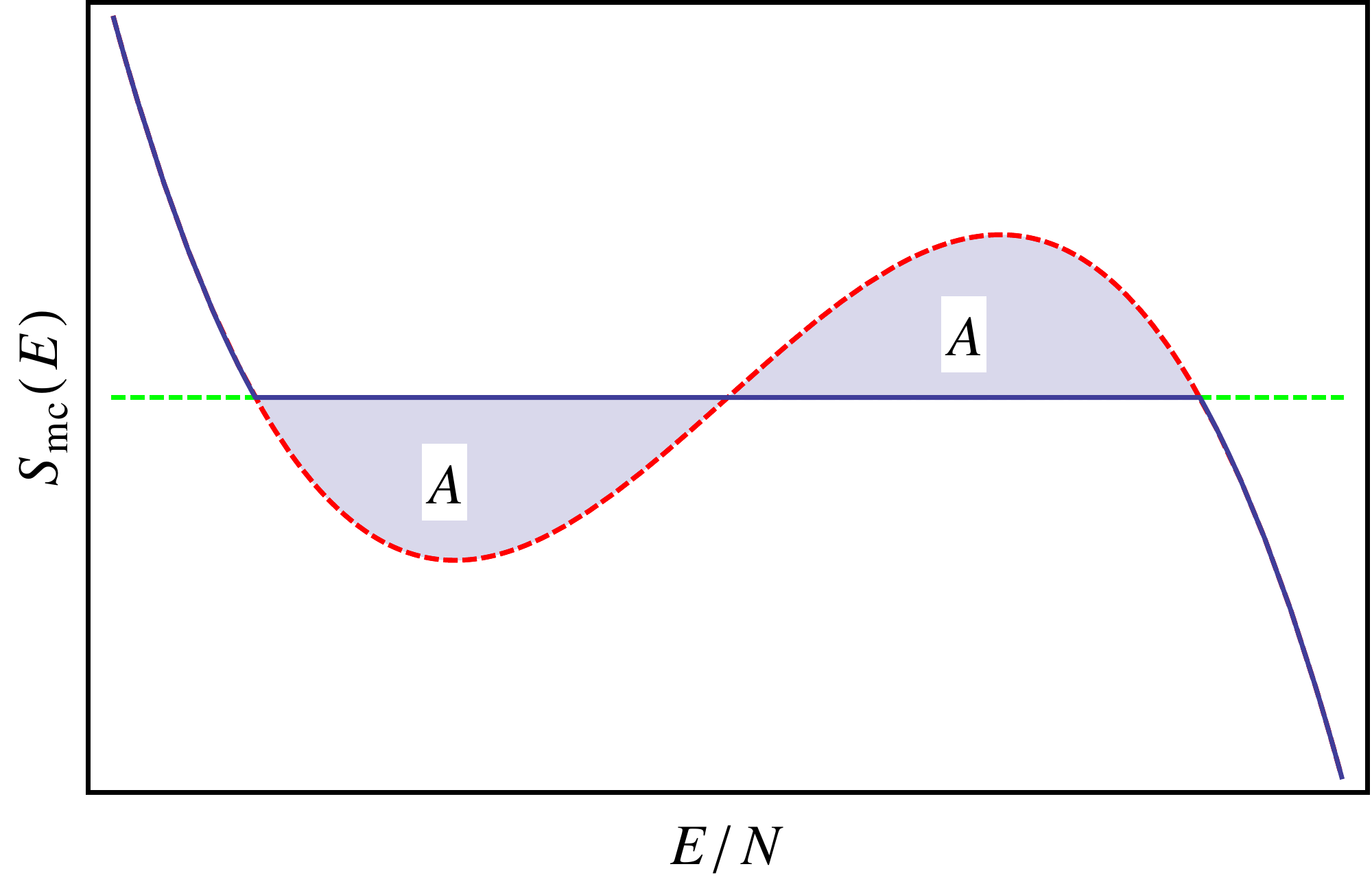}
\end{centering}

\caption{(Color online) The Maxwell construction for the
  microcanonical entropy. The red dashed line shows a typical finite
  size result, that is not convex. The best model of the large $N$
  limit is obtained by a Maxwell construction, where the two parts cut
  from the entropy curve are of equal area $A$, and we thus get the
  solid blue curve as the best possible equation of state. The size of
  the cut areas, $A$, will naturally scale with $1/\sqrt{N}$ [and thus
  scales with $\sqrt N$ as a function of $E$], and provides us with a
  measure of the interface free
  energy. \label{fig:MaxwellConstruction} }
\end{figure}

Of course in the thermodynamic limit we cannot have multiple
intersections.  Since $S_{\text{mc}}$ is convex, the best
approximation to the thermodynamic limit is obtained by a Maxwell
construction, Fig.~\ref{fig:MaxwellConstruction}. Here we draw a
straight line across the dip region, such that the area between the
curves between the first to second intersection equals that between
the second to third intersection.  This area, suitably scaled, is
actually just the interface free energy, which we expect to go to a
non-zero constant  if there is 
a first order transition. Since we plot $S_\text{mc}$ as a function of $E/N$, this
would mean that the area on the graph scales like $1/\sqrt{N}$, and
thus the interface free energy scaled with the interface length scales
as $F_{S}/\sqrt{N}=A$, with $A>0$.

\paragraph{Interface free energy}

\begin{figure}
\begin{centering}
\includegraphics[width=6cm]{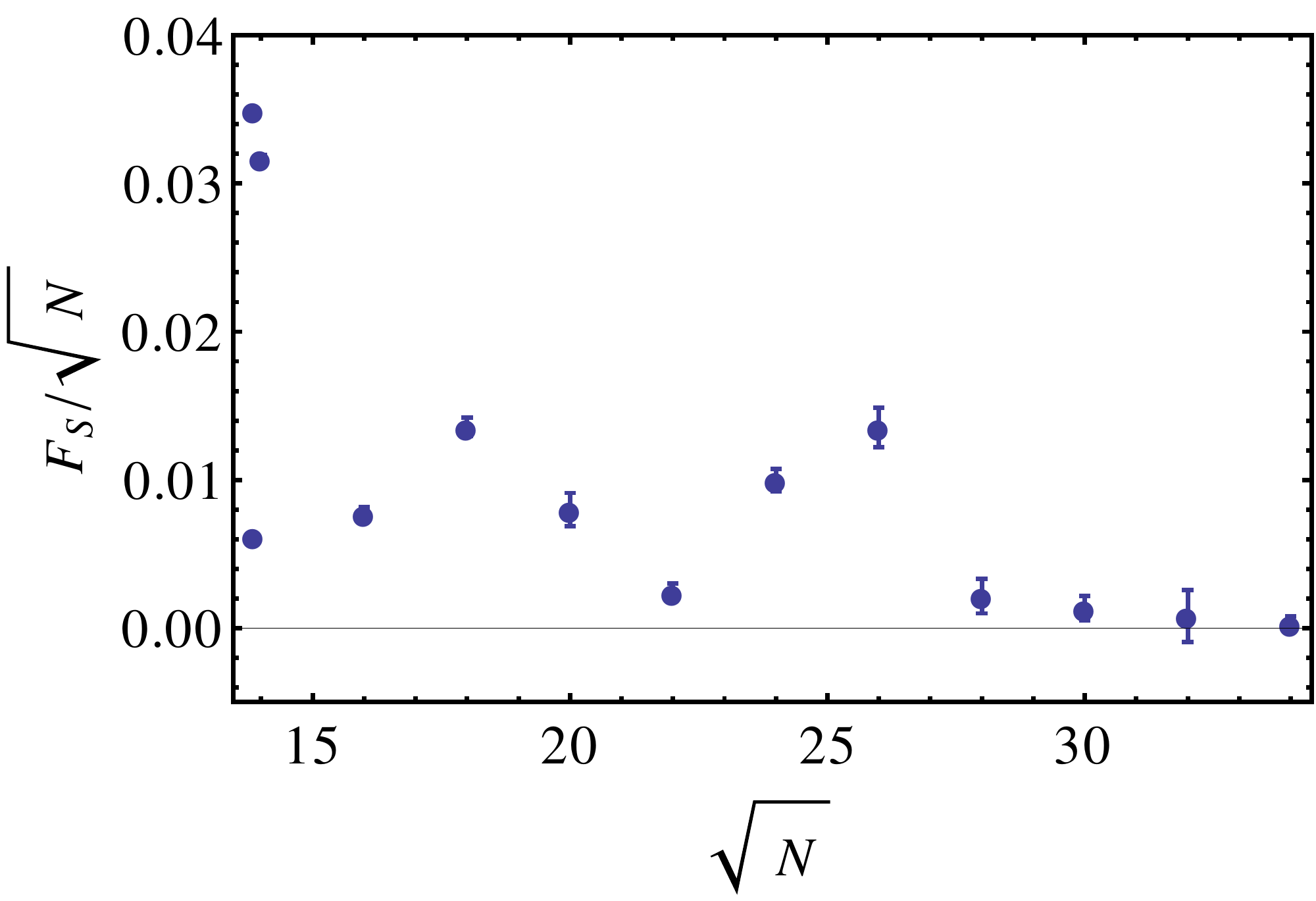}
\end{centering}
\caption{(Color online) Interface free energy per unit length as a
  function of the size of the system. The two additional high-lying
  points near $\sqrt{N}=14$ are $N_{x}\times N_{y}=16\times12$ and
  $12\times16$.\label{fig:Interface-free-energy} }
\end{figure}

The behavior we find is quite different -- we seem to be unable to
extract a reliable trend for the interface free energy from the
results as shown in Fig.~\ref{fig:Interface-free-energy}.  The results
for the largest systems seem to suggest that the interface free-energy
$A$ converges to zero, compatible with a total absence of a first
order phase transition for the larger system sizes, but with
substantial oscillatory behavior before that limit is reached.  So the
question is now: is there any evidence for a phase transition at all? The
results shown in Fig.~\ref{fig:TTtypical} suggest that there is one,
and reasonably likely a second, bottleneck in energy
diffusivity. Studies of the technique applied to spin models have
shown that such results are usually linked with a phase transition --
in most cases continuous transitions associated with diverging
correlation lengths and their associated diverging timescales. We
shall now argue that there is a clear indication that we have one and
possibly two continuous phase transitions in the narrow energy range
studied in Fig.~\ref{fig:The-microcanonical-entropy}.

\paragraph{Shear modulus and $C_{V}$}

\begin{figure}
\begin{center}
\includegraphics[width=7cm]{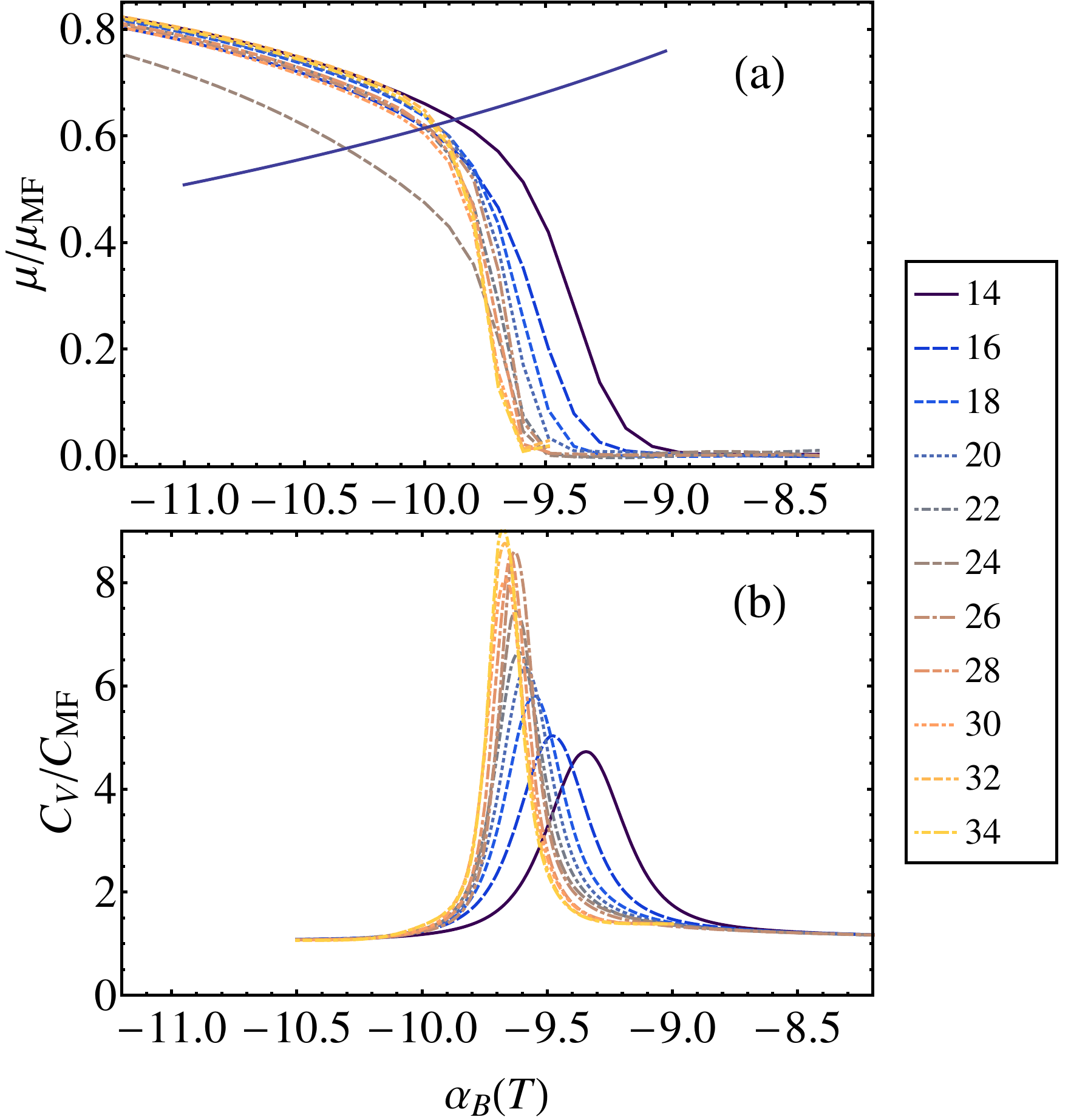}
\end{center}

\caption{(Color online) (a) The value of the shear modulus
$\mu$ divided by its mean-field value as a function of $\alpha_{B}(T)$
in the canonical ensemble. The solid blue line is the KT critical
line, see Eq.~(9) in Ref.~\onlinecite{vsavsik1994calculation}. The lowest $\alpha_B(T)$ results rely on an extrapolation of the
data, partially using the crystalline density of state [Appendix
\ref{app:cryst}] (b) Specific heat divided by its mean-field value as
a function of $\alpha_{B}(T)$ in the canonical
ensemble. As usual \cite{rosenstein2010ginzburglandau} we make the approximation that all other temperature dependencies are replaced by $T_c^{\text{MF}}$.\label{fig:Left:Shear_and_Cv}}
\end{figure}

\begin{figure}
\begin{center}
\includegraphics[width=6cm]{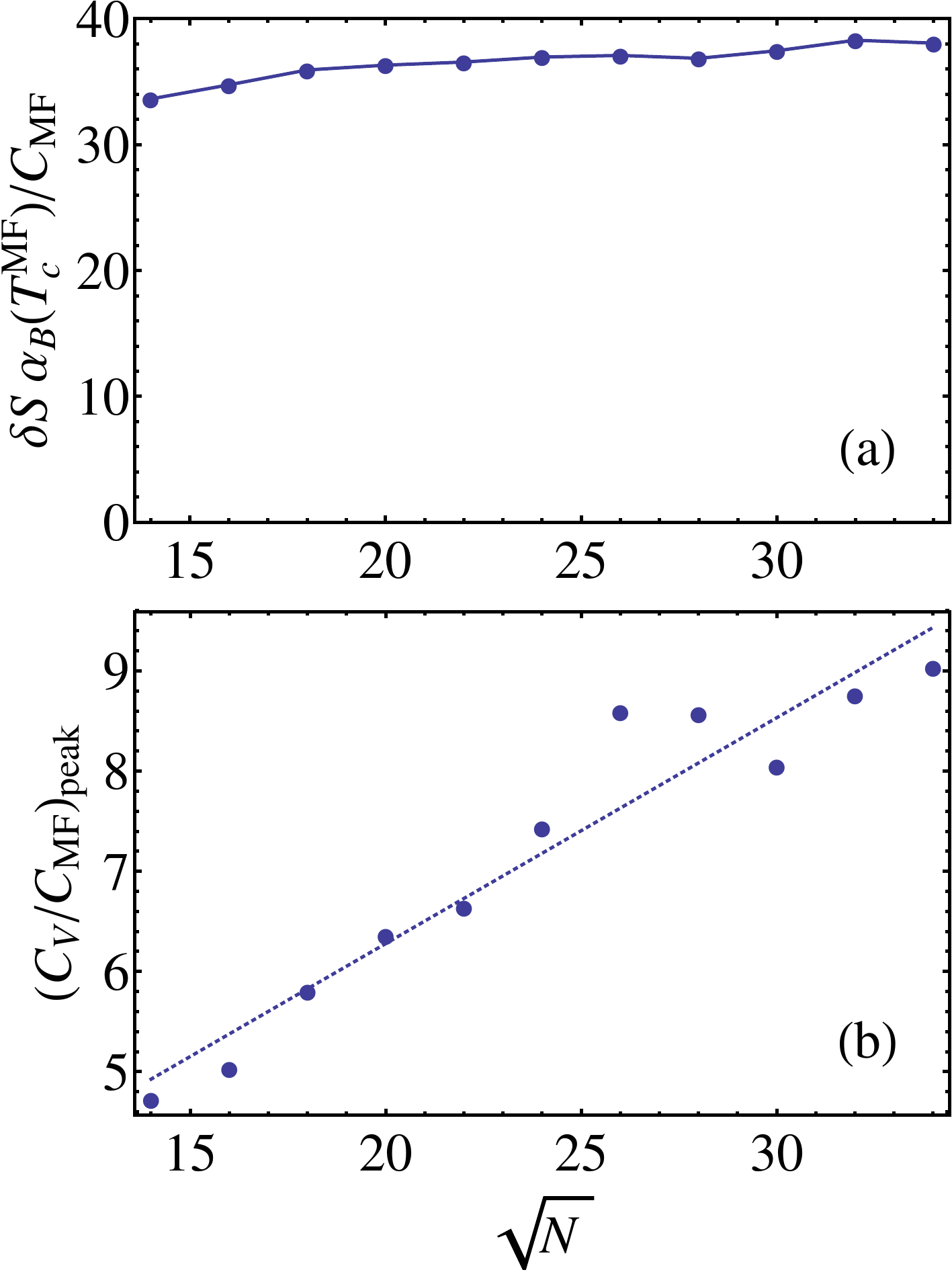}
\end{center}
\caption{(Color online)  Entropy  (a) and peak height (b) contained in the ``peak'' of
  the specific heat relative to a mean-field crystal
  background.  As usual \cite{rosenstein2010ginzburglandau} we make the approximation that all other temperature dependencies apart from $\alpha_B(T)$ are replaced by $T_c^{\text{MF}}$.
  \label{fig:Peak-height-and}}
\end{figure}

The natural alternative  to a first order phase  transition is
the    KTHNY    \cite{kosterlitz1973ordering,    halperin_theory_1978,
  nelson_dislocation-mediated_1979,  young1979melting}  scenario.  The
normal  liquid  to  hexatic  transition  is  supposed  to  be  in  the
universality class  of the Kosterlitz-Thouless (KT)  transition of the
XY model and is associated  with the binding of the free disclinations. 
In order to show a comparison with the classical picture of such transitions,
we have used our results, reported later in this paper, to obtain
 the   specific  heat   in   the   canonical  ensemble,   see
Fig.~\ref{fig:Left:Shear_and_Cv}. There is  a peak in  the specific
heat which seems  to grow with the number of  vortices. This is rather
similar     to     what    can     be     seen     for    the     XY
model\cite{vanhimbergen1981helicity}, but from  the data it is unclear
whether the peak in the specific heat saturates with $N$ -- as it does
for  the  XY model  --  or  continues to  grow  (as  it  would if  the
transition    were    first    order    as   discussed    in,    e.g.,
Refs.~\onlinecite{janke1990modelsof,bittner2005natureof}.   Another striking
feature of  the specific heat is  that the region  associated with the
peak is small  and there seem to be no noticeable  precursors of it on
the high-temperature  side of  the transition. It  seems to  spring up
from nowhere.  The critical region in $\alpha_B(T)$
associated with the  hexatic transition seems to be  very narrow.  The
entropy  in   this  peak,  obtained  by   subtracting  the  mean-field
contribution  and  integrating over  $\alpha_B(T)$  seems  to be  almost
independent     of     system    size     and     is    plotted     in
Fig.~\ref{fig:Peak-height-and}a. We calculate the entropy in the approximation
\begin{equation}\delta S=\int_{T_0}^{T_1} (C_V-C_{MF})/\alpha_B(T_c^\text{MF}) d\alpha_B(T).\end{equation}

The value  of the specific heat  at its peak  shows conflicting trends
(see Fig.~\ref{fig:Peak-height-and}b) as a function of system size.
For the  smaller systems the  peak height  increases roughly  as $\sqrt{N}$.
Such rapid growth is more  characteristic of a  system undergoing a
first-order  phase  transition. Indeed  if  we  only  had data  up  to
$\sqrt{N}  \le 26$ (see below for  further discussions of this point) 
 then  we might  have  concluded that  there was  a
genuine first order transition. But at the larger system sizes studied
there is evidence that the growth in the specific heat has stopped its
rise with $\sqrt{N}$ and is saturating to a finite value, which is the
expected behavior at the XY transition.

The other signature of KT transitions are jumps in a modulus from zero
to  a finite  value at  the transition\cite{vsavsik1994calculation}.
In  Fig.~\ref{fig:Left:Shear_and_Cv}a  data for  the shear  modulus is
displayed. The (universal) jump from zero to a finite value as the
temperature is reduced through the  hexatic phase into the crystalline phase
arises when the free  dislocations in the hexatic liquid  bind together.  Apart
from the case $N=24\times24$ (see  below for discussion of this case),
we  see the expected  convergence to  a finite  value at  the critical
point  and  to  zero  on  the  high  temperature  side  of  the  phase
transition.

\paragraph{Maxwell analysis}
\begin{figure}
\begin{centering}
\includegraphics[width=6cm]{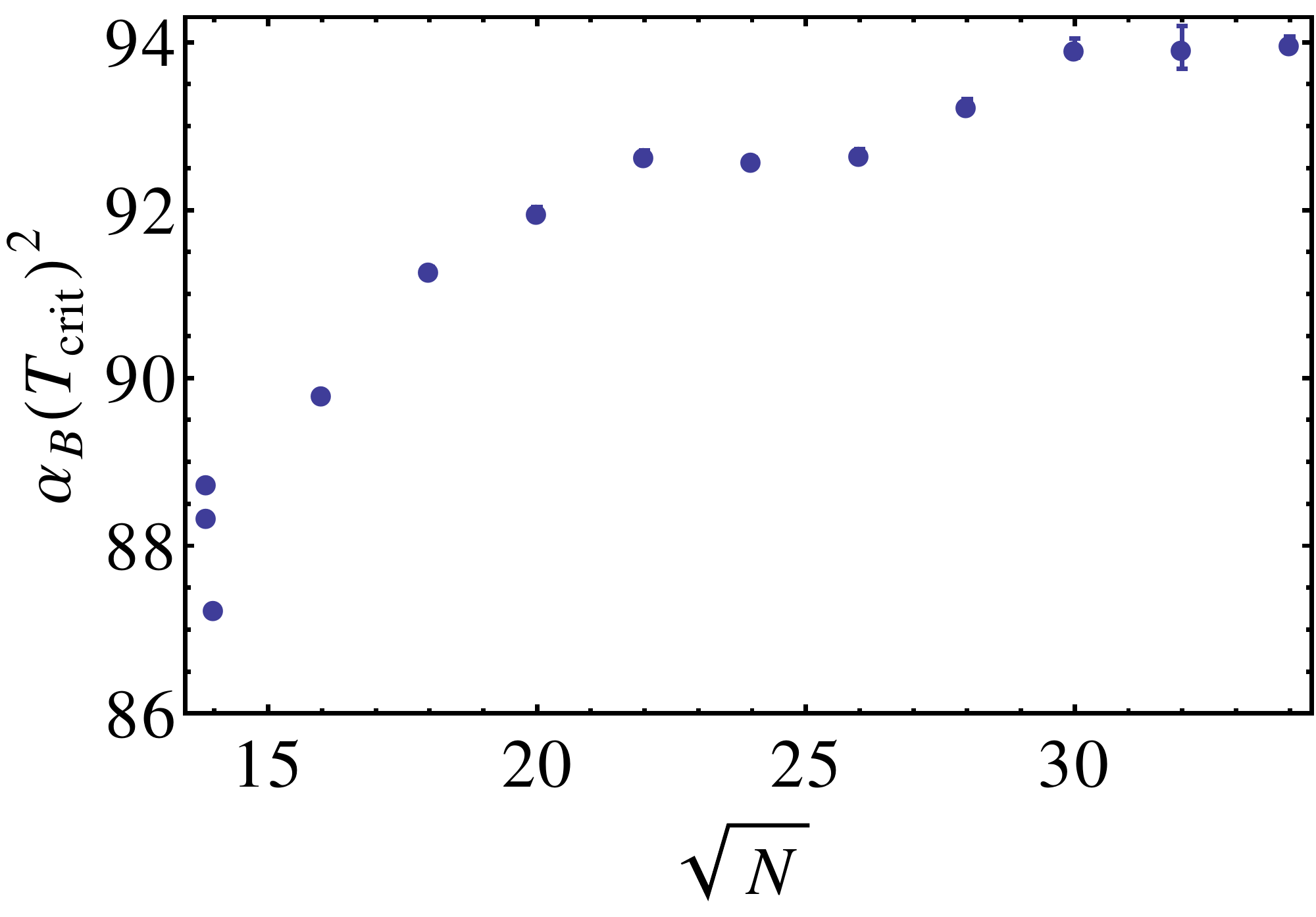}
\end{centering}
\caption{(Color online) A plot of the parameter $\alpha_{B}(T_{\text{crit}})^{2}$ for the temperature $T_{\text{crit}}$ where the two peaks in
$P_{\beta}(E)$ are of identical height.\label{fig:Tcrit}}
 \end{figure}

Now we have found some evidence that suggests the KTHNY scenario might
apply, let us analyze the Maxwell construction a little further. In
Fig.~\ref{fig:Tcrit} we have plotted the critical value of
$\alpha_{B}(T)$ where the two peaks in $P_{\beta}(E)$ are of the same
height. But there is more striking behavior of the intersection points
determined in the Maxwell construction and its uncertainty plotted in
Fig.~\ref{fig:Maxwell}.  The width of the Maxwell construction shrinks
rapidly (possibly to zero) with increasing $N$. This is not what would
be expected for a genuine first order phase transition, as seen for
instance in hard-disk systems \cite{bernard2011twostep}.  One would
have expected the width to saturate to the energy difference between
the liquid and crystal states.

\begin{figure}
\begin{centering}
\includegraphics[width=6cm]{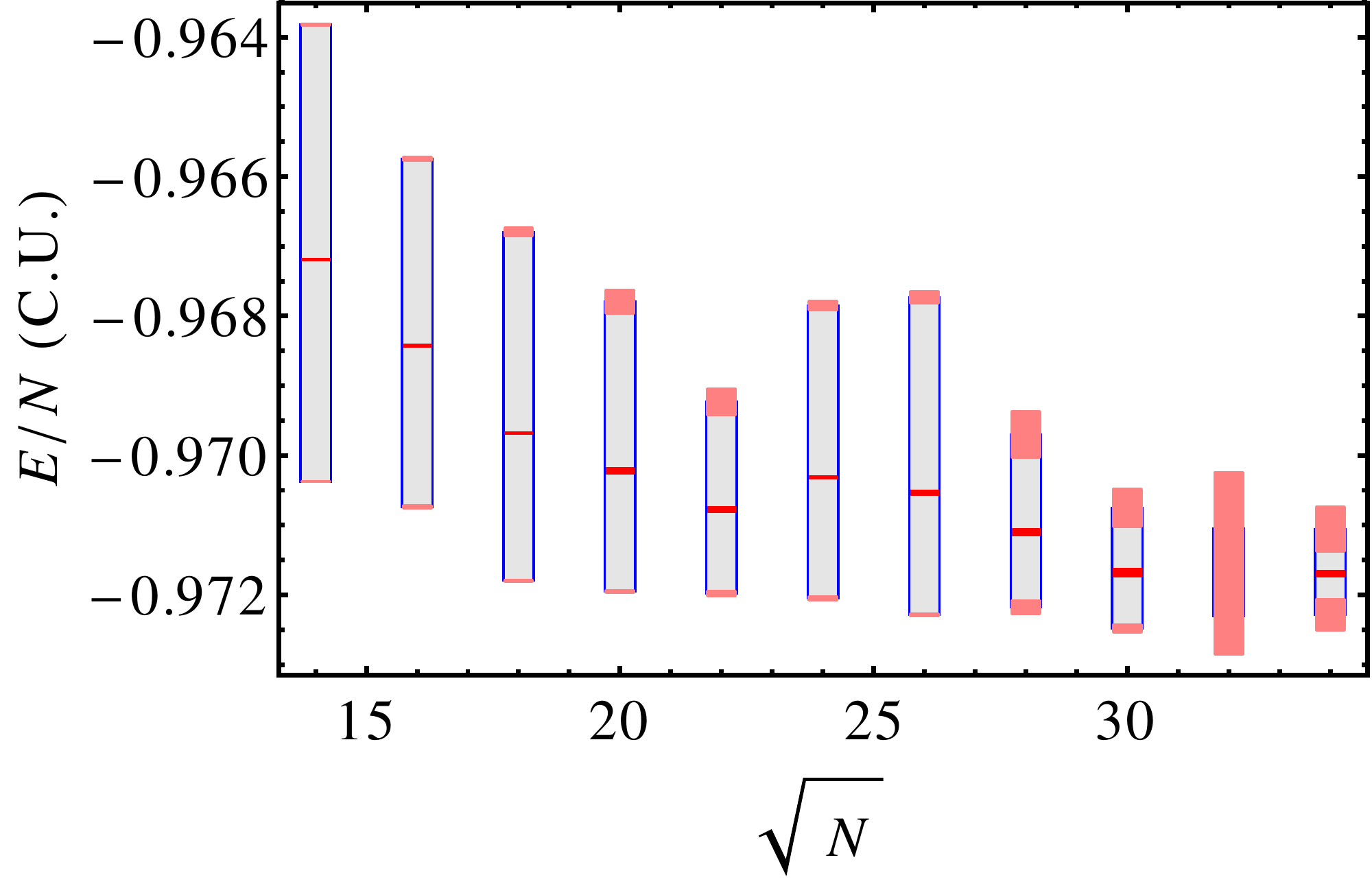}
\end{centering}
\caption{(Color online) Representation of the full
range of the Maxwell construction , with the midpoints and uncertainty
(red); full range (blue) and endpoints with range of uncertainty as
an orange bar.\label{fig:Maxwell}}
\end{figure}

\paragraph{Extrapolation}

\begin{figure}
\begin{centering}
\includegraphics[width=6cm]{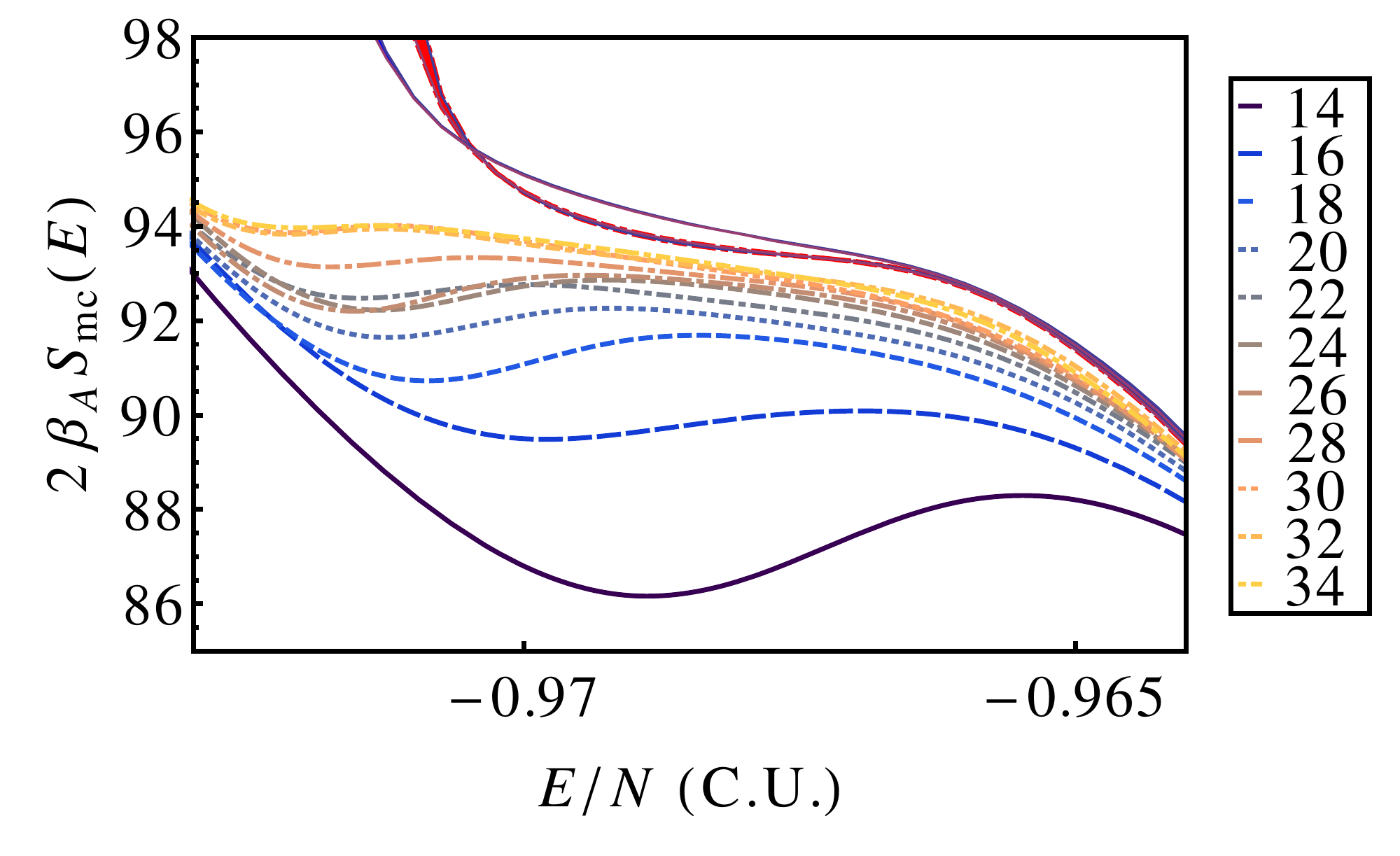}
\end{centering}
\caption{(Color online) Similar to Fig.~\ref{fig:The-microcanonical-entropy}b, with
  two naive $[1/1]$ Pad\'e extrapolations [see main text for details]
  to $N\rightarrow\infty$ (one including all data from $N=14^{2}$ and
  up, the other from $N=16^{2}$ and up). The width of the extrapolated
  curves is a measure of the uncertainty, obtained by excluding the
  case $N=34^{2}$.\label{fig:extrapollation}}
\end{figure}

One way to try and gain an understanding of the thermodynamic limit of
the entropy is by an approximate extrapolation. A naive extrapolation
using a $[1/1]$ Pad\'e approximant,
\begin{equation}
S_\text{mc}(E,\sqrt{N})=(a(E)+b(E)/ \sqrt{N})/(1+c(E)/ \sqrt{N}),
\end{equation}
fitted to the entropy curves, as in Fig.~\ref{fig:extrapollation}, shows
again that even though the set of curves is too limited
 to completely determine the
density of states in the phase transition region, a trend can nevertheless be
seen there. For lower energies (and for higher ones as
well, but we show no results here) the extrapolation is actually
quite stable.  Also, the results of a $[2/2]$ extrapolation do not
look appreciably different.

\paragraph{Understanding nonuniversal behavior}

\begin{figure}
\begin{center}
\includegraphics[width=6cm]{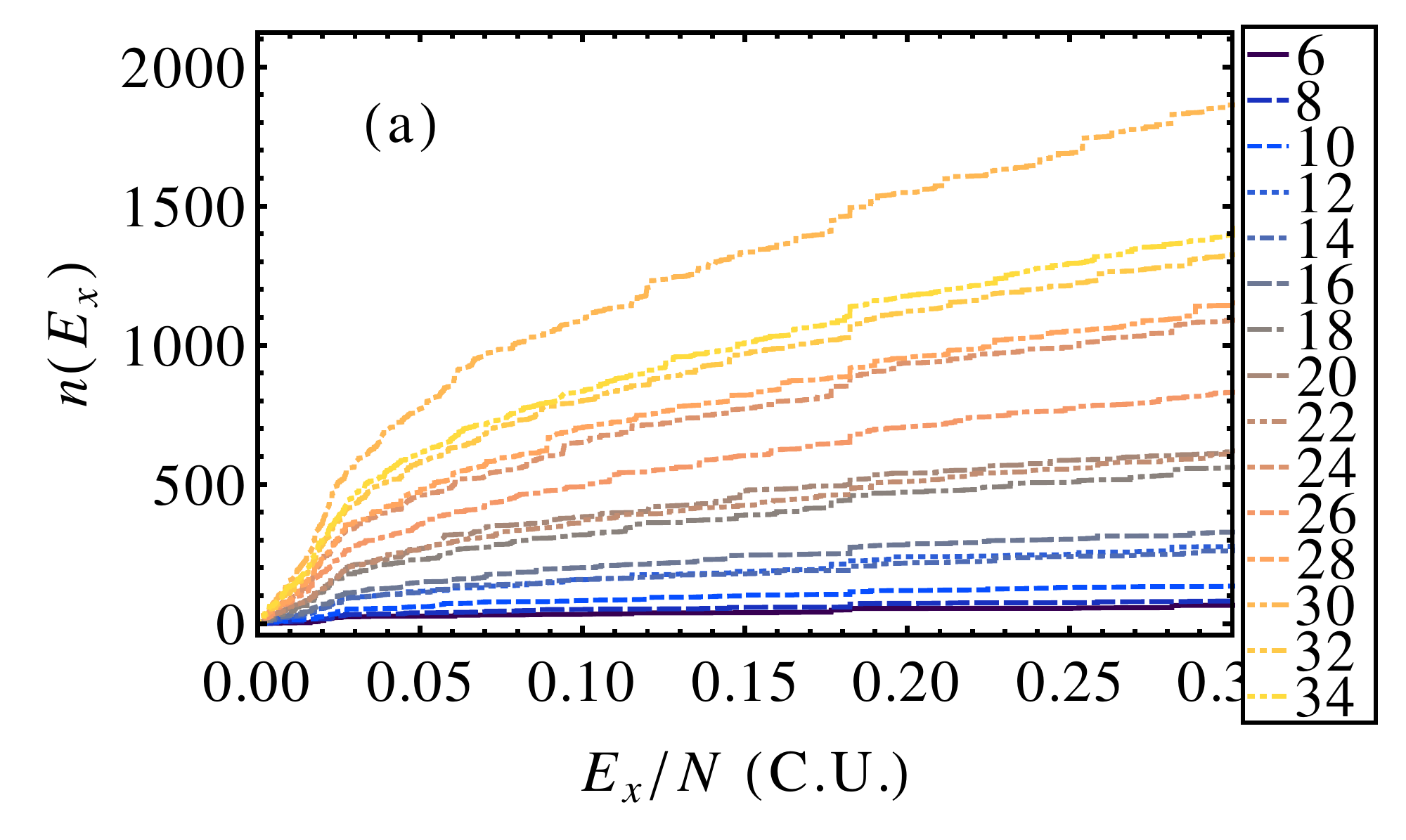}\\
\includegraphics[width=6cm]{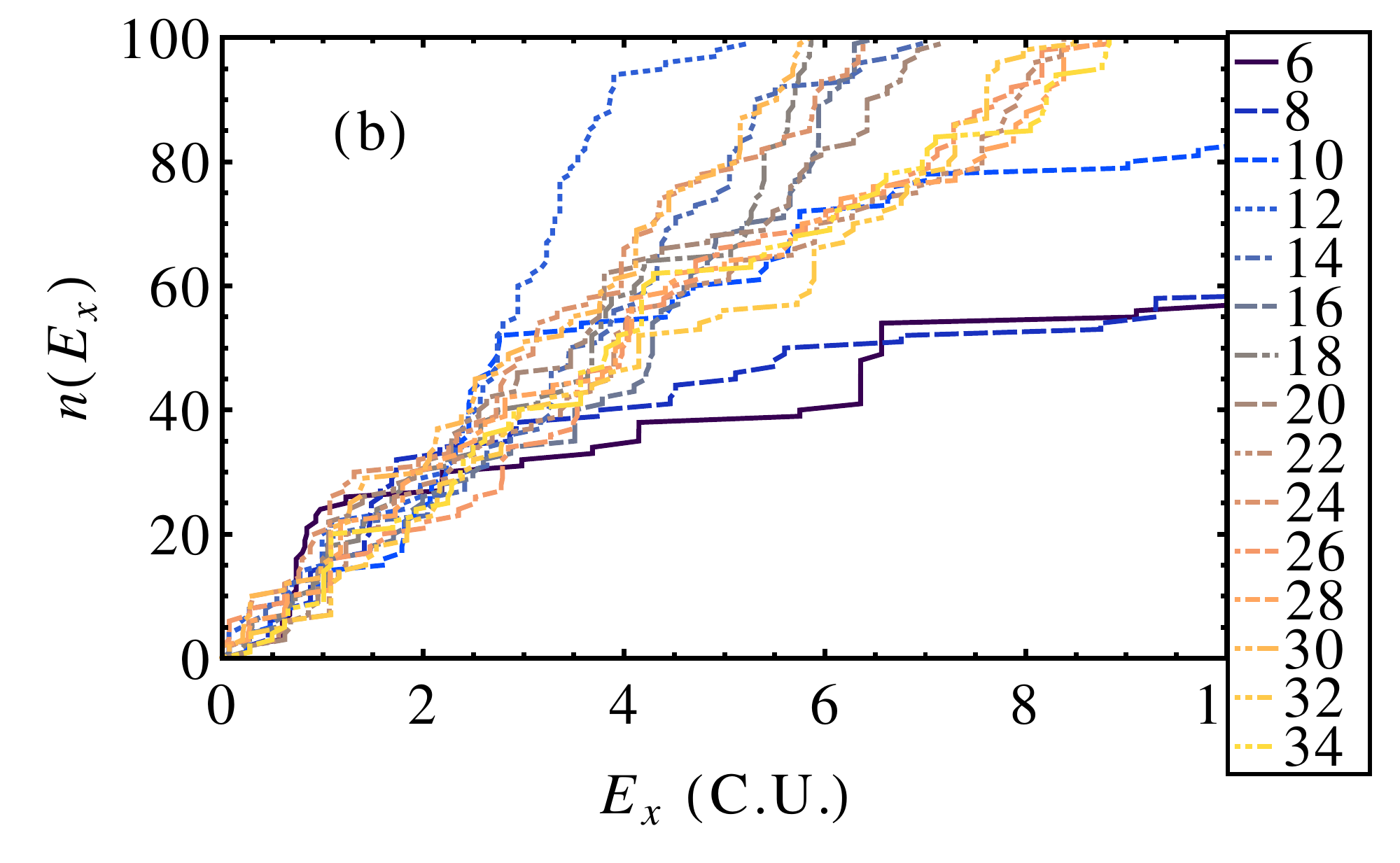}
\end{center}
\caption{(Color online) The number of crystalline states below the excitation energy $E_{x}$
on a torus of $\sqrt{N}l_{B}$ by $\sqrt{3N}l_{B}/2$ for various
values of $\sqrt{N}$. (a) shows the data as a function of the excitation energy per particle $E_x/N$ over a wide range; (b) shows the results as a function of $E_x$, which enters the Boltzmann factor.\label{fig:Excitation-energies-of}}
\end{figure}

One of the questions we need to ask ourselves, is why does the process
of increasing $N$ lead to such large changes in the apparent nature of
the phase transition from one size to the next as seen, e.g., in
Fig.~\ref{fig:Maxwell}. These differences we believe have a
topological origin, and are related to the occurrence of degenerate
ground states, which have some influence on the results. We now solve
the Diophantine equations (\ref{eq:diophantine}) with the help of
Mathematica, see Fig.~\ref{fig:Excitation-energies-of}.

\begin{figure}
  \centering{\includegraphics[width=6cm]{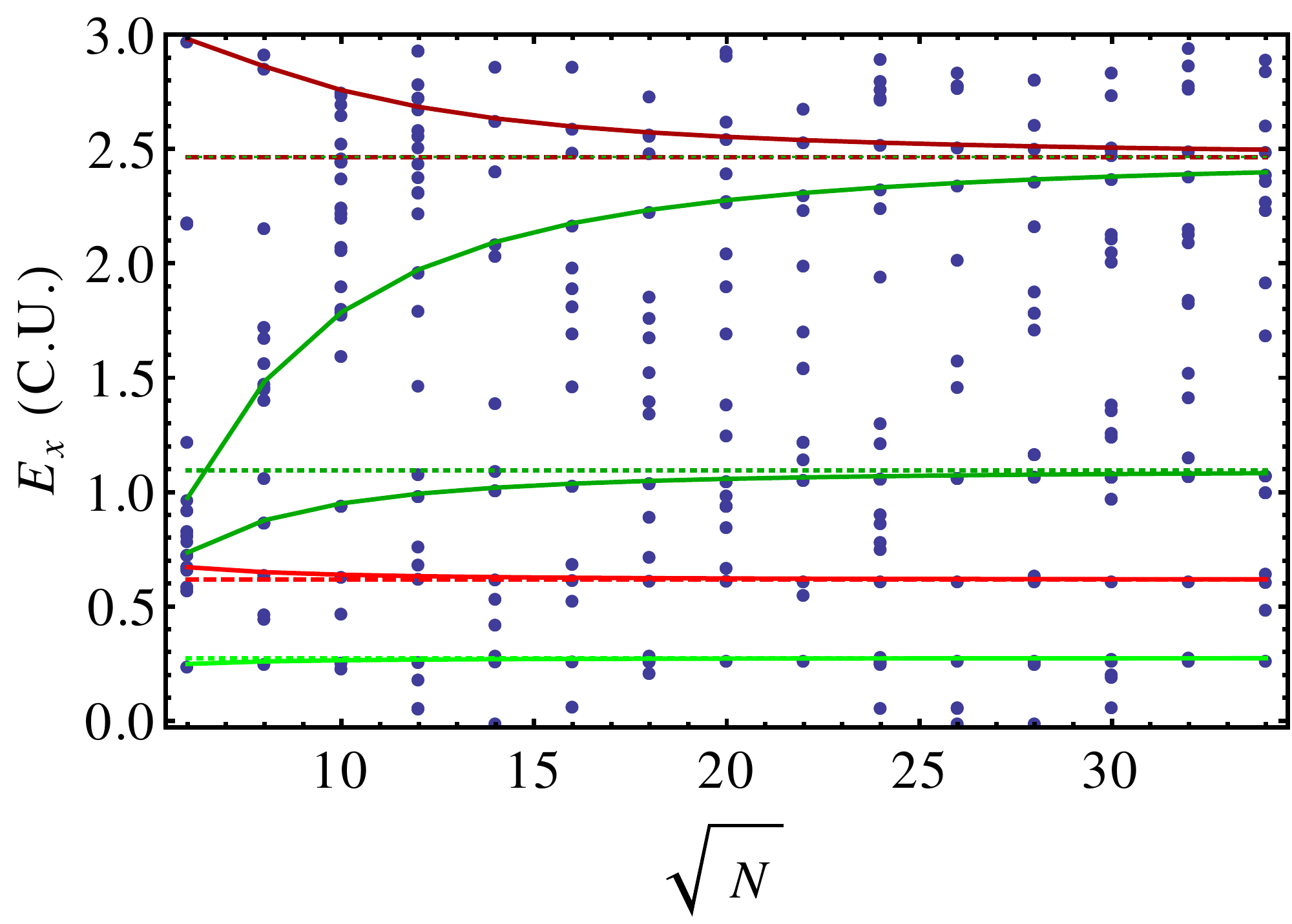}}

\caption{(Color online)
    Excitation energy of lowest energy crystalline states as a
    function of the number of vortices. The solid green and solid red
    lines connect excitations in two families that described small
    deformations of the triangular
    crystal; the dashed line are the thermodynamic limit of these energies. \label{fig:Energy-of-lowest}}
\end{figure}

\begin{figure*}
\begin{centering}
\begin{tabular}{ccc}
 & $14\times14$ & $16\times16$\tabularnewline
more liquid & \includegraphics[width=6cm]{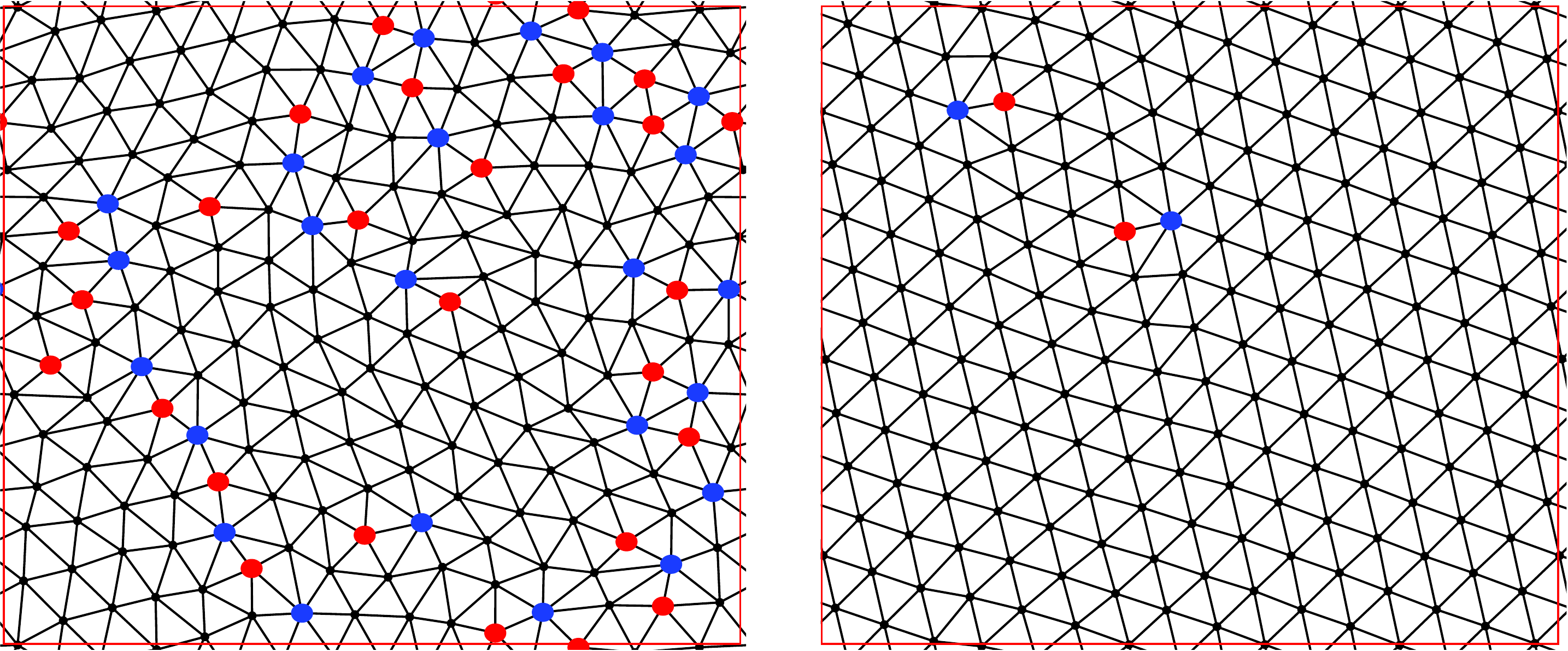} & \includegraphics[width=6cm]{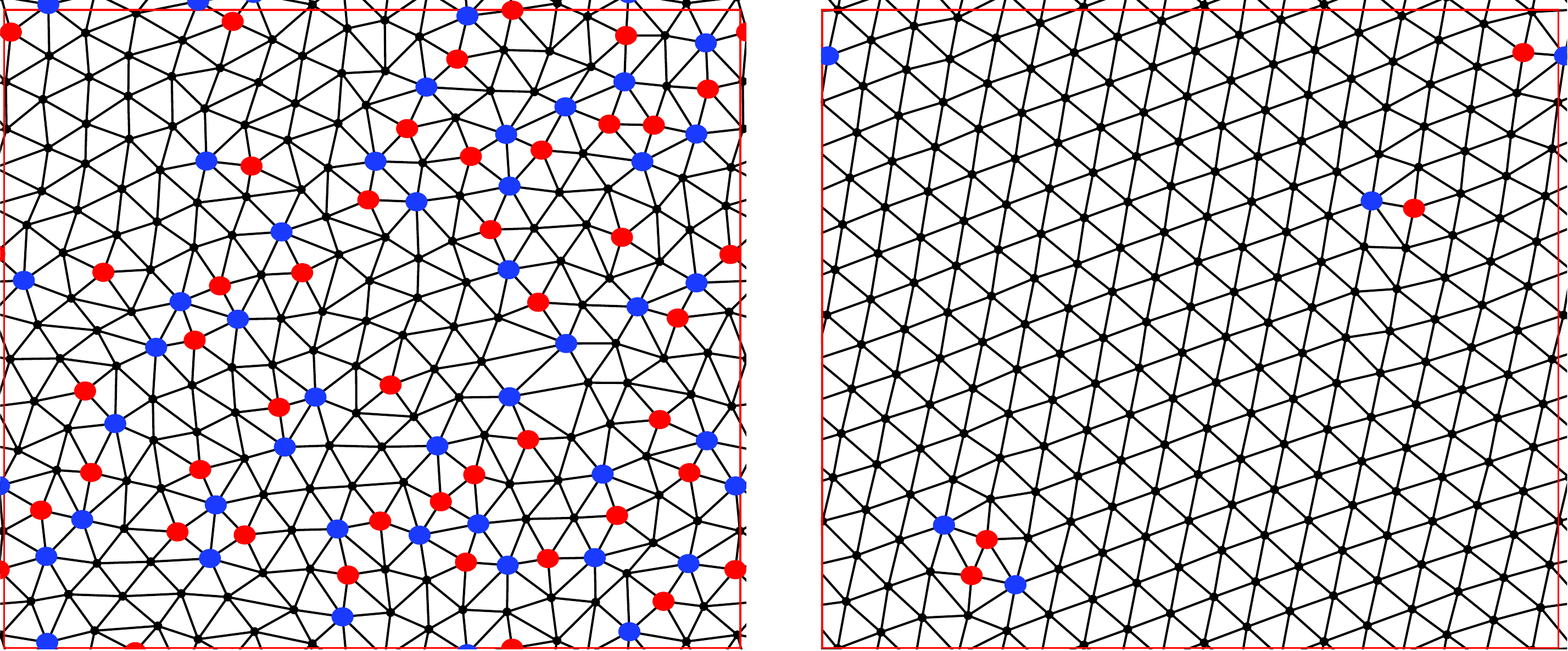}\tabularnewline
more crystalline & \includegraphics[width=6cm]{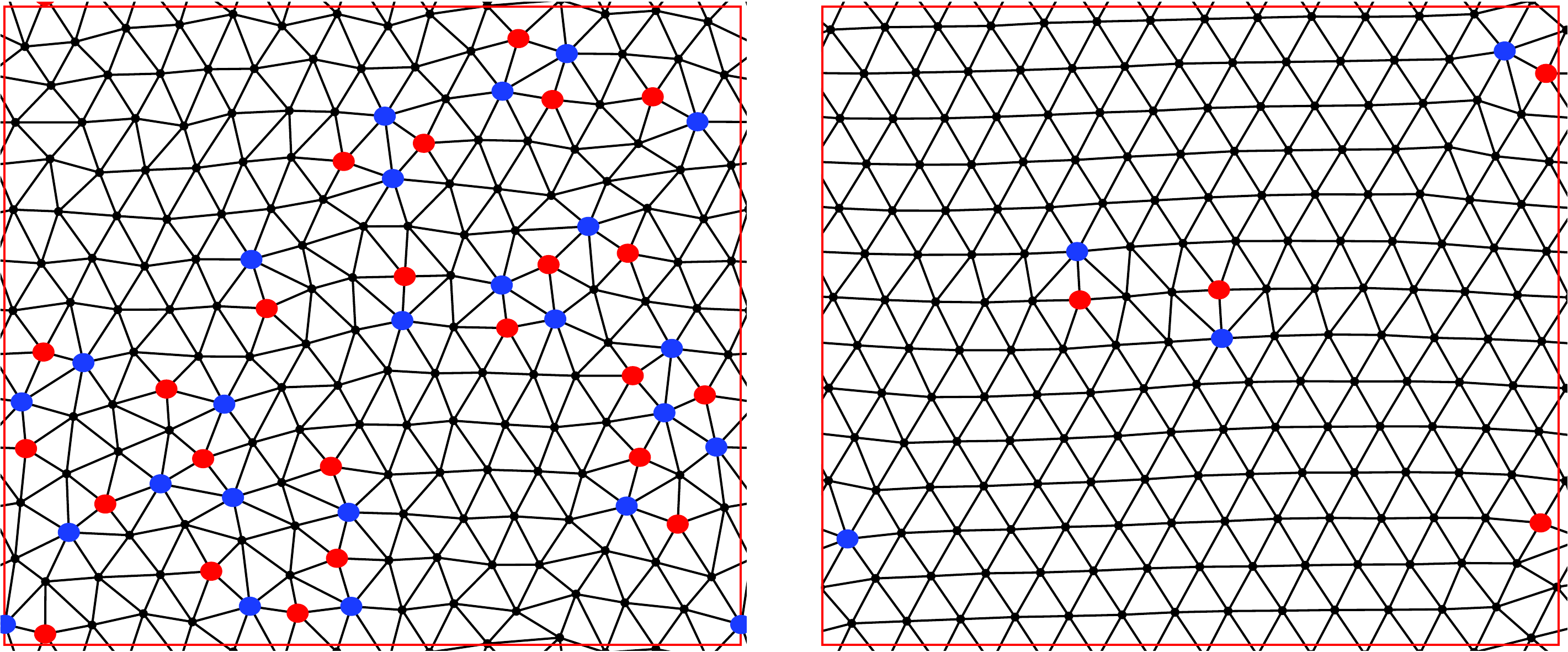} & \includegraphics[width=6cm]{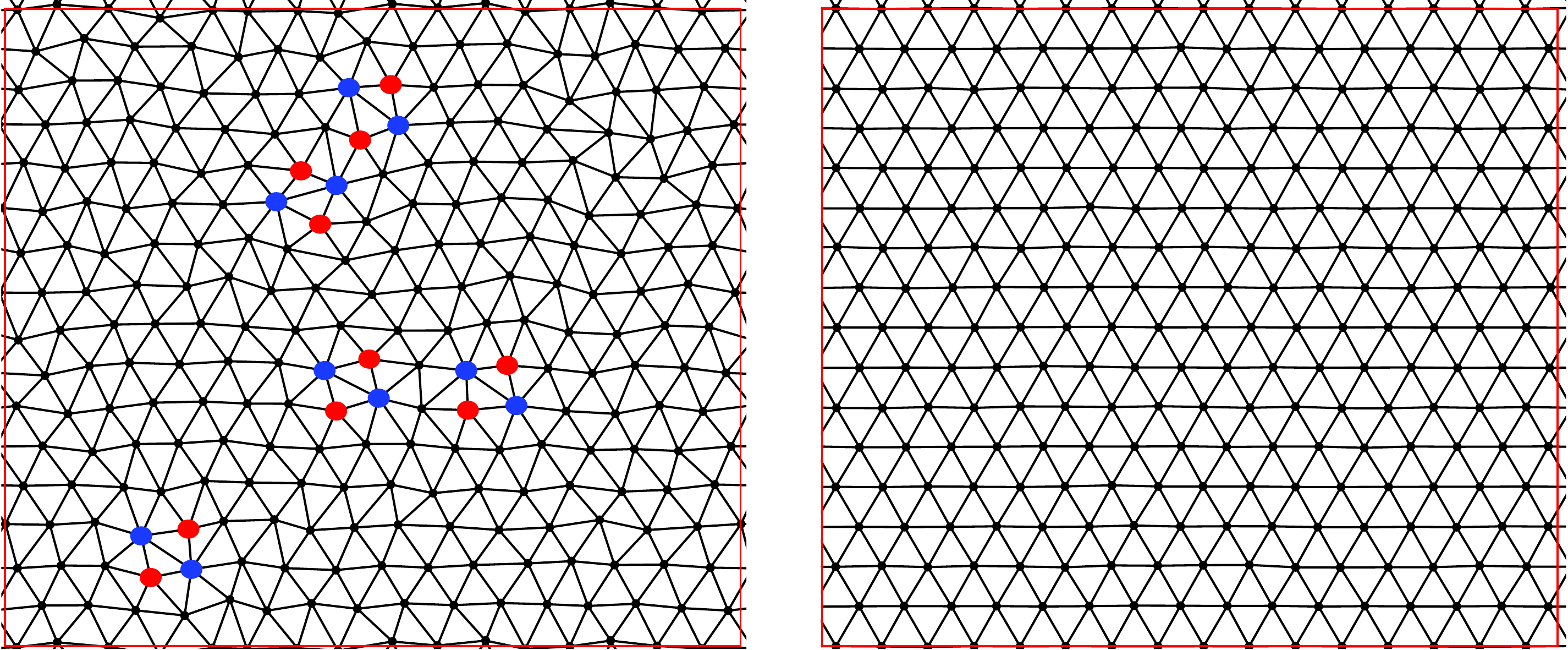}\tabularnewline
\end{tabular}
\end{centering}
\caption{(Color online) Snapshots and steepest descent minimization: a
  few typical examples.  Left column for $14\times14$ vortices, right
  $16\times16$. The top row is at a slightly higher energy (more
  liquid like). Each pair of pictures shows two related Delaunay
  diagrams of the vortex lattice.  The left picture is a direct image
  from a snapshot (a single configuration in a simulation); the right
  picture is the result from a steepest descent minimization based on
  the snapshot. In each case a red dot shows 5-fold coordination, a
  blue dot 7-fold.\label{fig:defect1}}
\end{figure*}

The picture that emerges seems rather complex.  We see a density of
states that roughly increases with system size, but there are obvious
exceptions. The low-energy spectrum, which is the only part with any
relevance to the simulations, is highly complex.  The most telling
analysis is shown in Fig.~\ref{fig:Energy-of-lowest}.  There we see
that for $\sqrt{N}=14,26,28$ we have a degenerate ground state
(actually, since the additional groundstates come in pairs
inequivalent under reflection, the degeneracy is three).  The
neighboring points, $\sqrt{N}=12,16,24,30$ have exceptionally low
energy states ($E_x=0.065$ and $0.066\ \text{C.U.}$ for $\sqrt N=12$, $0.071$ for $\sqrt N=16$, $0.067$  for $\sqrt N=24$, and $0.070$ for  for $\sqrt N=30$,
respectively) that almost behave like an additional ground state. For
$\sqrt N =26$ we have both a degenerate ground state, and a close-by
set of states at $E_x=0.067$ and $0.070\ \text{C.U.}$. The green and
red solid lines in Fig.~\ref{fig:Energy-of-lowest}  connect families of states that
describe small deformations of the original equilateral lattice (the
dashed lines shows the large-$N$ limit). If we restrict ourselves to
the case $N_{x}=N_{y}=\sqrt{N}$, we find that in this limit these
typical excitations energies are given by
\begin{equation}
E=-N+\delta_{i}\, k^{2},
\label{distortenergy}
\end{equation}
with $k=1,2,\ldots$ and $\delta_{1}=0.274$ and $\delta_{2}=9/4\,\delta_{1}$. (These are the horizontal dashed lines in Fig.~\ref{fig:Energy-of-lowest}).
Also note the enormous growth in the number density of crystalline states
as $N$ increases revealed in Fig.~\ref{fig:Excitation-energies-of}.

\paragraph{Defects}

One of the  very important aspects of the KTHNY  mechanism is the role
of defects, and their (un)binding. Early on in our work, we decided to
look  at  the  topological  nature  of  some  of  the  states  in  the
simulations,  by looking  at configurations  (snapshots),  and letting
these  relax  through a  steepest  descent  minimization to  determine
better their topological content. We produced a Delaunay triangulation
to   examine    the   topological   content   of    the   states.   In
Fig.~\ref{fig:defect1}, we present a  few interesting examples of this
work.  The snapshots  themselves mainly  show grain  boundaries, apart
from the lowest energy  $16\times16$ picture, which shows bound defect
pairs.  What we  see  in the  crashes  is rather  typical:  we have  a
``misaligned  crystal''  due to  a  few  bound  defect pairs. 

\begin{figure}
\begin{centering}
\begin{tabular}{cc}
\includegraphics[width=4cm]{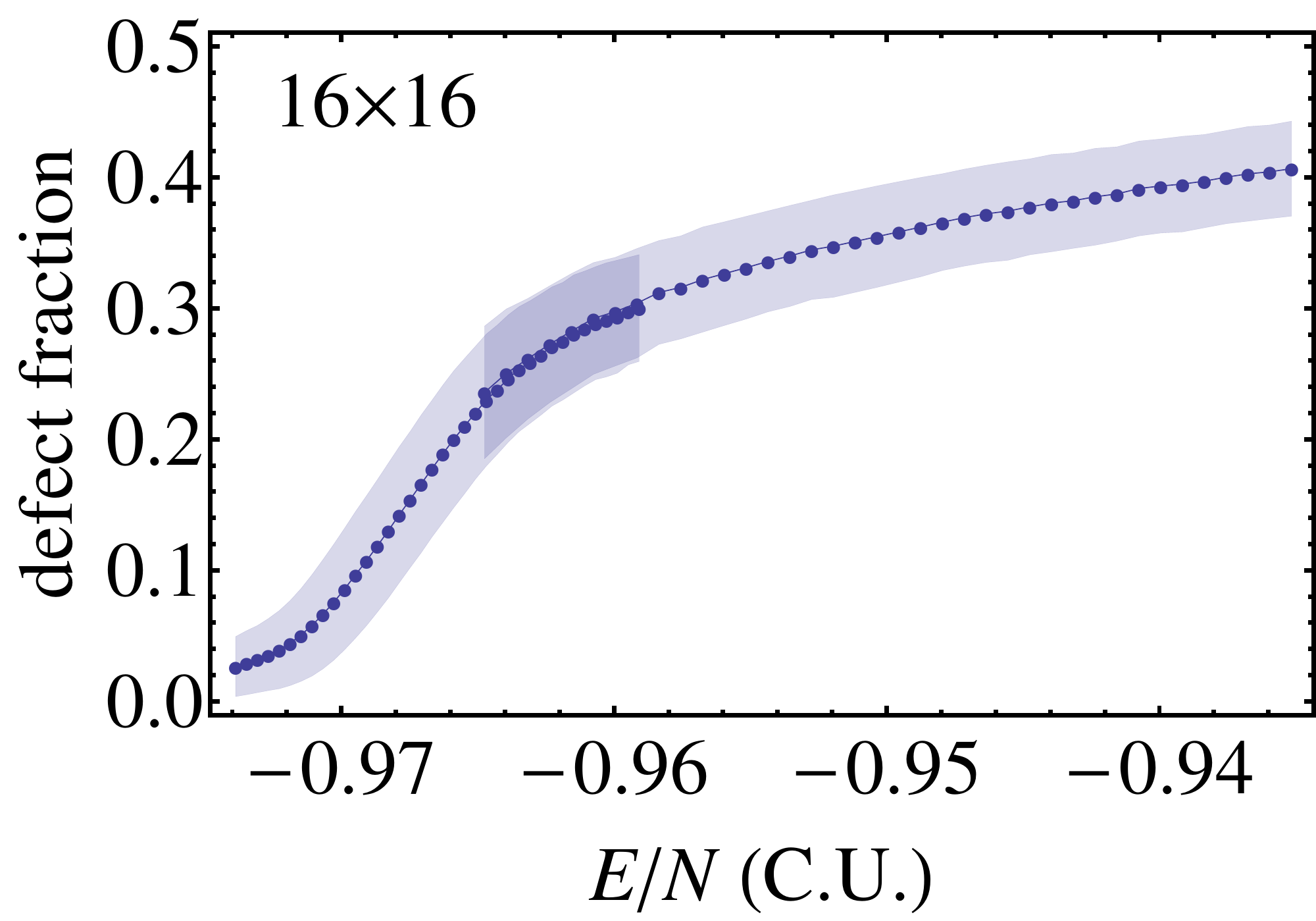} & \includegraphics[width=4cm]{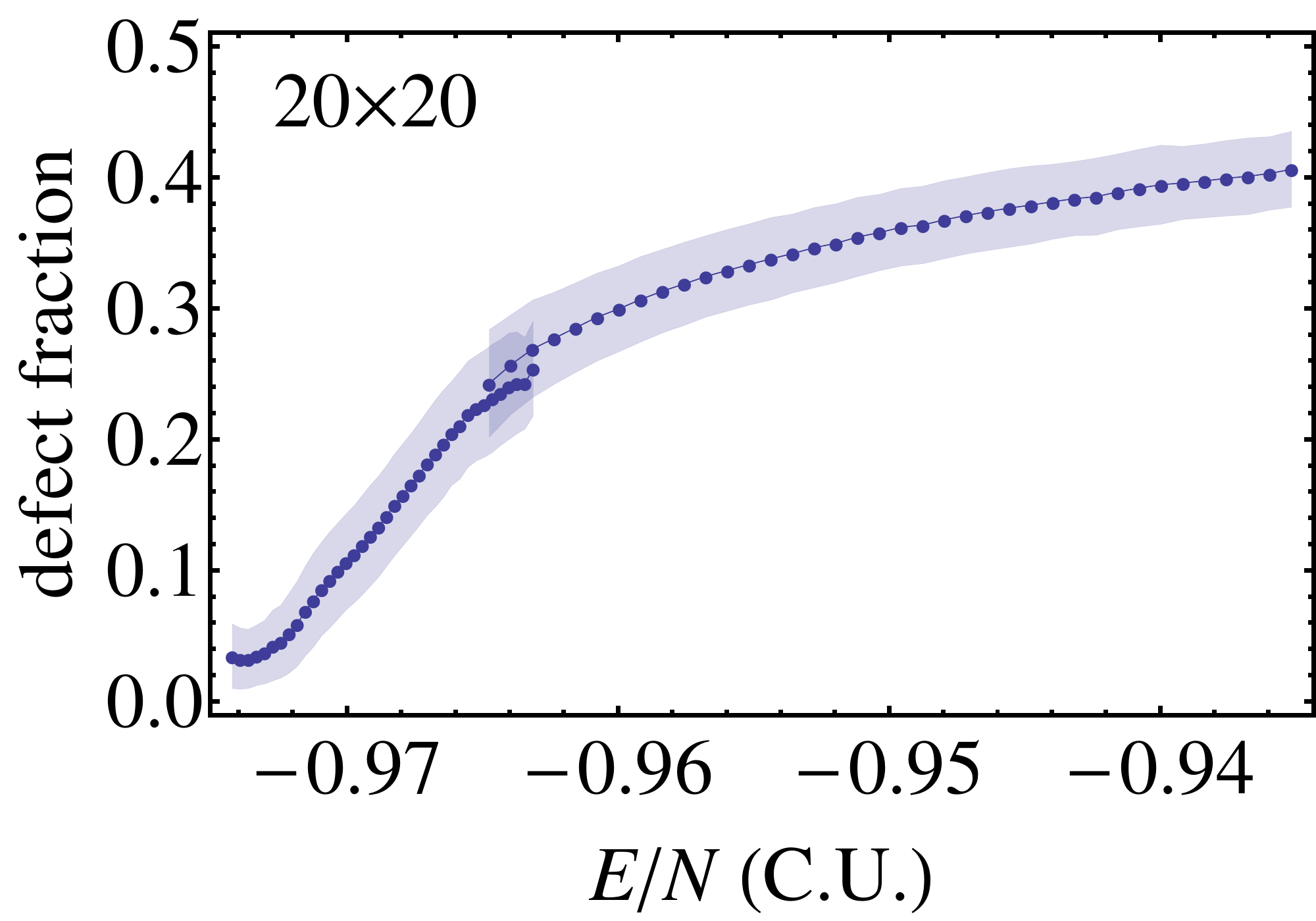}\tabularnewline
\includegraphics[width=4cm]{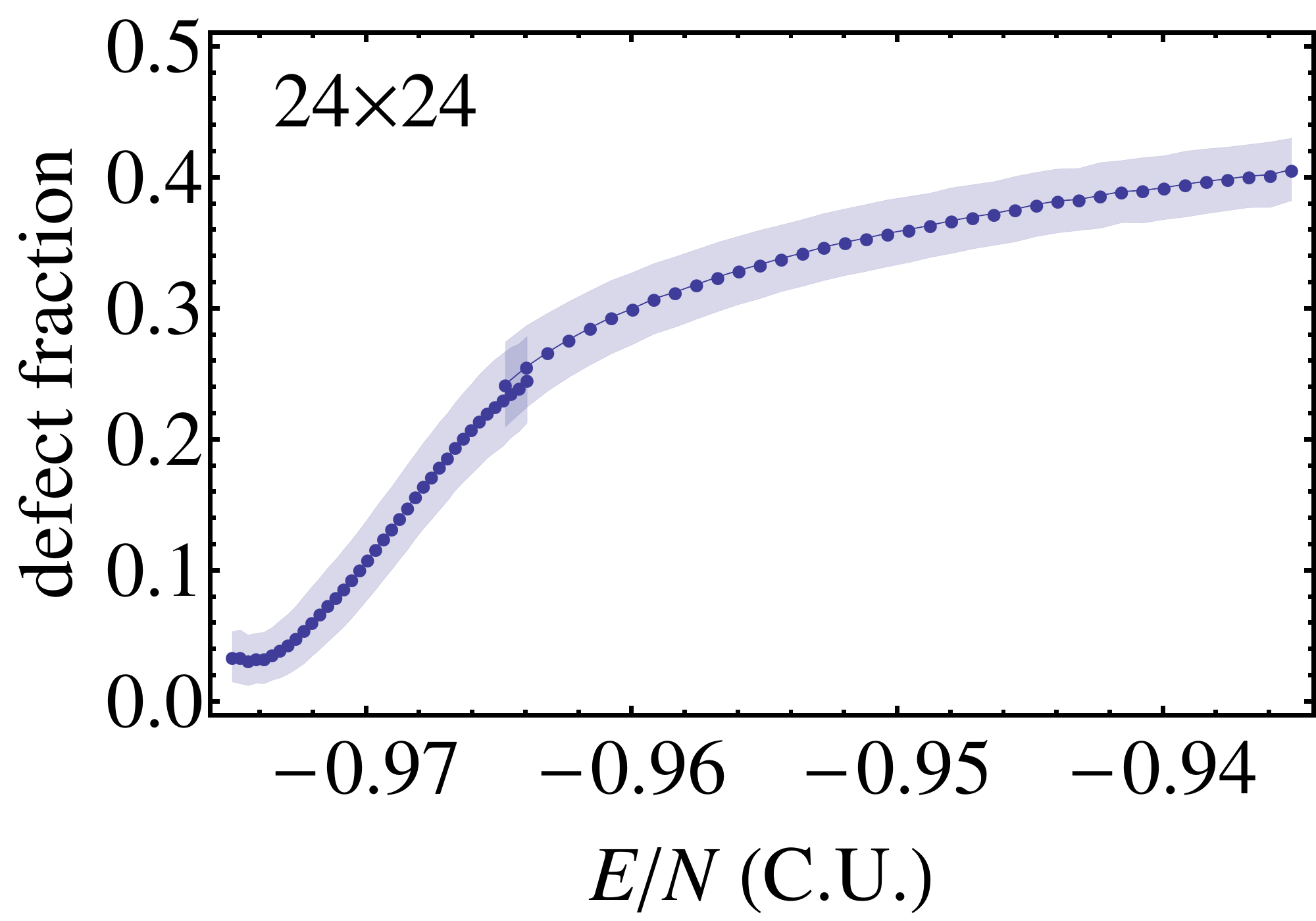} & \includegraphics[width=4cm]{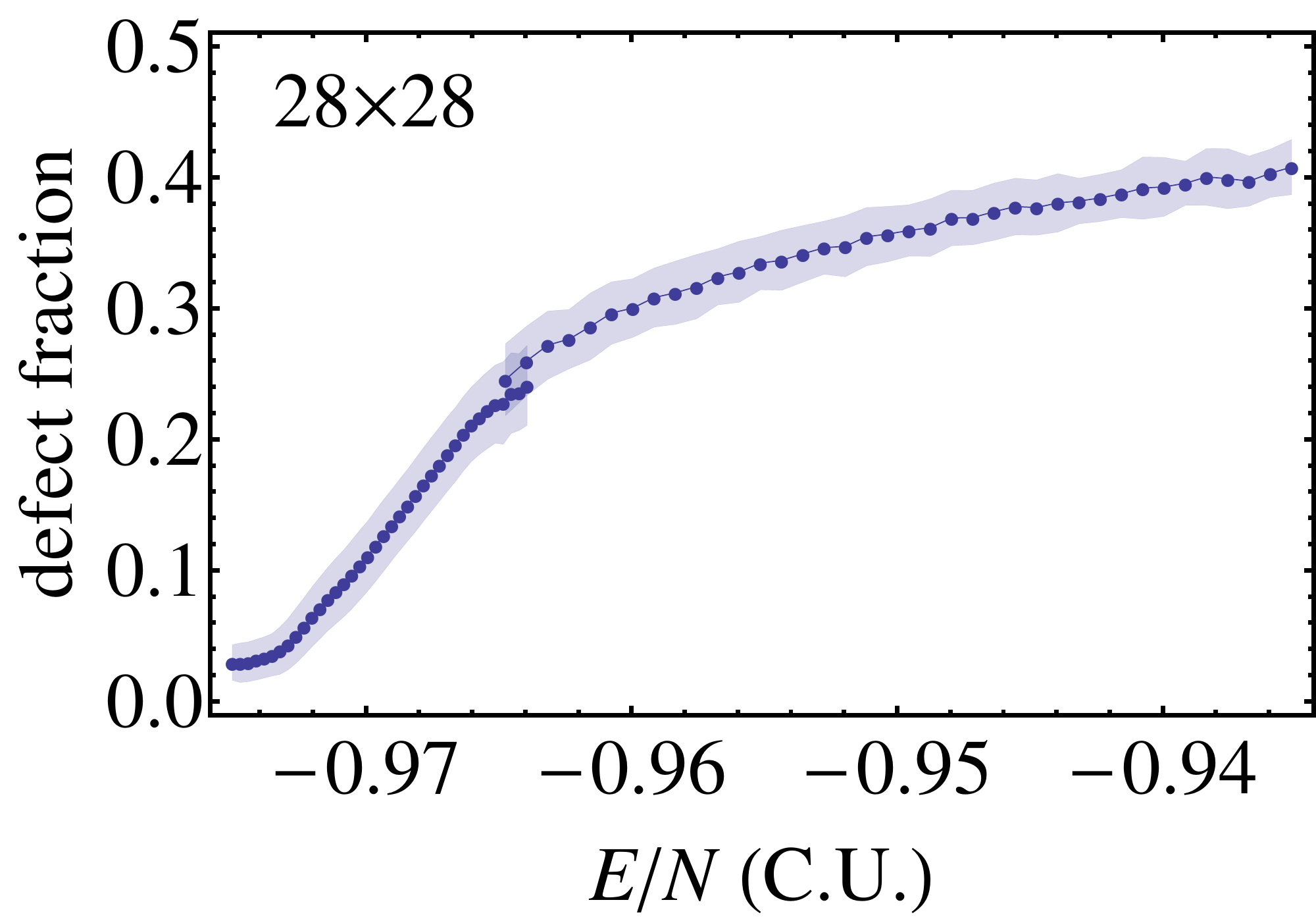}\tabularnewline
\end{tabular}
\par\end{centering}

\caption{(Color online) Density of defects, defined as the fraction of
  points in the Delaunay diagram without 6-fold coordination, as a
  function of energy per vortex $E/N$ for four different system sizes, labeled as
  $N_x\times N_y$.  The width of each band is a measure for the
  standard deviation in the answers. There appears to be clear
  evidence that apart from at extremely low energies, where we may be
  seeing a sharp feature developing, the density of defects is
  largely independent of $N$. (Breaks in the lines are caused by
  overlaying two separate simulations).\label{fig:defect2}}
\end{figure}

As we shall explain below, we have done a large number of calculations
of the vortex positions. Unfortunately, a visual analysis in terms
of Delaunay diagrams is practically impossible, but one thing we can
do is plot the density of defects (any point that is not 6-fold coordinated)
as a function of energy, as in Fig.~\ref{fig:defect2}. We find a sharp
fall at low energies in the simulations. The conclusion appears
to be that for energies down to $E/N=-0.96$ the phase-diagram is dominated
by grain boundaries; at lower energies we find (mostly bound) pairs of defects. There
appears to be some (but rather insubstantial) evidence of a further potential
phase transition at the lowest possible energies, $E/N<-0.97$.

\paragraph{Density-Density Correlations}

\begin{figure}
\begin{centering}
\begin{tabular}{cccc}
$N=16^{2}$ & $N=20^{2}$ & $N=24^{2}$ & $N=28^{2}$\tabularnewline
\includegraphics[width=2cm]{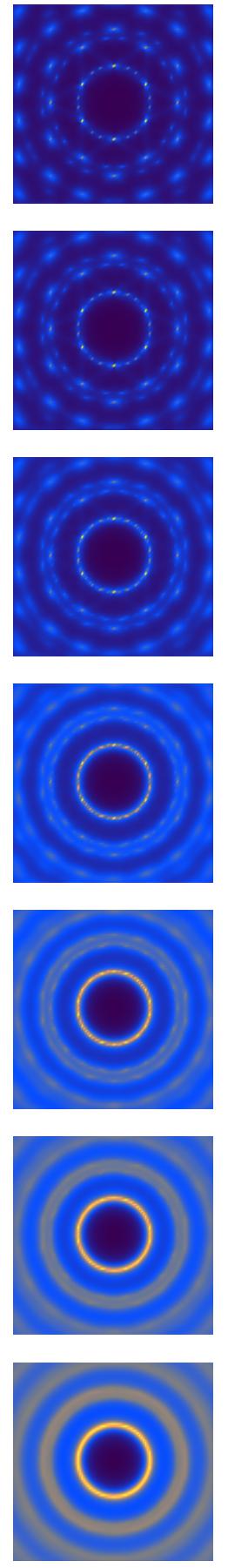} & \includegraphics[width=2cm]{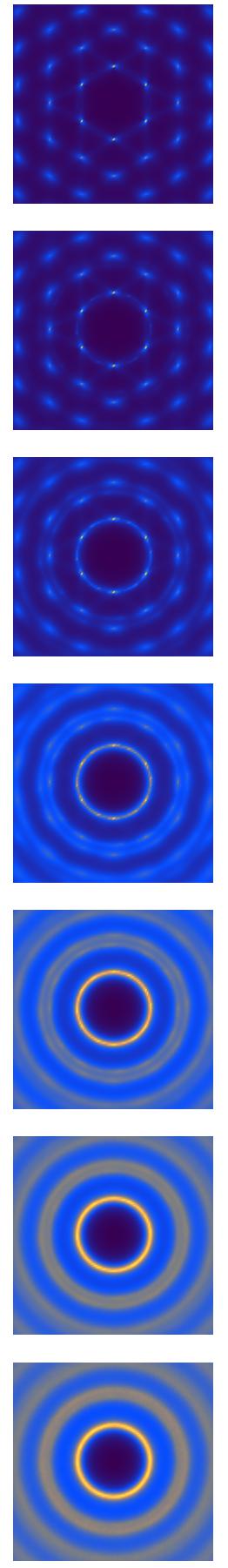} & \includegraphics[width=2cm]{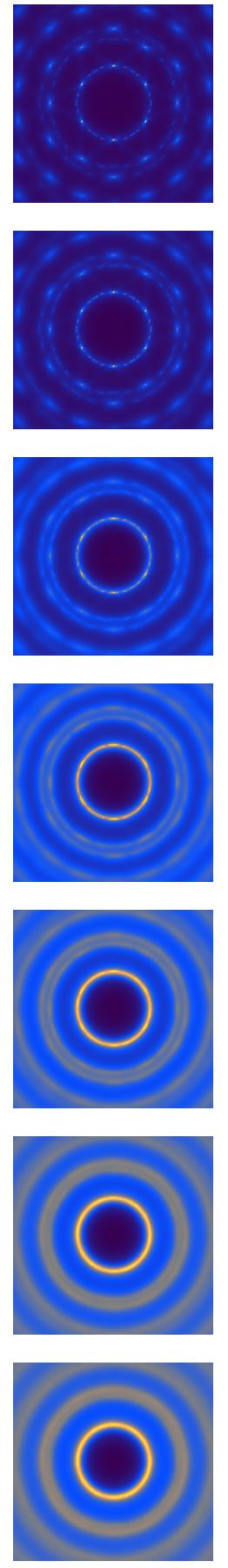} & \includegraphics[width=2cm]{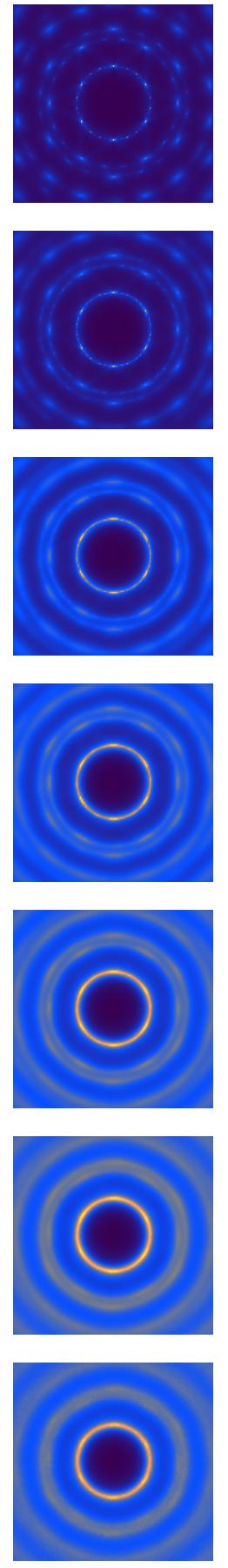}\tabularnewline
\end{tabular}
\end{centering}
\caption{(Color online) Density-density correlation functions for four
  different system sizes ($16^{2}$, $20^{2}$, $24^{2}$, and
  $28^{2}$). The energy increases from top to bottom,
 but all values are in the flat region of the entropy
  curve (first row: $E/N=-0.9738$,
  second: $-0.9723$, third: $-0.9706$, fourth: $-0.9691$, fifth:
  $-0.9675$, sixth: $-0.9658$, and last row:
  $-0.9643$.)\label{fig:DDcorrelations}}
\end{figure}

A convenient measure for the extent of crystalline order in the liquid
state can be obtained by the density-density correlations discussed in
Sec.~\ref{sec:Delta}. We  have plotted a  few selected images  for the
four standard system sizes  in Fig.~\ref{fig:DDcorrelations}.  We see  in that
figure that as  we increase the energy the system  goes from a crystalline
Bragg-like  pattern to   a typical  liquid  pattern with  rotational
invariance. The  Bragg-like patterns will not consist of
true delta   function   Bragg   peaks    because  we   can   have   only
quasi-crystalline order  in 2D  systems according to  the Mermin-Wagner
theorem.

Notice the  differences even at  the lowest energies  between $N=20^2$
and  the  other  sizes.  The   other  sizes  are  much  less  strongly
modulated. This could  be due to the fact that all  of the other sizes
in  that   figure  have  either  an   anomalously  low-lying  deformed
crystalline state, or even a degenerate ground state.  This could have
important consequences for correlations at the lowest energies used in
Fig.~\ref{fig:rescaled}. But  there are also other  mechanisms at play
which act to restore rotational invariance.

There is a  competition between the  ordering into the
perfect  low-energy  structure(s)  favored  by the  periodic  boundary
conditions and the misalignments induced by dislocations (bound $7-5$ defect
pairs).  The energy cost  of defects can be
partially  balanced by  the  entropy  gain from  the  large number  of
positions available  for them.   The width of  the peaks in  $P(E)$ in
Fig.~\ref{fig:rescaled}  shows that many  types of defects  could be
present  and as a  consequence several  mechanisms can  be at  work in
restoring the  rotational symmetry of  the liquid state at  the system
sizes  we  can  study.   But  at  energies  which  correspond  to  the
crystalline  and hexatic  states, the  central ring  is  clearly being
modulated, indicating that at these  energies the defects have not yet
fully  restored the rotational  invariance. For  the hexatic  state at
least, one expects that in  the thermodynamic limit there will be full
rotational invariance; it is a liquid (see also Sec.~\ref{discussion}).

\begin{figure}
\begin{center}
\includegraphics[width=4cm]{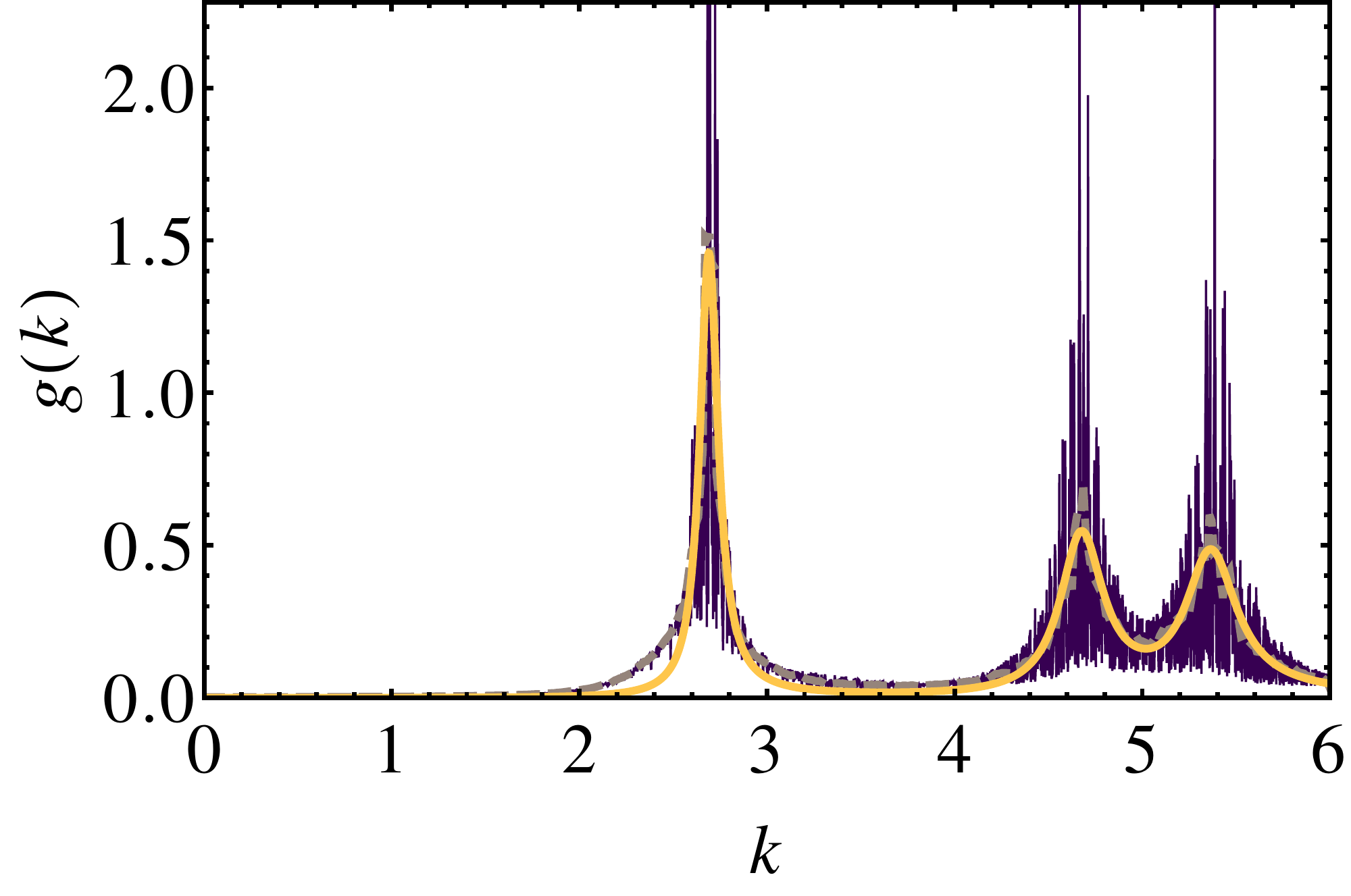}\includegraphics[width=4cm]{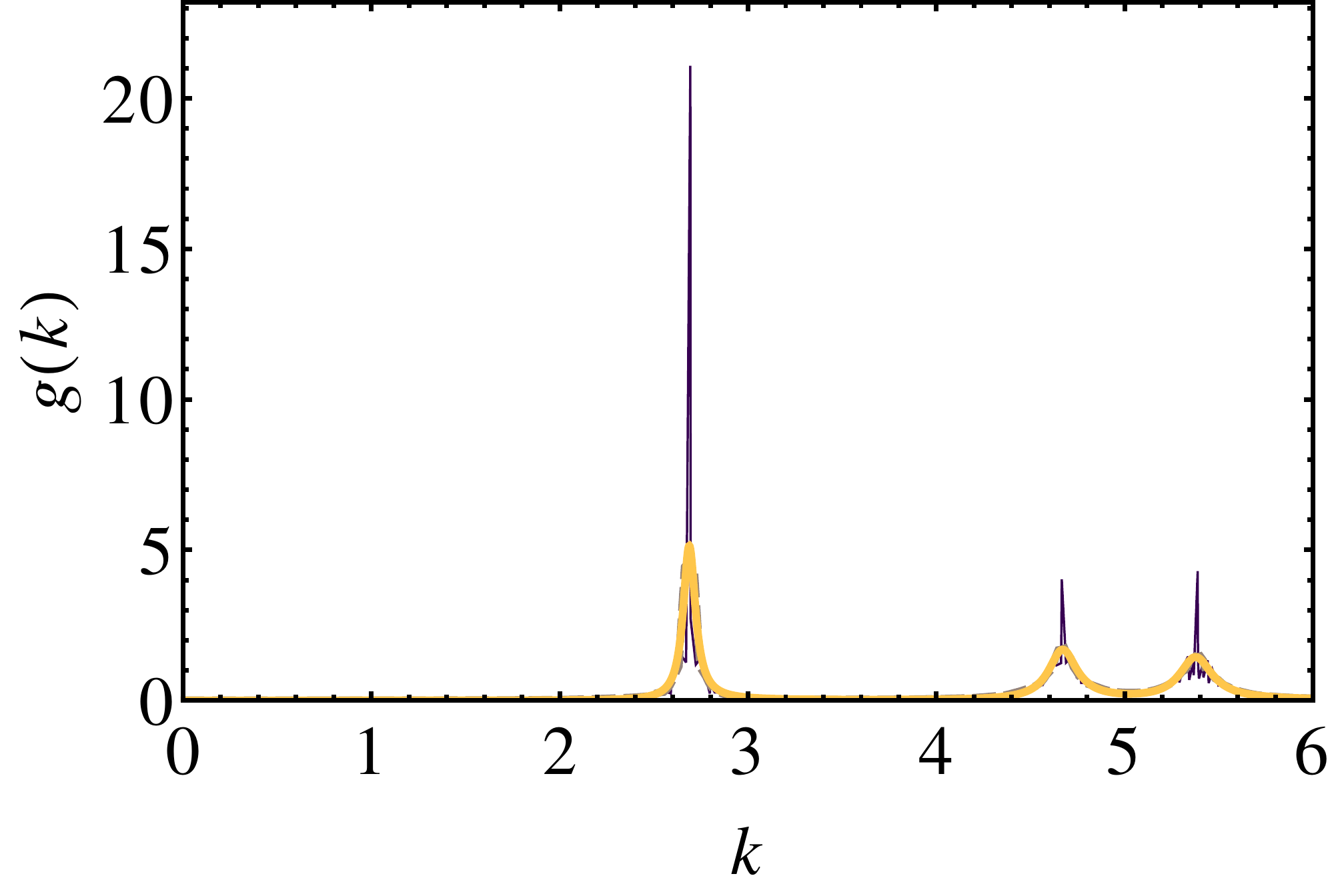}\\
\includegraphics[width=4cm]{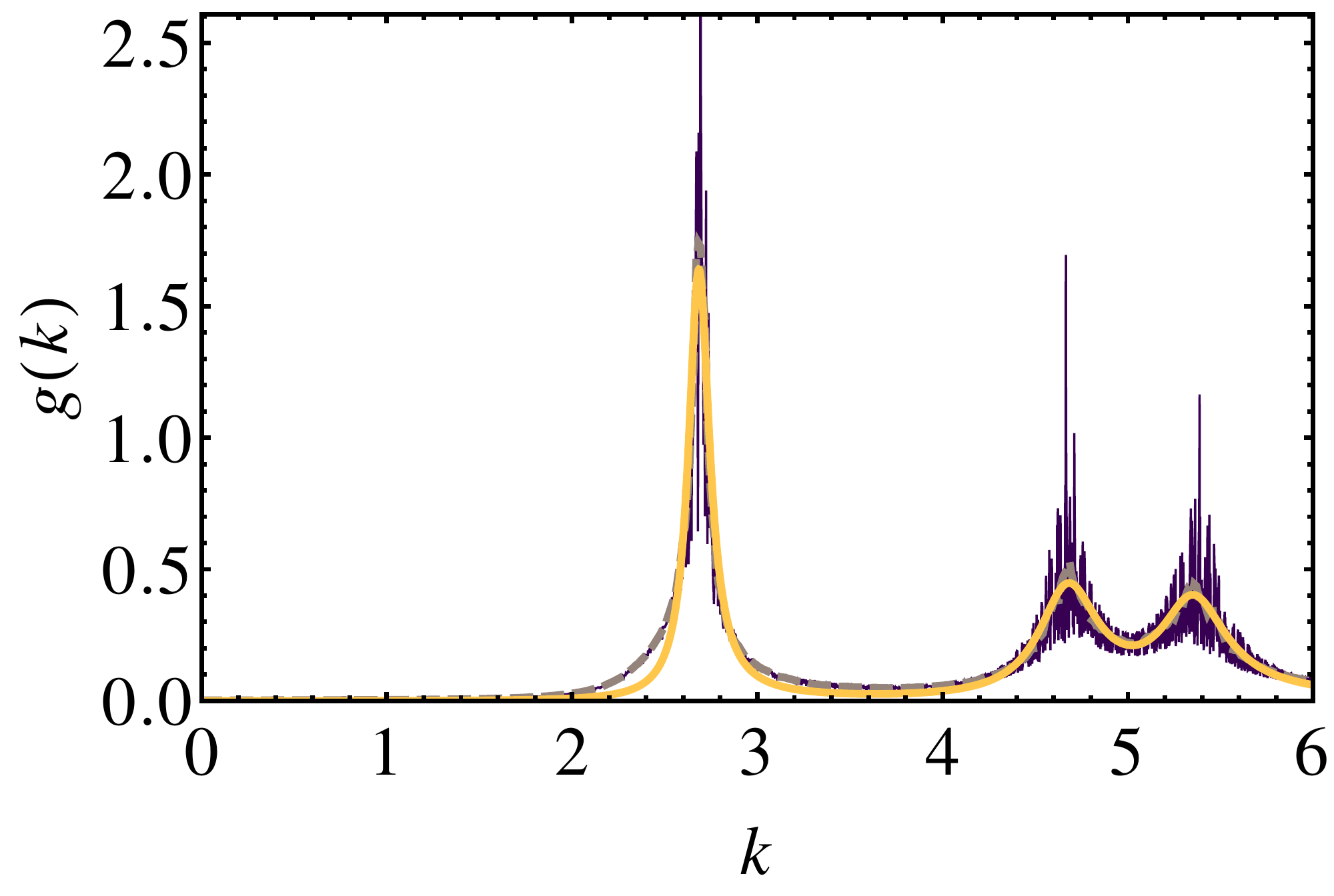}\includegraphics[width=4cm]{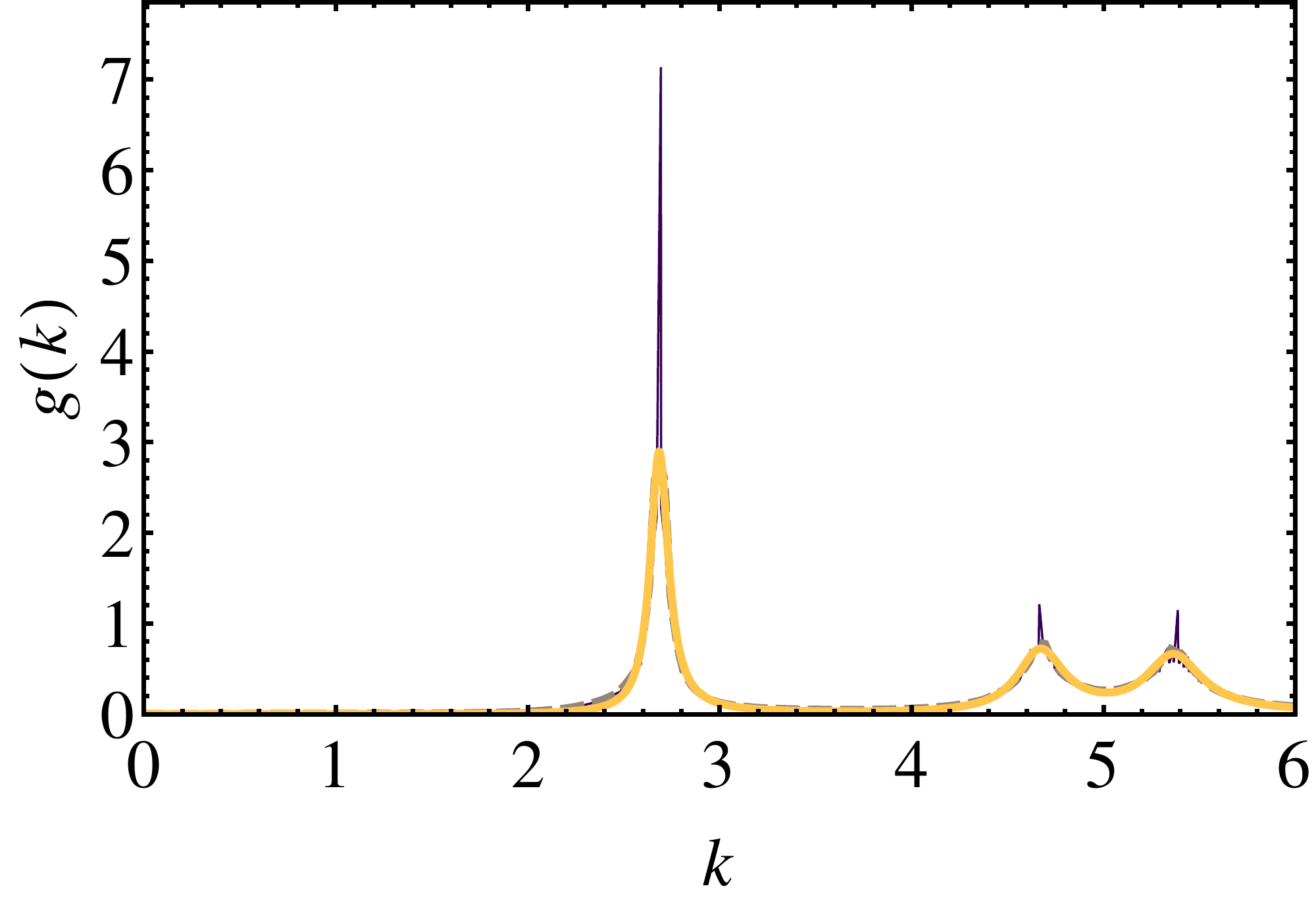}\\
\includegraphics[width=4cm]{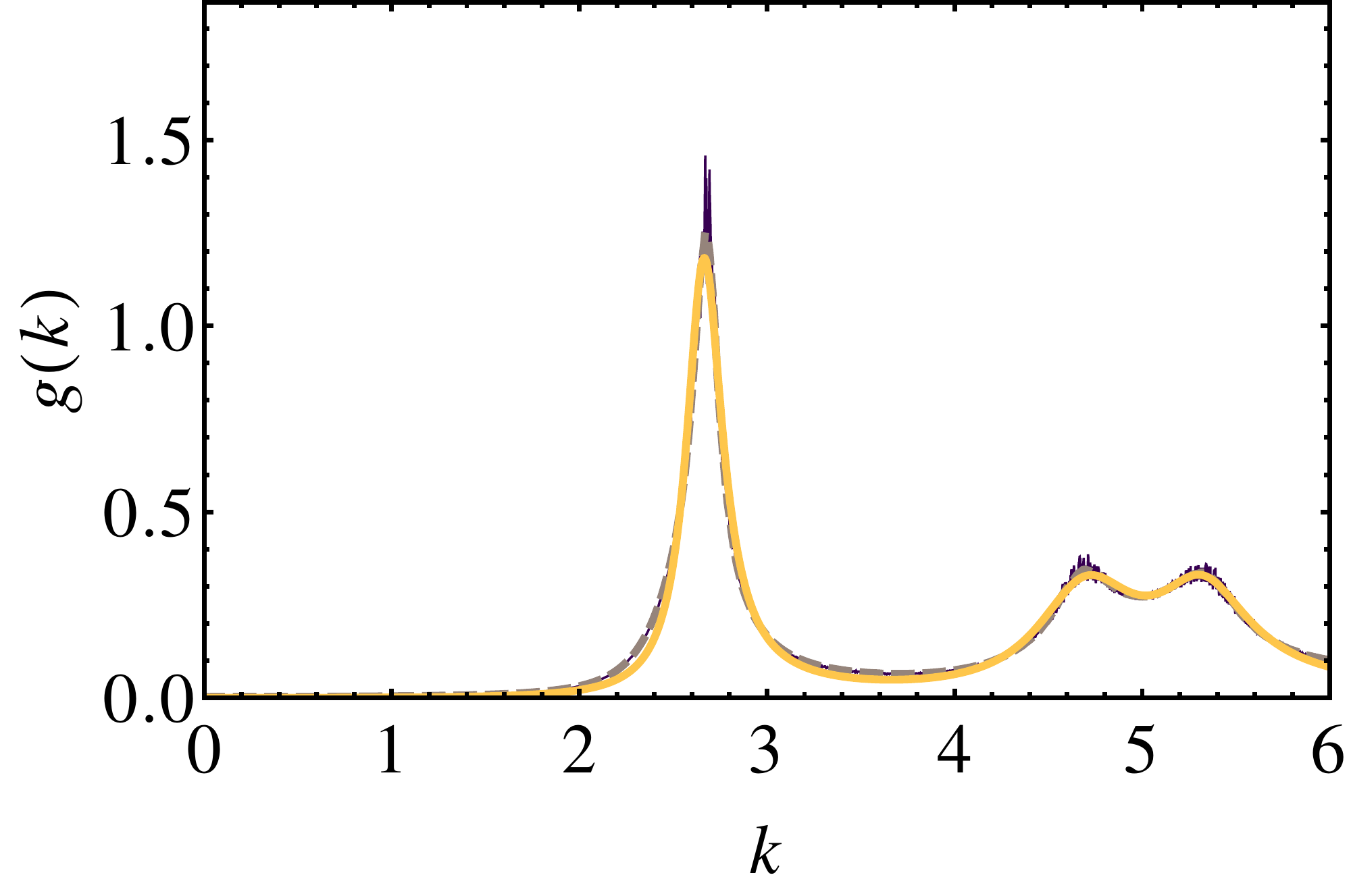}\includegraphics[width=4cm]{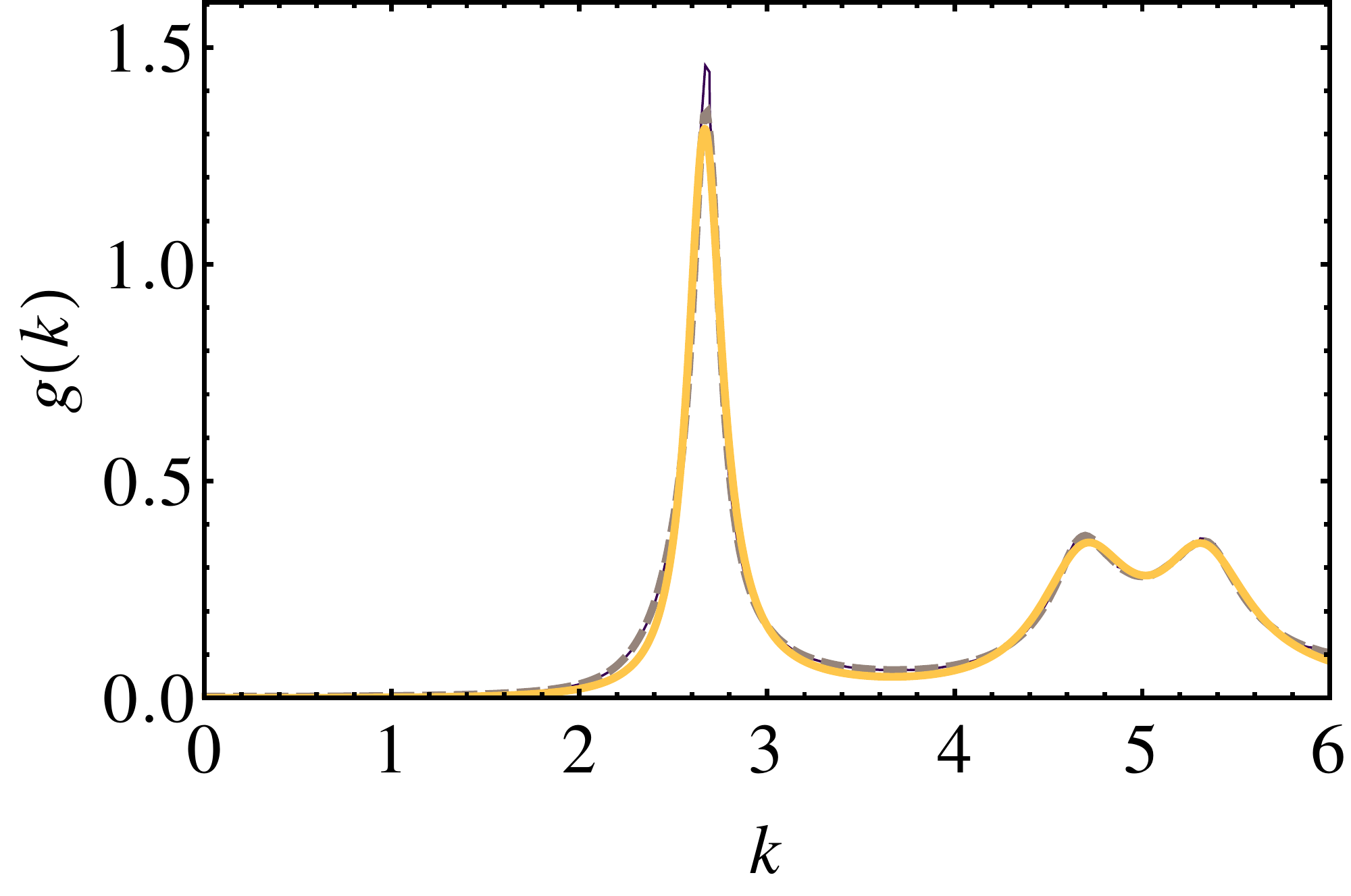}\\
\includegraphics[width=4cm]{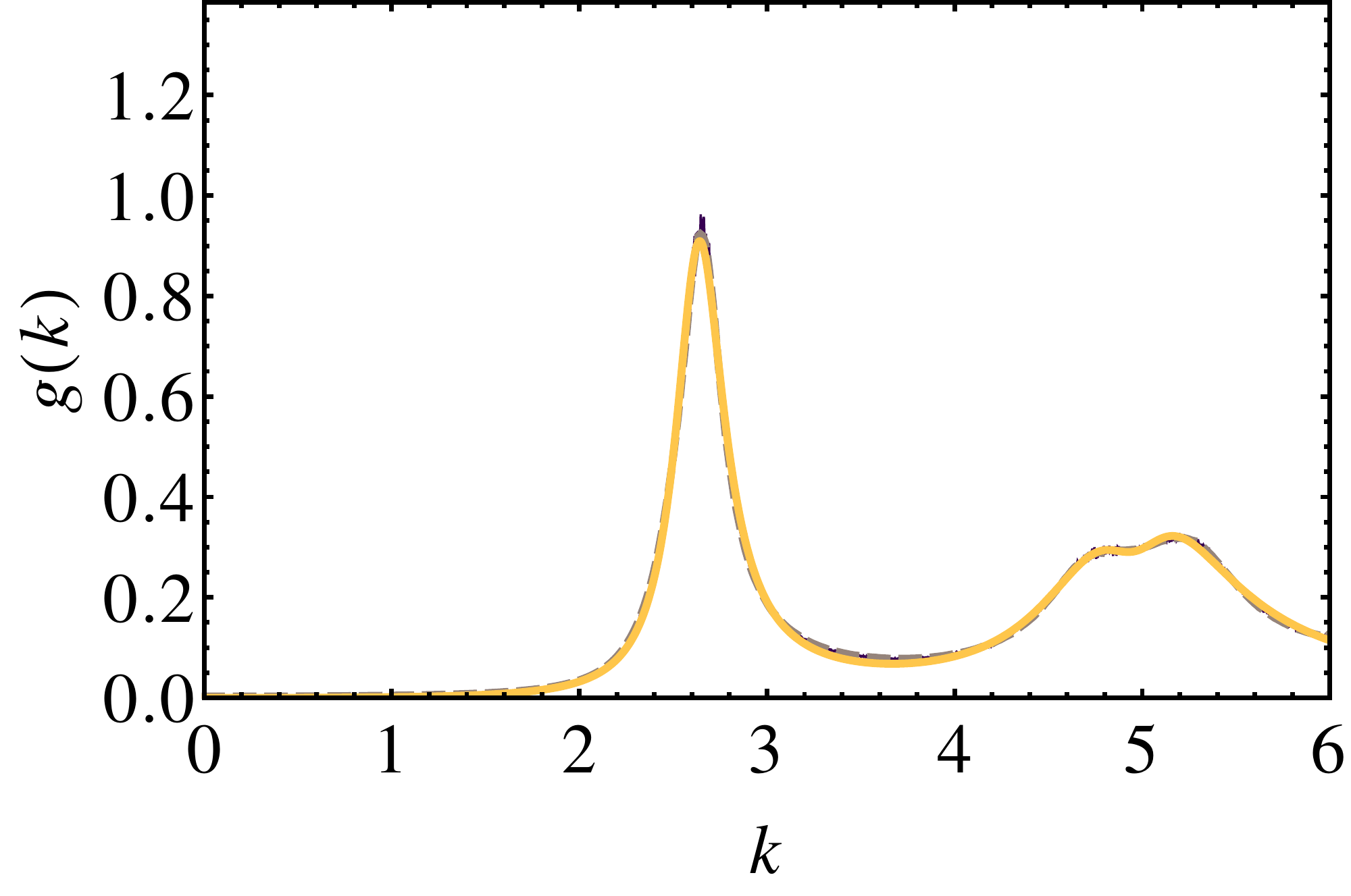}\includegraphics[width=4cm]{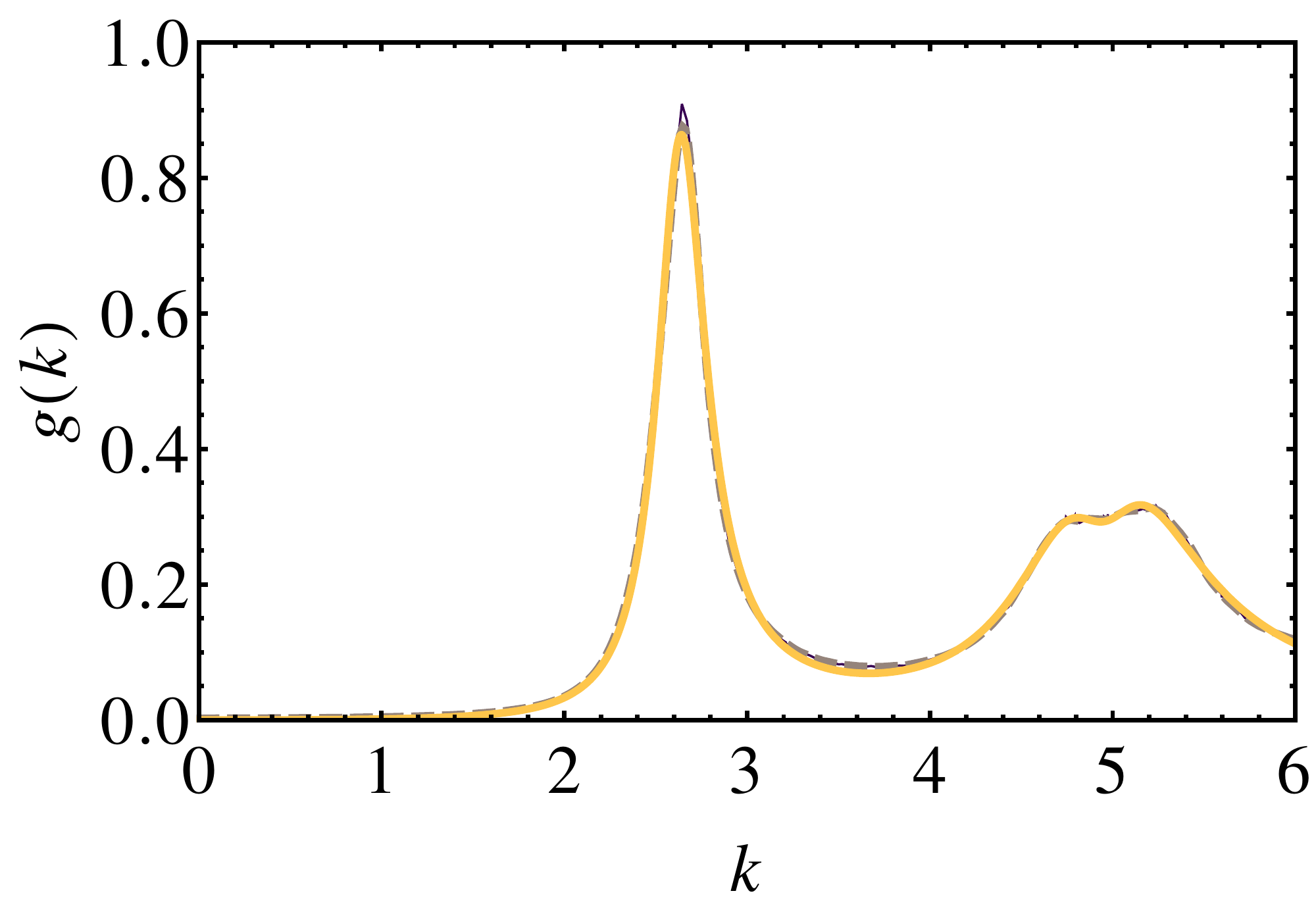}
\end{center}
\caption{(Color online) Correlation length-sample evaluation for
  $N=26\times26$, and energies (from top to bottom) $E/N=-0.974431$,
  $-0.971131$, $-0.967831$, $-0.964531$. The left column shows the
  angular average of the density-density correlations function; the
  right column shows the same for a narrow ray of width
  $\delta\theta=\pi/30$ about the crystalline axes. In each curve the
  thin dark purple curve is the data, and the thick orange curve is
  the result with a fit of a smoothed version of the data with a sum
  of Lorentzians multiplied with $k^2$ to take care of the correlation
  hole. Note that for the column at the right we use the width at the
  top, rather than the Lorentzian fit, to determine the correlation
  length.\label{fig:Correlation-length-sample-evaluation}}
\end{figure}

\paragraph{Correlation lengths}

\begin{figure*}
\begin{centering}
\includegraphics[width=16cm]{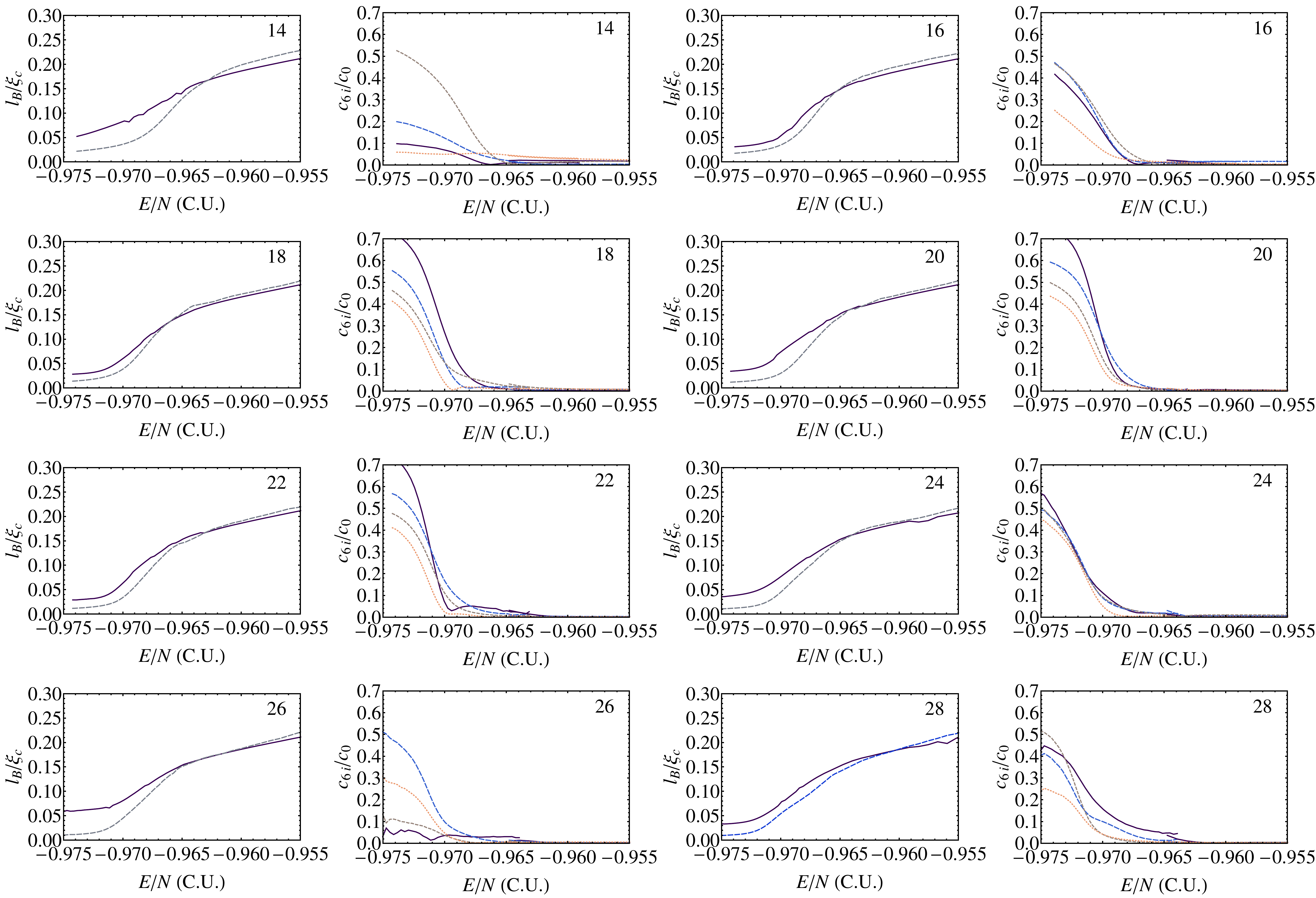}
\end{centering}
\caption{(Color online) Inverse crystalline correlation lengths (solid line: bulk and dashed
line: on-axis) and ``first-ring'' Fourier components ($c_{6i}$:
$i=1$ solid black, $i=2$: dashed gray, $i=3$: long-dashed blue,
$i=4$: dotted orange).\label{fig:first-ring correlationson}}
\end{figure*}

We find it hard to extract a crystalline correlation
length from the simulation data -- especially near the crystalline states.
If we collapse the data shown in Fig.~\ref{fig:DDcorrelations}
onto the radial axis by performing an angular integration -- which in
is just a sum over a finite number of grid points --  we typically
obtain data such as that shown in the left column of Fig.~\ref{fig:Correlation-length-sample-evaluation}.

One approach which we use to extract a correlation length is to fit
this data with a sum of Lorentzians, multiplied with a factor of
$k^{2}$ to take into account the correlation hole.  We then take the
length scale for the inverse width of the lowest peak as an estimate
of the crystalline correlation length.  This choice seems rather
obvious, but gives what is likely to be an overestimate of the width
(and thus an underestimate of the correlation length) at energies at
which the the rings are modulated.  An alternative definition of the
correlation length, would be to consider correlations only along (or
very near) the three ``natural'' crystal axes, aligned with the
simulation cell.  We then fit a Lorentzian peak through 5 points at
the top of the first peak of the correlation function, and we get a
correlation length that is substantially larger at low energies than
that obtained from the angular average, but rather similar at higher
energy. We shall call this the ``on-axis correlation length'', and use
this as our preferred measure of a crystalline correlation
length. Both definitions of the correlation length are plotted in
Fig.~\ref{fig:first-ring correlationson}.

\paragraph{Dependence of the correlation length on energy }

\begin{figure}
\begin{centering}
\includegraphics[width=6cm]{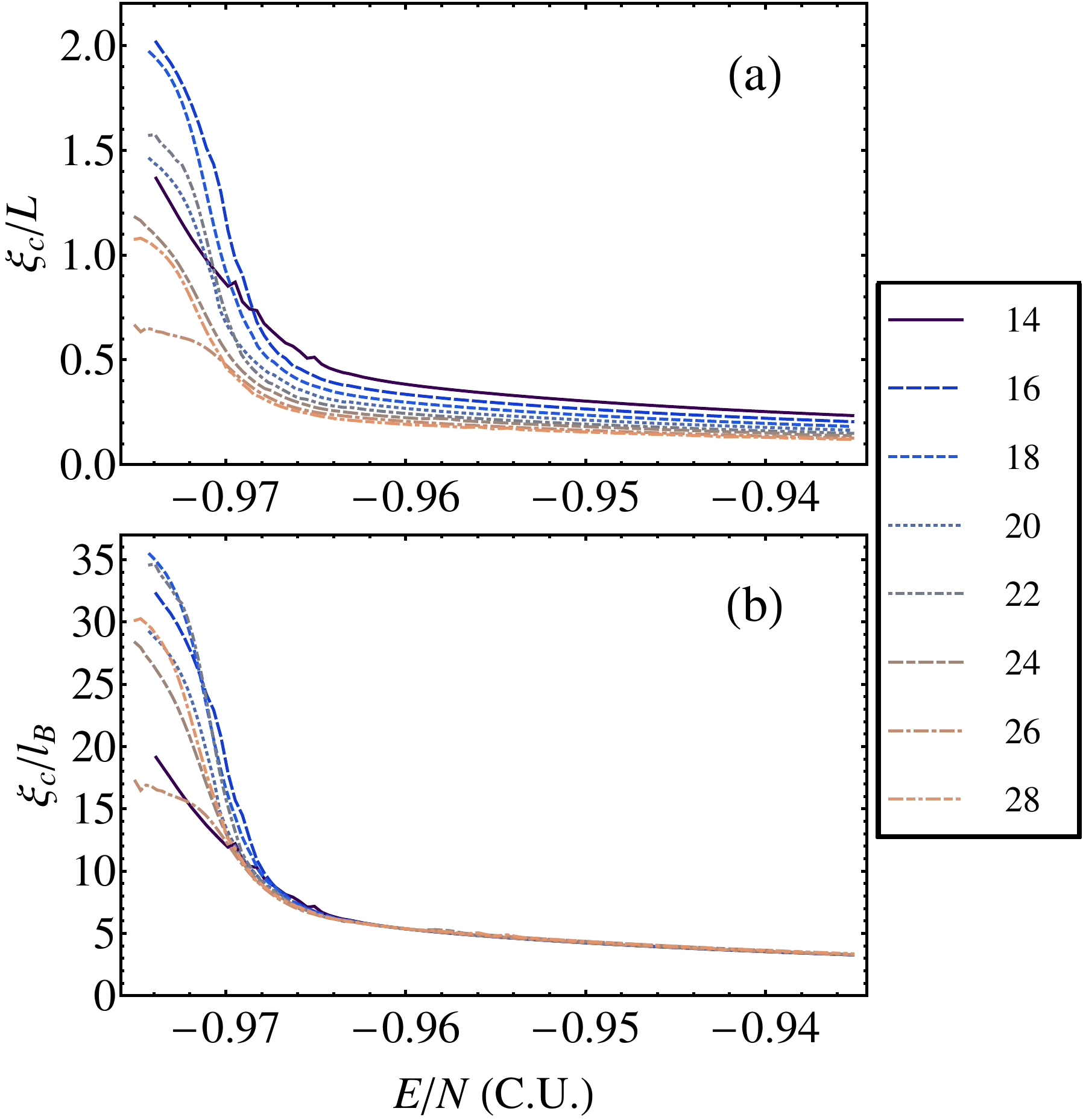}
\end{centering}
\caption{(Color online) Crystalline correlation lengths as a function of energy; (a): in units of the system size, $L=\sqrt{N}l_{B}$, (b): in units of the magnetic length $l_{B}$.\label{fig:Crystalline-correlation-lengths}}
\end{figure}

\begin{figure}
\begin{centering}
\includegraphics[width=6cm]{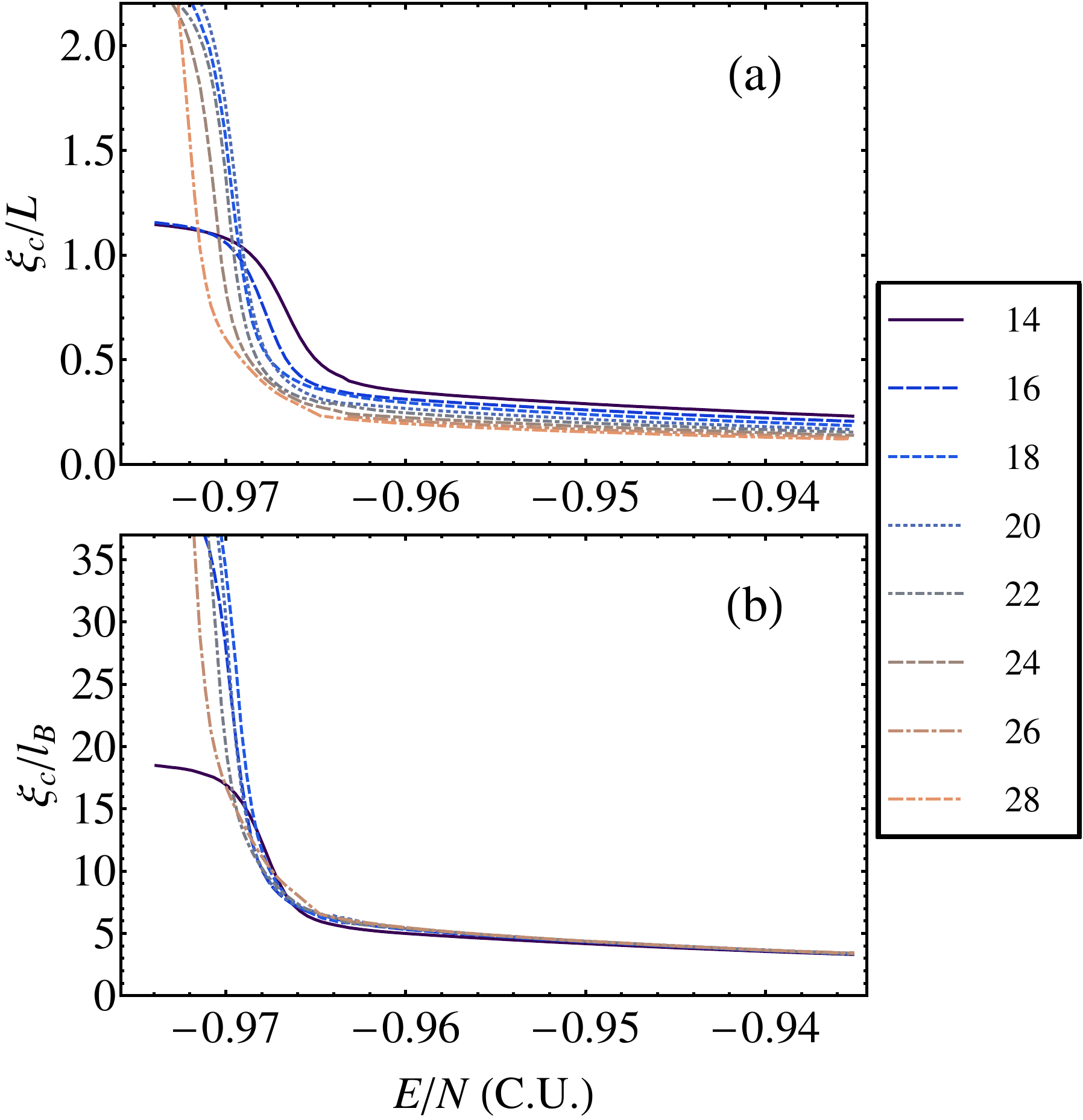}
\end{centering}
\caption{(Color online) On-axis crystalline correlation lengths as a function of energy; (a): in units of the system size, $L=\sqrt{N}l_{B}$, (b): in units of the magnetic length $l_{B}$.\label{fig:On-axis-crystalline-correlation}}
\end{figure}

We now can try to analyze  the behavior of the correlation lengths as
a function of energy. In  the left column of Fig.~\ref{fig:first-ring
  correlationson} we see that  the correlation lengths decrease quickly in
the  liquid as the energy $E/N$ increases, and  behaves  at  the crystal  end -- very  much like  the
behavior of the  density of defects shown in Fig.~\ref{fig:defect2}. 
 It once again shows
three regions  of behavior: liquid behavior,  which we shall
argue  below   is  linear  in   energy;  a  sudden  change   at  about
$E/N=-0.965$, and  a second change for  $E/N\simeq-0.972$. This latter
change is  hard to  interpret, because at  that point  the correlation
length is much larger than the system size, as we can see more clearly
in       Figs.~\ref{fig:Crystalline-correlation-lengths}      and
\ref{fig:On-axis-crystalline-correlation}.

 We  have  also   examined  the  modulation  of  the   first  ring  by
 constructing  the  Fourier  analysis  along  the first  ring  of  the
 density-density correlation  function. In other words  we have looked
 at  the  azimuthal intensity  profile  $I(\phi)=\sum_nc_{6n} \cos  6n
 \phi$;  see Fig.~\ref{fig:first-ring correlationson}.  The modulation
 sets in  around an energy $E/N \approx - 0.965$, but the  form it takes
 seems to  be sensitive to  the value of $\sqrt{N}$,  indicating again
 that low-lying crystalline states might be playing a significant
 role.

\paragraph{Scaling analysis}


One possible way to see whether there is any sign of a phase transition
in the crystal data is to perform a scaling analysis, and plot the
data as a function of the system size, taken as $L=\sqrt{N}l_{B}$.
As we can see in Figs.~\ref{fig:Crystalline-correlation-lengths}
and \ref{fig:On-axis-crystalline-correlation} neither of the definitions
scales as one would expect if there was a  phase transition, when plots of $\xi_c/L$ should cross at the critical value of the energy.  


We can perform a similar analysis for the hexatic correlation length.
These are expensive to calculate, and we have thus limited the results
to a few examples. In each case we have applied the contour finding
and improvement method on the data for the order parameter
field. Depending on energy, we had to reject between 15-25\% of our
results due to lack of or misidentification of (usually one) vortex,
and there thus is a possibility of a bias in the data. From the
positions we can through standard ways determine the hexatic
correlation length (essentially, the width of the peak at
$\vec{k}=0$.)

\begin{figure}
\begin{centering}
\includegraphics[width=6cm]{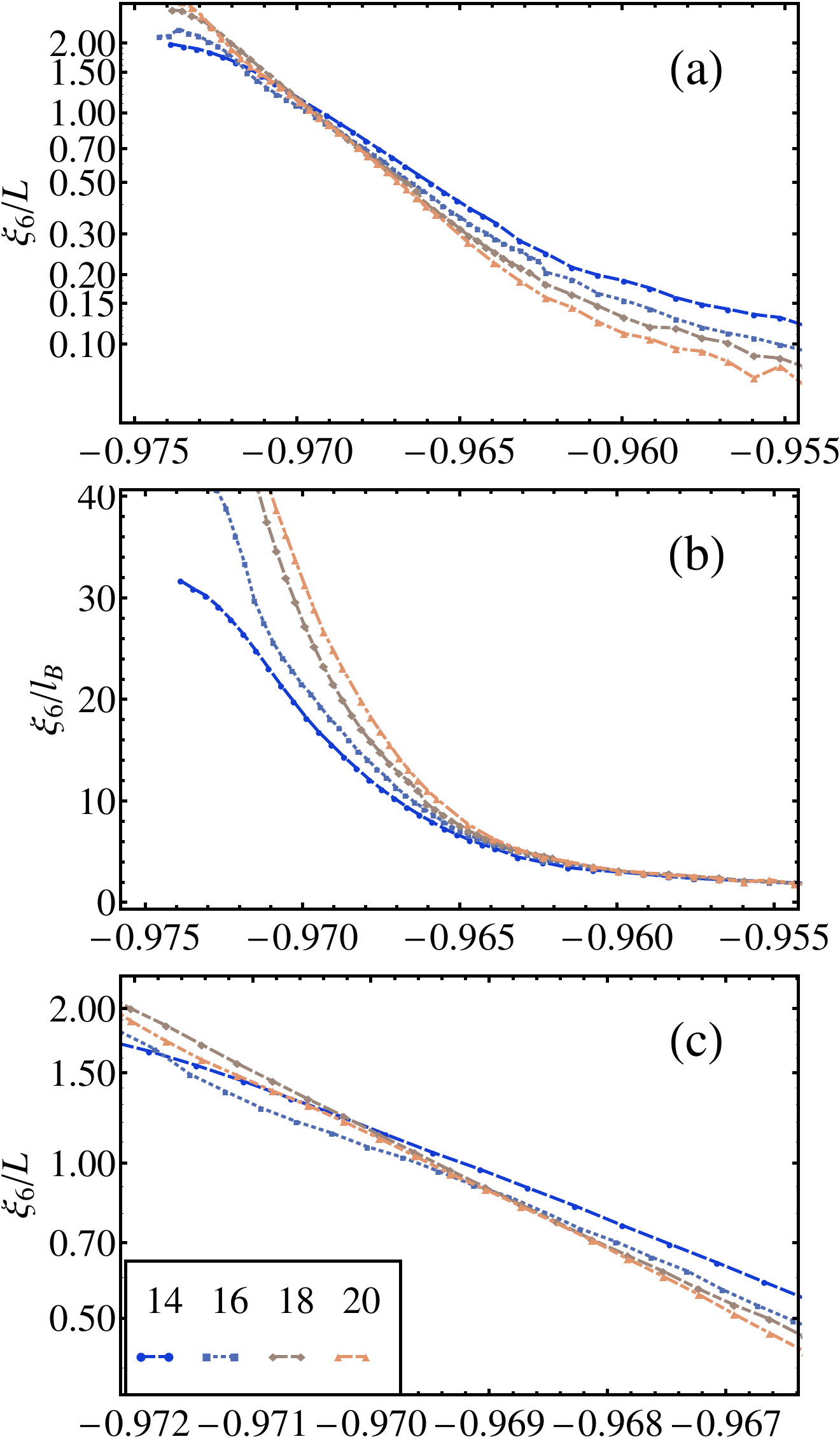}
\end{centering}
\caption{(Color online) Hexatic correlation lengths $\xi_6$ as a function of
  energy; (a): divided by $L=\sqrt{N}l_{B}$, (b): divided by the
  magnetic length $l_{B}$; (c) zooms in on the detail of the crossing in (a).\label{fig:hexatic}}
\end{figure}

As  we  can see  in  Fig.~\ref{fig:hexatic},  the hexatic  correlation
lengths do show  some of the behavior expected  for a phase transition
between liquid and hexatic liquid,  which is in the universality class
of the  XY model; they  even cross at  values of $\xi_6/L$  similar to
those of the XY model.  Clearly, we have not reached convergence and
we would benefit from additional results for larger systems, which are
unfortunately prohibitively computationally expensive.

\paragraph{Extrapolation from the liquid}

\begin{figure}
\centering{}\includegraphics[width=8cm]{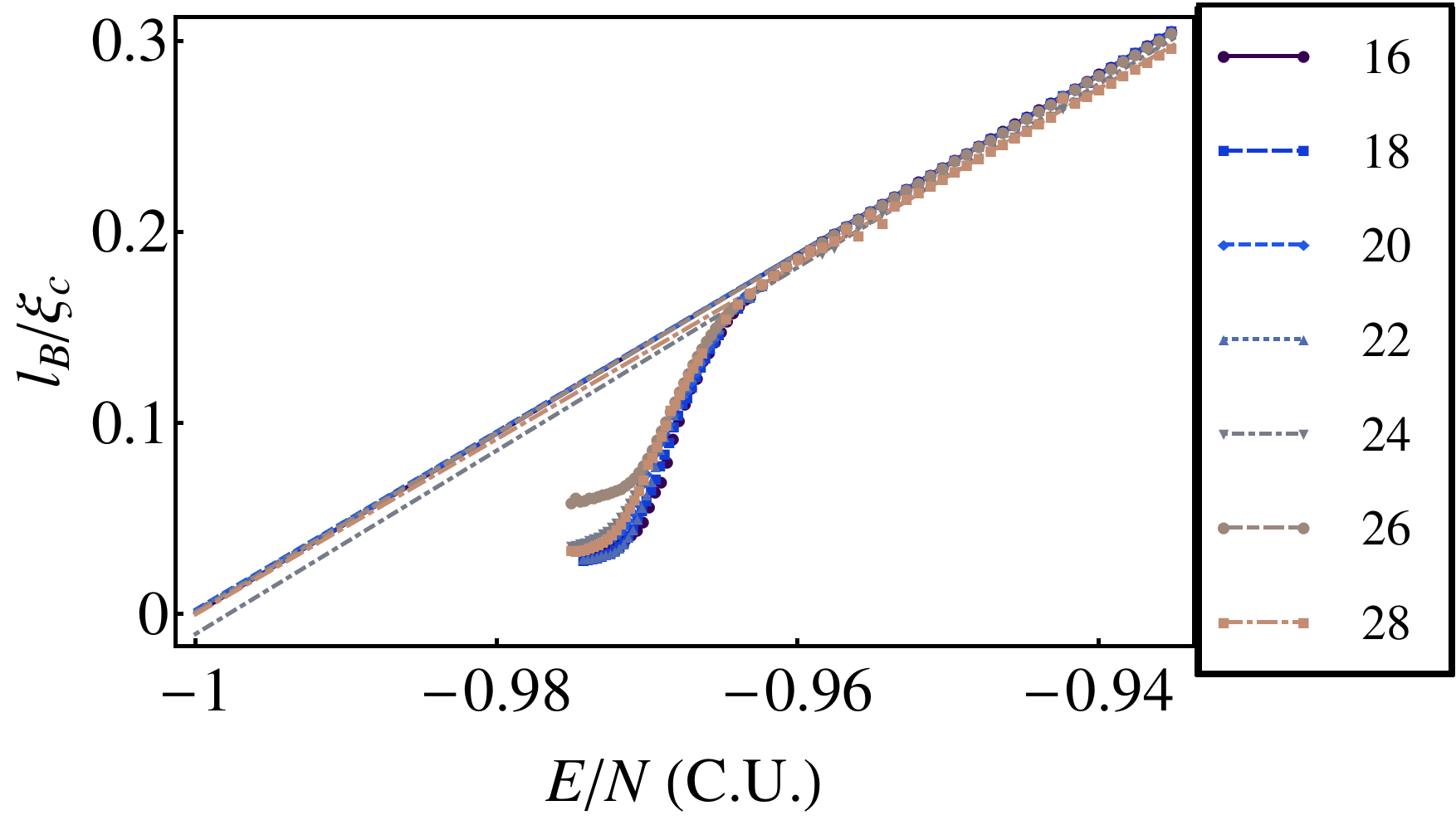}
\caption{(Color online) Crystalline correlation lengths: Extrapolation of the liquid behavior. The symbols show the various data sets; the lines are the extrapolation from the higher energy part of the data. 
\label{fig:liquid-extrapollations}}
\end{figure}

It          has          been          argued          in          the
past\cite{dodgson_vortices_1997,lee1994montecarlo,oneill_monte_1992}
that the  liquid might persist all  the way down  to zero temperature,
when it  forms the  Abrikosov crystal. If  we extrapolate  the inverse
correlation  length $l_B/\xi_c$ in  the liquid  regime (which  seem to
satisfy a naive linear relation $l_{B}/\xi_{c}=c(E/N-E_{0})$), we find
that all bar one of the  cases extrapolate with $E_{0}=-1$ -- and that
the  one  exceptional case  (at  $\sqrt{N}=24$)  is  still close,  see
Fig.~\ref{fig:liquid-extrapollations}.   The  simulation  data at  the
lower energies deviate rather strongly  from the fit, showing a change
of behavior for  an energy around $E/N=-0.965$, presumably due  to the
transition to  the hexatic phase.   In the earlier simulations  on the
sphere  the  correlation length  was  found  to  grow as  $\xi_c  \sim
|\alpha_B(T)|  l_B$ within  the  canonical ensemble.  Alas, we  cannot
confirm this  result as we  only collected data in  the microcanonical
ensemble  over the  energy range  close  to the  transitions, but  the
results might  be consistent  with such a  growth of $\xi_c$,  at least
until the  topological defects start to bind.   Notice that departures
from the  linear relation set  in rather sharply, indicating  that the
critical region of the hexatic transition is  small.

\paragraph{Shear Modulus (microcanonical) }

\begin{figure}
\begin{centering}
\includegraphics[width=6cm]{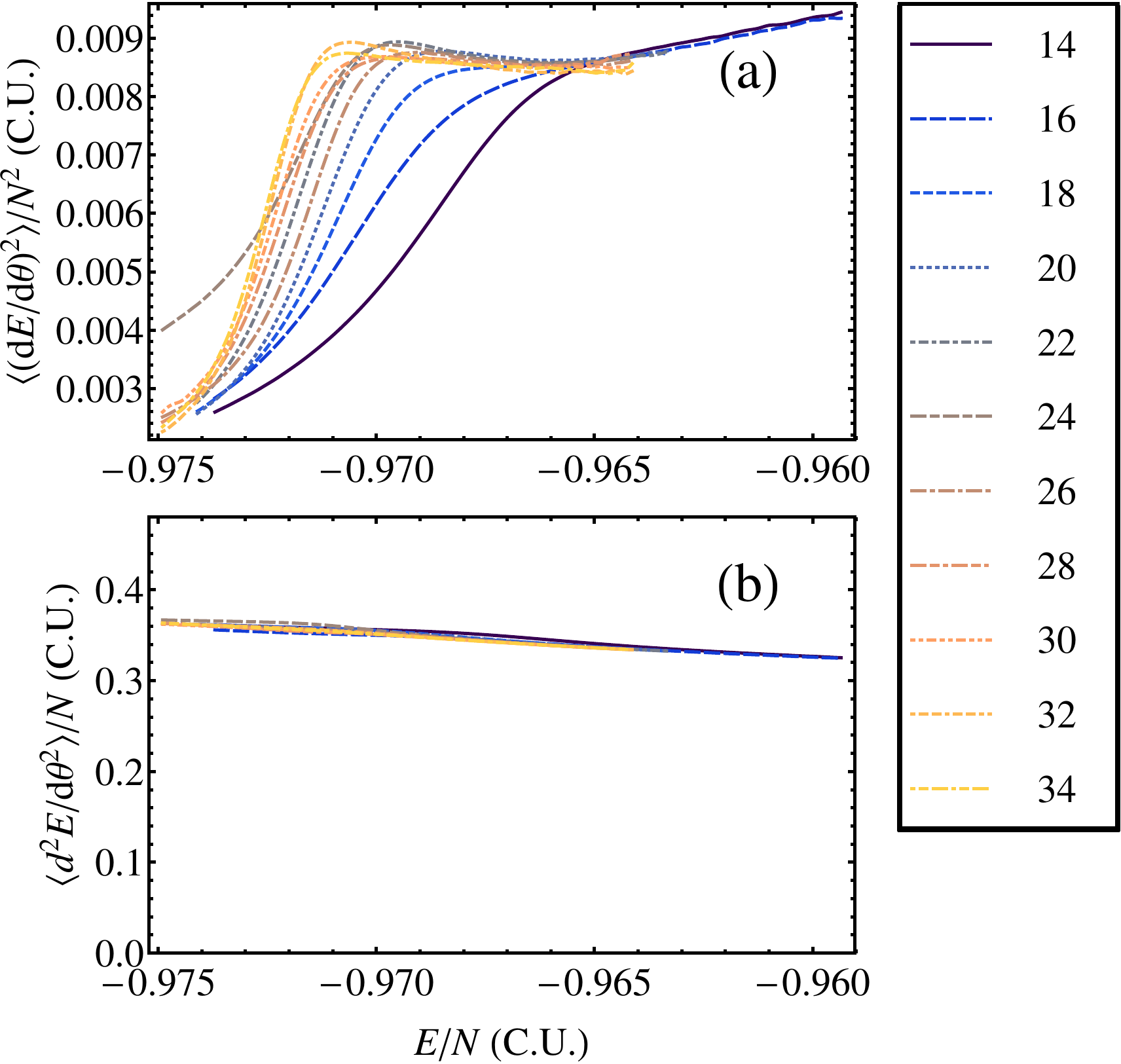}
\end{centering}
\caption{(Color online) The micro-canonical ingredients to the calculation of the shear modulus
as a function of energy, for values of $\sqrt{N}$ from $14$ to $34$. In (a) we show the shear susceptibility, in (b) the shear stiffness.
\label{fig:micro-canonical-shear}}
\end{figure}

In Fig.~\ref{fig:micro-canonical-shear} we show the microcanonical
ingredients that enter the calculation of the shear modulus. Clearly
all of the structure in the shear modulus is caused by the shear susceptibility
$\langle(dE/d\theta)^{2}\rangle$. Its behavior is once again suggestive
of two places of changing behavior. The first change of behavior
occurs again at $E/N\approx-0.965$, where the slope of the curve
changes sign; the second change occurs at energy probably below $E/N=-0.9715$,
where we see a sharp drop, and a change to crystalline behavior (where
the shear susceptibility decays to zero). The first of these changes
is much smoother than the second.

\section{Discussion and Conclusions}
\label{discussion}

Our main conclusion is that  the old simulational evidence for a first
order phase  transition to  the crystal state  from the  vortex liquid
state is just an artifact of  the rather small system sizes which were
previously studied.  The evidence for  this statement is  contained in
Fig.~\ref{fig:Interface-free-energy}. If  we had  only  studied sizes
up to $\sqrt{N}=26$, there  would have been good evidence  for a first
order  transition.  However,  the   interface  free  energy  shown  in
Fig.~\ref{fig:Interface-free-energy}  collapses  towards zero  for  larger
systems, implying that the transition is just not first-order.

The apparent first-order transition reported in earlier numerical
studies with quasi-periodic boundary conditions
\cite{kato1993firstorder,kato1993montecarlo,hu1994correlations,
  vsavsik1994calculation,vsavsik1995phasecoherence, li2003fluxlattice}
is, we suspect, both a finite size effect and a consequence of the use
of periodic boundary conditions. When the vortices move on the surface
of a sphere there was no evidence of a first-order transition
\cite{dodgson_vortices_1997,lee1994montecarlo,oneill_monte_1992}. We
now suspect that the transition between crystal and hexatic liquid is
continuous, and this will occur according to the KTHNY picture when
the bound dislocations in the crystalline state unbind.  However, for
quasi-periodic boundary conditions the system can prematurely lower
its energy by going to the crystalline state (probably by an amount
not of order $N$ but just of magnitude
$k_BT^{\text{MF}}_c\alpha_B(T)^2$, see Eq.(\ref{distortenergy})), and
this small amount of energy is sufficient to upset the delicate
balance of free energies of liquid and crystal and produce the
apparent first-order transition, seen for small systems.  It is an
interesting question which set of boundary conditions,
quasi-periodic or spherical, converges  to the thermodynamic limit faster,
but the evidence of this paper is that at least in the liquid regime,
spherical boundary conditions might have the advantage. 

In the KTHNY scenario there should be two transitions: The first is
the one from a normal liquid to an hexatic liquid phase, and the
second transition is that from the hexatic liquid to the crystal. We
believe that we have strong evidence for the liquid-hexatic
transition, see Fig.~\ref{fig:hexatic}. The crossing of the curves of
$\xi_6/L$ occurs just as would be expected for a transition in the XY
universality class. Alas, the equivalent curves of $\xi_c/L$ for the
hexatic to crystal transition fail to show a similar crossing.  This
is at least partially due to the magnitude of the crystalline
correlations for low energies, which substantially exceed the system
size. Fortunately the shear modulus (see
Fig.~\ref{fig:micro-canonical-shear}a), provides strong evidence for
the expected jump in its value at the transition.

We suspect that the failure to  see crossing of the plots of $\xi_c/L$
is  another  consequence  of  finite  size  effects,  which  are  very
noticeable in  the hexatic phase. The  density-density correlations as shown in
Fig.~\ref{fig:DDcorrelations}  
depends  on the  orientation of $\vec{q}$ with respect to the
boundaries of the simulational cell and a detailed examination of this
modulation is  given in Fig.~\ref{fig:first-ring  correlationson}.  In
the hexatic phase Peterson and Kagener\cite{peterson_diffraction_1994}
showed  that   will  be  no  such  modulation   in  the  thermodynamic
limit. They  derived a formula for  the length scale  $L^*$ which, for
sizes $L>L^*$,  the modulation will  disappear.  In our  notation $L^*
\sim \xi_c  \exp[c \alpha_B(T)^2]$, but the numerical  constant $c$ is
not  known.    The  results  in   Figs.~\ref{fig:DDcorrelations}  and
\ref{fig:first-ring  correlationson} indicate that  in our  studies of
the hexatic  phase we have not  reached this limit  and that therefore
finite size effects might still play a significant role.

Taking all of the evidence together we estimate the crystal to hexatic
transition to be at $E/N=-0.9725\pm0.0005\text{ C.U.}$ and the hexatic
to normal liquid one at $E/N=-0.9645\pm0.0005\text{ C.U.}$. The latter
corresponds to $\alpha_B(T_{hl})\approx-9.5$, in agreement with our
calculations of the canonical specific heat and shear modulus. We
can only put an upper bound on the hexatic to crystal value,
$\alpha_B(T_{ch})< -9.7$.

One striking feature  of our simulations is the  pronounced but narrow
peak       in       the        specific       heat       seen       in
Fig.~\ref{fig:Peak-height-and}. Given its prominence, it is surprising
that     experiments     like      that     of     Urbach     \emph{et
  al.}\cite{urbach_specific_1992}  failed  to see  it.  We would  urge
further  experiments  to understand  this  discrepancy,  which we  are
inclined to attribute to variations in the thickness of their films, which would tend to smear out the peak in the specific heat.

Finally we  make a  comment on  the Monte Carlo  simulations. It  is a
tacit  assumption of  simulation  that  the sizes  one  can reach  are
sufficiently  large that  one can  make useful  statements  about what
happens in the thermodynamic  limit. The old simulations with periodic
boundary conditions  were mainly  done for $\sqrt{N}  \leq 16$  and so
were  naturally  reported as  providing  evidence  for  a first  order
transition.  They  could not  anticipate the trend  which sets  in for
$\sqrt{N}>  26$. By  the same  token,  large finite  size effects  are
clearly still present  in our work, and going to  larger values of $N$
might still  conceivably produce a  different story.  We  already have
substantially improved  upon the sizes which  were previously studied,
not  only  by   using  better  computers,  but  mainly   by  a  better
algorithm.  Without  the  adoption  of the  optimal  energy  diffusion
algorithm   we  would   never   have  been   able   to  converge   the
simulations. The main bottleneck to  increasing the size of the system
being simulated is the barrier to energy diffusion.  So in order to do
such simulations  a better  algorithm will have  to be  considered. We
believe the  most likely  place improvements could  be made is  in the
basic Monte Carlo step, but we  have no alternative to propose at this
stage.

\begin{acknowledgments}
  Part of the computational element of this research was achieved
  using the High Throughput Computing facility of the Faculty of
  Engineering and Physical Sciences, The University of Manchester. We
  would also like to thank Victor Martin-Mayor for his comments and
  for carrying out simulations to determine the extent of the XY
  universality class.
\end{acknowledgments}
\appendix

\section{Quartic terms in the interactions\label{app:quartic}}
\begin{widetext}
To calculate the quartic interaction terms in the Ginzburg-Landau 
functional we need to calculate the
integral of  $|\Psi|^{4}$ over the simulational cell. We again expand in the basis (\ref{eq:phi2}).
First we integrate over $y$,
\begin{equation}
\int_{0}^{L_{y}}dy\, g_{j_{1},s_{1}}^{*}(y)g_{j_{2},s_{2}}^{*}(y)g_{j_{3},s_{3}}(y)g_{j_{4},s_{4}}(y)=\frac{1}{L_{y}}\delta_{j_{1}+j_{2}+N(s_{1}+s_{2}),j_{3}+j_{4}+N(s_{3}+s_{4})}\,,
\end{equation}
and then simplify the calculation  over $x$ using
the Kronecker delta from the $y$ integral above to
\begin{equation}
\int_{0}^{L_{x}}dx\, f_{j_{1},s_{1}}^{*}(x)f_{j_{2},s_{2}}^{*}(x)f_{j_{3},s_{3}}(x)f_{j_{4},s_{4}}(x)=\frac{e^{-\frac{4\pi\left(n_{p_1}^{2}+n_{p_2}^{2}\right)}{\sqrt{3}N_{y}^{2}}}}{\pi l_{B}^{2}}\int_{0}^{L_{x}}dx\,\exp\left(-\frac{2N_{x}^{2}}{\sqrt{3}\pi N^{2}}\left(\theta_{y}+2\pi(n_{s}-Nx/L_{x})\right)^{2}\right),\label{eq:4fs1}
\end{equation}
where
\begin{align}
n_{p_1} & =j_{1}-j_{2}+N(s_{1}-s_{2}),\nonumber \\
n_{p_2} & =j_{3}-j_{4}+N(s_{3}-s_{4}),\nonumber \\
2n_{s} & =j_{1}+j_{2}+j_{3}+j_{4}+N(s_{1}+s_{2}+s_{3}+s_{4})
=2(j_{1}+j_{2}+N(s_{1}+s_{2})).\label{eq:nsnp}
\end{align}
We now perform the sum over the variables $s_{i}$ which is required in
each of the basis functions, see Eq.~(\ref{eq:phi2}). Changing the
summation over $s_i$ to the variables $s_{p_1}$, $s_{p_2}$ and
$s_{s}$, where we write
\begin{equation}
n_{\alpha}=\left\lfloor n_{\alpha}\right\rfloor + 2Ns_{\alpha}
\end{equation}
and $\left\lfloor n_{\alpha}\right\rfloor $ is an integer ranging from
$0$ to $2N-1$, and the label $\alpha=p_1,p_2,s$. The variables $s_{\alpha}$
are three independent integers. We also see that  $n_{p_1}$,
$n_{p_2}$, $n_{s}$ are  all integers, and that their pairwise sums must all be
even.
If we now perform the sum over $s_{s}$ in the expression (\ref{eq:4fs1}),
we find that we can write
\begin{align}
\sum_{s_{s}=-\infty}^\infty\int_{0}^{L_{x}}dx\, f_{j_{1},s_{1}}^{*}(x)f_{j_{2},s_{2}}^{*}(x)f_{j_{3},s_{3}}(x)f_{j_{4},s_{4}}(x) & =\frac{e^{-\frac{\pi\left(n_{p_1}^{2}+n_{p_2}^{2}\right)}{\sqrt{3}N_{y}^{2}}}}{\pi l_{B}^{2}}L_{x}\int_{-\infty}^{\infty}d\xi\,\exp\left(-\frac{2N_{x}^{2}}{\sqrt{3}\pi N^{2}}\left(\theta_{y}+2\pi(\left\lceil n_{s}\right\rceil -N\xi)\right)^{2}\right)\nonumber \\
 & =\frac{e^{-\frac{\pi\left(n_{p_1}^{2}+n_{p_2}^{2}\right)}{\sqrt{3}N_{y}^{2}}}}{\pi l_{B}^{2}}\sqrt{\frac{\pi}{2}}l_{B}
 =\frac{e^{-\frac{\pi n_{p_1}^{2}}{\sqrt{3}N_{y}^{2}}}e^{-\frac{4\pi n_{p_2}^{2}}{\sqrt{3}N_{y}^{2}}}}{\sqrt{2\pi}l_{B}}.
\end{align}
Taking all of this together we get
\begin{multline}
\int dxdy\,\phi_{j_{1}}^{*}(x,y)\phi_{j_{2}}^{*}(x,y)\phi_{j_{3}}(x,y)\phi_{j_{4}}(x,y) \nonumber \\
 =\delta_{\left\lceil j_{1}+j_{2}-j_{3}-j_{4}\right\rceil ,0}\sum_{s_{p_1}s_{p_2}=-\infty}^\infty\frac{e^{-\frac{4\pi\left(\left\lceil n_{p_1}\right\rceil +2Ns_{p_{1}}\right)^{2}}{\sqrt{3}N_{y}^{2}}}e^{-\frac{4\pi\left(\left\lceil n_{p_2}\right\rceil +2Ns_{p_{2}}\right)^{2}}{\sqrt{3}N_{y}^{2}}}}{\sqrt{2}\sqrt[4]{3}\pi l_{B}^{2}N_{y}}\nonumber \\
  =\delta_{\left\lceil j_{1}+j_{2}-j_{3}-j_{4}\right\rceil ,0}\frac{\sqrt[4]{3}N_{y}}{4\sqrt{2}\pi l_B^{2}N^{2}}\theta_{3}\left(\frac{n_{p_1} \pi}{2N}\middle|e^{-\frac{\sqrt{3}N_{y}^{2}\pi}{4N^{2}}}\right)\theta_{3}\left(\frac{n_{p_2} \pi}{2N}\middle|e^{-\frac{\sqrt{3}N_{y}^{2}\pi}{4N^{2}}}\right).
\end{multline}
The quartic term in the  GL free energy is given by
\begin{align}
E_4(\{c\})&=
\frac{\pi^2}{2^{5/2}3^{1/4}\text{ }N_{y}}\sum_{n_{s}=0}^{2N-1}\left|\sum_{n_{p}=0}^{2N-1}\delta_{n_{s}+n_{p},\text{even}}\left[\sum_{s_{p}=-\infty}^{\infty}e^{-\frac{\pi(n_{p}+Ns_{p})^{2}}{\sqrt{3}N_{y}^{2}}}\right]c_{\left\lceil (n_{p}+n_{s})/2\right\rceil }c_{\left\lceil (n_{p}-n_{s})/2\right\rceil }\right|^{2}\nonumber\\
&=\frac{\pi^2}{2^{5/2}3^{1/4}\text{ }N_{y}}\sum_{n_{s}=0}^{2N-1}\left|Q_{n_{s}}\right|^{2}.\label{eq:E4}
\end{align}
Here $Q_{n_s}$ is as  given in Eq.~(\ref{eq:Qns}).

\subsection{Shear Modulus\label{app:shear}}

To calculate the shear modulus we need to calculate the $\theta$ dependent
energy.  We expand in the basis Eq.~(\ref{eq:phitheta}). The $y$
integral gives the usual Kronecker delta, and it is really
straightforward to show that the quadratic term in the energy is
unchanged, whereas in the quartic term we can use the technique shown
above to find
\begin{align}
\sum_{s_{s}=-\infty}^\infty\int_{0}^{L_{x}}dx\, f_{j_{1},s_{1}}^{*}(x,\theta)f_{j_{2},s_{2}}^{*}(x,\theta)f_{j_{3},s_{3}}(x,\theta)f_{j_{4},s_{4}}(x,\theta) & =
 \frac{e^{-\frac{(\pi-i\tan\theta/2) n_{p_1}^{2}}{\sqrt{3}N_{y}^{2}}}e^{-\frac{(\pi+i\tan\theta/2) n_{p_2}^{2}}{\sqrt{3}N_{y}^{2}}}}{\sqrt{2\pi}l_{B}}.
\end{align}
Thus the quartic term in the energy is as  given in Eqs.~(\ref{eq:Etheta},\ref{eq:QNstheta}).
\end{widetext}

\section{Evaluation of density-density correlation function\label{app:dens-dens}}

Here we evaluate the density-density correlation function. We start from 
\begin{align}
g(\vec{q)} & \equiv\int d^{2}r\, d^{2}r'e^{i\vec{q}.(\vec{r}-\vec{r}')}\nonumber \\
 & \quad\times\left[\left\langle \left|\Psi(r)\right|^{2}\left|\Psi(r')\right|^{2}\right\rangle -\left\langle \left|\Psi(r)\right|^{2}\right\rangle \left\langle \left|\Psi(r')\right|^{2}\right\rangle \right]/\mathcal{N}^{4}\nonumber \\
 & =\left\langle \left|\rho(\vec{q})\right|^{2}\right\rangle -\left|\left\langle \rho(\vec{q})\right\rangle \right|^{2},
\end{align}
with 
\begin{align}
\rho(\vec{q}) & =\left\langle \int d^{2}r|\Psi(r)|^{2}e^{i\vec{q}\cdot\vec{r}}\right\rangle /\mathcal{N}^{2}.
\end{align}
We now multiply by the normalization constant squared, since we are only
interested in the relative magnitude of correlations.
We now evaluate $\rho(\vec{q})$. We choose 
\begin{equation}
q_{y}=k_{y}/l_{B}=m_{y}2\pi/L_{y}=\frac{2\sqrt{\pi}m_{y}}{\sqrt[4]{3}l_{B}N_{y}},
\end{equation}
 and find that that the $y$-integral in this quantity can easily
be evaluated as
\begin{equation}
\int_{0}^{L_{y}}dy\, e^{iq_{y}y}g_{j_{1},s_{1}}(y)^{*}g_{j_{2}s_{2}}(y)=\delta_{n_{1}-n_{2},n_{y}}=\delta_{n_{p},m_{y}},
\end{equation}
where $n_{i}=j_{i}+Ns_{i}$ is the ``unrestricted'' summation variable.
Once again we can replace the summation variables by $n_{p}=n_{1}-n_{2}$,
$n_{s}=n_{1}+n_{2}$. 

This is also helpful with the $x$-integral, which can now be rewritten
as 
\begin{equation}
\int_{0}^{1}dt\,\frac{2N_{x}\exp\left(-\frac{4\pi\left(n_{p}^{2}/4+(n_{s}/2-Nt)^{2}\right)}{\sqrt{3}N_{y}^{2}}+\frac{2i\sqrt{\pi}lN_{x}q_{x}t}{\sqrt[4]{3}}\right)}{\sqrt[4]{3}},
\end{equation}
where $t=x/L_{x}$. We still need to sum this result over all $n_{s}$;
we write $n_{s}=j_{s}+N_{x}N_{y}s_{s}$, (where $j_{s}$ is still
half-integral, but $s_{s}$ is integer) and sum over all $s_{s}$.
Using the fact that we can combine $t$ with each $s_{s}$, which
we can rewrite as a shift in the integration boundaries by $s_{s}$,
we find we can replace the sum over $s_{s}$ with a change in the
integration boundaries to 
\begin{equation}
\int_{-\infty}^{\infty}dt\,\frac{2N_{x}\exp\left(-\frac{4\pi\left(n_{p}^{2}/4+(j_{s}/2-Nt)^{2}\right)}{\sqrt{3}N_{y}^{2}}+\frac{2i\sqrt{\pi}lN_{x}q_{x}t}{\sqrt[4]{3}}\right)}{\sqrt[4]{3}},
\end{equation}
with 
\begin{equation}
q_{x}=k_{x}/l_{B}=2\pi m_{x}/L_{x}=\frac{\sqrt[4]{3}\sqrt{\pi}m_{x}}{l_{B}N_{x}}.
\end{equation}
The result of this integral is 
\[
\exp\left(\frac{i\pi j_{s}m_{x}}{N_{x}N_{y}}-\frac{\sqrt{3}\pi m_{x}^{2}}{4N_{x}^{2}}-\frac{4\pi n_{p}^{2}}{\sqrt{3}N_{y}^{2}}\right).
\]
Now combining $x$ and $y$ integrals with the coefficients in expansion
of the wave function
\begin{eqnarray}
\rho(\vec{q}) & = & \sum_{j_{s}=0}^{2N-1}c_{\left\lceil \left(j_{s}+n_{y}\right)/2\right\rceil }^{*}c_{\left\lceil \left(j_{s}-n_{y}\right)/2\right\rceil }\nonumber \\
 &  & \quad\times\exp\left(-\frac{\sqrt{3}\pi m_{x}^{2}}{4N_{x}^{2}}-\frac{\pi m_{y}^{2}}{\sqrt{3}N_{y}^{2}}\right)\exp\left(\frac{i\pi m_{x}j_{s}}{N}\right)\nonumber \\
 & = & \exp\left(-\frac{1}{4}\left(k_{x}{}^{2}+k_{y}{}^{2}\right)\right)\nonumber \\
 &  & \times\sum_{j'=0}^{N-1}c_{j'}c_{\left\lceil j'+m_{y}\right\rceil }^{*}\exp\left(\frac{2i\pi m_{x}(j'+m_{y}/2)}{N}\right)\nonumber \\
 & = & \exp\left(-\frac{1}{4}\left(k_{x}{}^{2}+k_{y}{}^{2}\right)\right)\exp\left(\frac{ik_{x}k_{y}}{2}\right)\nonumber \\
 &  & \times\sum_{j'=0}^{N-1}\left(c_{j'}\exp\left(\frac{2i\pi m_{x}j'}{N}\right)\right)c_{\left\lceil j'+n_{y}\right\rceil }^{*}.
\end{eqnarray}

\section{Monte Carlo techniques\label{app:Monte-Carlo-techniques}}

\subsection{``Normal'' Monte Carlo Calculations}

\subsubsection{Prior art}

The classical way to tackle the problem at hand has been to use a
standard Metropolis algorithm to sample the free energy at a given
value of $\alpha_B(T)^{2}$, which plays a role similar to the usual
$\beta$ parameter in statistical physics problems. Such techniques rely
on a probabilistic acceptance criterion for a ``move'' between two
configurations, which is usually of the bi-local form
\begin{equation}
P(E\rightarrow E')=\min\left(1,w(E')/w(E)\right).\label{eq:Ptrans}
\end{equation}
The standard form is to use the Boltzmann factor, 
\begin{equation}
w(E)=e^{-\beta E},
\end{equation}
 or in our case 
\begin{equation}
w(F)=e^{-F_{GL}/k_BT_c^{MF}}.
\end{equation}

Simulations of such a nature
\cite{kato1993firstorder,kato1993montecarlo,hu1994correlations,vsavsik1994calculation,vsavsik1995phasecoherence,
  li2003fluxlattice} usually find that in a certain range of ``inverse
temperatures'' $\alpha_{B}(T)$ -- with our definition of the coupling strength $-10\lessapprox \alpha_{B}(T)\lessapprox -9$ -- where we find coexistence of long lived states.

\begin{figure}
\centering{}\includegraphics[width=6cm]{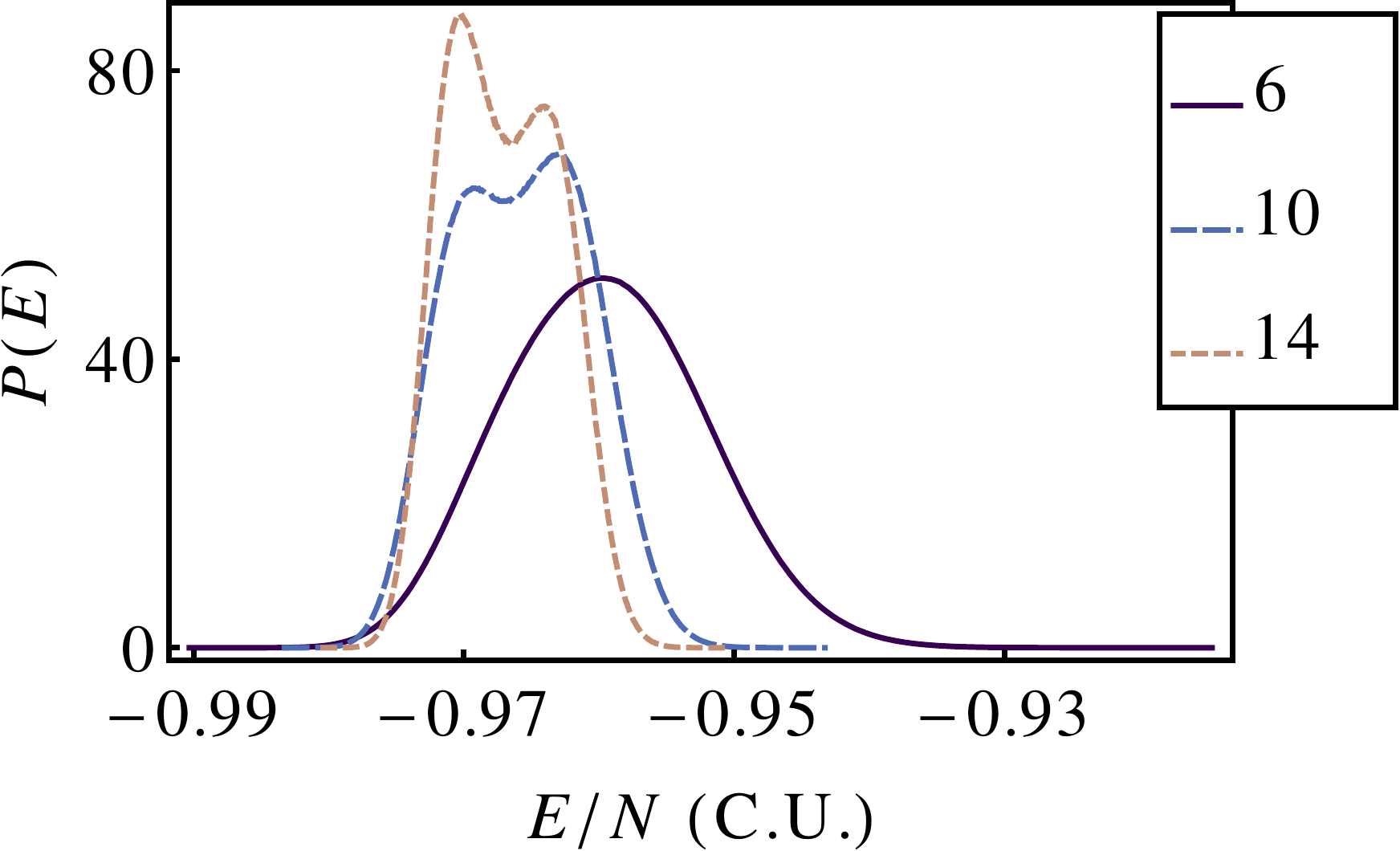}
\caption{(Color online) A plot of the energy probability distribution
  near phase coexistence for Metropolis Monte Carlo simulations for
  systems containing $6^2$, $10^2$ and $14^2$
  vortices.\label{fig:twopeak}}
\end{figure}

For a given temperature, it is instructive to look at the distribution
of energies contributing at that temperature. A typical example is
shown in Fig.~\ref{fig:twopeak}, where we see the  double
peak structure which is taken to be indicative of a first order phase
transition. We can understand this as follows: We rewrite the partition
function as an integral over a probability that the system has energy
$E$, 
\[
Z(\beta)=\left\langle e^{-\beta E}\right\rangle =\int dE\, g(E)\, e^{-\beta E}=\int dE\, P_{\beta}(E).
\]
This function $P_{\beta}(E)$ has extrema at an energy where the microcanonical
entropy takes the value $\beta$,
\begin{equation}
S_{\text{mc}}(E)=\frac{d}{dE}\ln g(E)=\beta.\label{eq:rhobeta}
\end{equation}
In the thermodynamic limit a first order phase transition thus occurs
when there is a range of energies that satisfies (\ref{eq:rhobeta}),
i.e., $S_{\text{mc}}(E)$ has a flat region. It is well-known that this
not what is seen in finite system: In the finite size precursor we see
a change in the slope of $S_{\text{mc}}(E)$ from negative to positive
for a sort while, violating the convexity rules $S$ must satisfy in
the thermodynamic limit. This gives rise to three intersections,
corresponding to the two maxima and one minimum we see in
Fig.~\ref{fig:twopeak}. The best approximation of the convex
thermodynamic limit can then be obtained from a Maxwell-type
construction, where we replace the curve by a flat segment, for which the
area above and below the curve must be equal, see
Fig.~\ref{fig:MaxwellConstruction}. The size of this area is an
estimate for the interface free energy for the first order phase transition, and
since this is related to the area of the interface between the two
phases, it should grow for two dimensions as the linear
dimension of the system.

\subsection{Broad histogram techniques}

There are various broad-histogram techniques that sample the energy
landscape directly \cite{wang2001determining,trebst2004optimizing}.
An advantage of these is that we are able to get a more detailed view
of the density of states, which reflects the nature of any phase transition
more directly. In this work we  concentrate on two
such methods, the Wang-Landau technique\cite{wang2001determining}
and its variations and the Optimal Energy Diffusion (OED) Method
\cite{trebst2004optimizing}. Both of these methods describe
the statistical physics in the microcanonical ensemble, where the Monte Carlo aspect is a random walk in energy--and potentially a few other degrees of
freedom, a case which we shall not consider here. The diffusion through
the energy landscape -- which we can of course completely describe as
a high-dimensional configuration space -- we prefer to parametrize by
a few collective coordinates, one of which is energy. This low-dimensional
projection can have all the complications, such as caustics, etc.,
we know from the topology of such reductions.

In all cases we shall assume that the transition probability used in
the random walk will be chosen to depend on the projection coordinate
(energy) only. The broad histogram that is used in these methods is an
energy histogram. In the case of a continuous problem, this is
obtained by dividing the energy into bins. If the energy is bounded
from both above and below we can use the full range of energies, even
though that is not always efficient. If the energy of the system is
not bounded, or if there are certain areas in the range of energies
that deserve more attention, we can carve out a restricted interval, and
only look at energies within it. There is a serious danger if we make
such intervals too small, however: suppose there is an (important)
energy barrier just above the interval of interest. In that case, we
would not be sampling all states for the energies we are considering,
and may obtain an incorrect estimate for the density of states.

\subsubsection{Wang-Landau}

In the Wang-Landau algorithm \cite{wang2001determining} and its variants
the random walk is driven  directly by the logarithm of the density
of state -- the attempt is made to make the sampled histogram flat in all
the projected directions. It is very easy to convince oneself that
this is the case if the acceptance probability is given by the inverse
of the (best estimate of the) density of states itself, i.e., $w(E)=1/g_{\text{est}}(E)$.
For the continuous system considered here -- most of the initial applications
were to systems with discrete energy spectra -- we divide the range of the 
interesting energy into bins, and start with an estimate of the density
of states.%
\footnote{For a continuous system, one can either use a ``staircase'' type
piecewise constant function, or a linear interpolation across the
interval, or probably an even fancier parametrization. As long as
there is an update rule that is consistent with this choice, these
are all allowed.%
} Then, when the random walk visits an energy in a given bin, we update
the estimate of $g$ in that bin by a factor $f>1$, 
\begin{equation}
\ln g_{\text{est}}\rightarrow\ln g_{\text{est}}+\ln f.\label{eq:WL1}
\end{equation}
When the histogram, the number of times we have visited each energy
bin is suitably flat, we reduce the factor 
\begin{equation}
f\rightarrow\sqrt{f},
\end{equation}
or a similar reduction by another suitably chosen power-law. When
$f$ becomes very small, $g_{\text{est}}$ will have converged
to $g$. 

There are various refinements one can make to this process \cite{wu2005overcoming,reynal2005fastflathistogram,morozov2007accuracy,belardinelli2007fastalgorithm,cunha-netto2008improving,zhou2008optimal}.
In general, it has been shown that more complex patterns of choosing
$f$ can lead to faster convergence. Also, we can choose to divide
up the energy range into many small intervals that are easier to deal
with -- but this suffers from the risk that there will be important
barriers that we are not treating well, and must always fail for a
sufficiently complex energy landscape.

It is  believed that the end-to-end transmission time, the simulation-time
it takes for the random walk to move form the lowest part of the energy
space, or vice-versa, is an important measure of the likelihood of
success of the simulation. As has been shown in Ref.~\onlinecite{dayal2004performance}
there can be a serious issue with this end-to-end tunneling rate in
real systems, with substantial slowing down of the process if there
are funnels or barriers in the problem. Also, the use of Eq.~(\ref{eq:WL1})
precludes the use of standard parallelization techniques usually applied
to Monte Carlo simulations, since we update $\rho_{\text{est}}$ continuously.

\subsubsection{Optimal energy diffusion}

In the optimal energy diffusion method, one tries to optimize the
current of random walkers between two extremal points in energy space
\cite{trebst2004optimizing,wu2005overcoming,bauer2010optimized},
by first building a simple model of this diffusion process, and choosing
optimal parameters based on that model. The extrema can be the real
bounds on energy, or the bounds of an interesting region -- this must
of course always be the case for continuous systems with energy unbound
from above.  We create two populations of walkers, one moving from
the lowest to highest energy, the other from highest to lowest; in
practice as soon as a walker reaches its final target, it is converted
to a walker going in the opposite direction. If we take this as 
starting a new walker, we see that this is effectively an absorbing
boundary condition. %
\footnote{Strictly speaking, we need to release a random representative of the
states occurring at this energy, if there is a degeneracy. For practical
reasons, this is not taken into account in our simulations.%
} Walkers reaching the opposite boundary, the one that does not correspond
to their target, are simply ``reflected'', and keep their label.
We record the position at each walker in two variables, $n_{w\pm}(E)$,
where the $+$($-$) indicates the walker was released from the maximal
(minimal) energy. The simulation process is started by releasing a
walker from the boundary (if the states are known) or by starting
with a walker without a label, which is absorbed and restarted when
it reaches a boundary.

We now define the fraction of the walkers diffusing from the upper
boundary as 
\begin{equation}
f(E)=\frac{n_{w+}(E)}{n_{w}(E)},\quad n_{w}(E)=n_{w+}(E)+n_{w-}(E).
\label{eq:nw}
\end{equation}
Clearly $f(E_{\text{min}})=0$ and $f(E_{\text{max}})=1$. The basic
premise of the method is that we can use the simple model that in
steady state, where the current $j$ of walkers will be independent
of energy, we have a diffusion process for the current of $+$ walkers
(which is equal in magnitude but opposite in sign to the current of
$-$ walkers) described by the local process 
\begin{equation}
j=D(E)n_{w}(E)\frac{df}{dE},\label{eq:energydif}
\end{equation}
where $D(E)$ is an unknown function describing the local energy diffusivity.
The expression (\ref{eq:energydif}) is a model of the process 
based on insight from diffusion, but clearly has no deep justification.

If we now optimize the current of walkers by the method of Lagrange
multipliers, we find that it is maximal when 
\begin{equation}
n_{w\text{opt}}(E)=w_{\text{opt}}(E)g(E)=\frac{df_{\text{opt}}}{dE}\propto\frac{1}{\sqrt{D(E)}}.
\end{equation}
We now wish to use a standard Metropolis dynamics to find the optimal
solution, which we take as sampling with the optimal choice of the
weights $w$. We start from an initial set of weights $w(E)$, and
use a standard Metropolis simulation to evaluate
\begin{equation}
D(E)\propto\left(n_{w}(E)\frac{df}{dE}\right)^{-1},
\end{equation}
using 
\begin{equation}
\frac{n_{w\,\text{opt}}(E)}{n_{2}(E)}=\frac{w_{\text{opt}}(E)}{w(E)}\propto\frac{1}{n_{w}(E)}\frac{1}{\sqrt{D(E)}}\propto\sqrt{\frac{df/dE}{n_{w}(E)}},
\end{equation}
we find that up to a constant an improved estimate for the weights
should be given by 
\begin{equation}
\ln w_{\text{opt}}(E)=\ln w(E)+\frac{1}{2}\ln\frac{df}{dE}-\frac{1}{2}\ln n_{w}(E).\label{eq:TTupd}
\end{equation}
Of course that relies on both the model and the linearity in all the
parameters -- neither of which are perfectly true. 

Since the Metropolis algorithm only depends on ratios of $w$'s, we
do not need to know the value of the constant. At the same time, we
see that the corresponding best estimate is
\begin{equation}
\ln g(E)=-\ln w(E)+\ln n_{w}(E).
\end{equation}
Evidence seems to suggest that this approach is very powerful indeed.
One of its main advantages is that it is a standard Metropolis calculation,
which can be computationally optimized in the normal way for such
approaches, including parallelization. The Wang-Landau algorithm,
where weights change while we are simulating is much harder to fine
tune in such a way. Also, as we have also seen in our work one can get
stable results with weights that are not perfectly optimal.

\subsubsection{Implementation}

Since we are particularly interested in the derivative of the density
of states (\ref{eq:rhobeta}), we have a slightly more complex task
at hand than normally considered in the flat-histogram approaches.
For all methods, we have  to take numerical first
and second derivatives of statistically fluctuating estimates of physical
quantities, such as the density of states. For the Wang-Landau technique
the approach is relatively straightforward; if we assume a limited
correlation between the fluctuations in different bins, we can either
apply a running average or fit a smooth curve to the weights to obtain
a sensible estimate of the logarithm of the density of states, which
can then be differentiated. The real problem with the Wang-Landau
algorithm and its many variants is the difficulty in driving it to
convergence; the simple decrease  of the size of the update
$\ln f$ in Eq.~(\ref{eq:WL1}) by a factor is not very effective
at convergence. There are a number of alternative approaches where
we increase as well as decrease these updates in an suitable pattern
\cite{belardinelli2007fastalgorithm, cunha-netto2008improving, komura2012difference, morozov2007accuracy, reynal2005fastflathistogram, wu2005overcoming, zhou2008optimal}, many of which we attempted.

For the optimal energy diffusion method the smoothing task is more
difficult, but also more critical: we need a sensible form for $f(E)$
to be able to differentiate it. We have typically applied this method
to what we shall call the ``interesting'' region, containing the
energy where the potential phase transition occurs. In that region
we need to make a smooth approximation to the function $f(E)$ -- as
can be seen from the work of Ref.~\onlinecite{bauer2010optimized}, that is no trivial
task. We have chosen a rather different approach than that used in that reference,
and used a sigmoid-like approximation for $f$,
\begin{equation}
f=\sum_{i=1}^{N_{\text{sig}}}\frac{a_{i}}{\left(1+\exp\left(-b_{i}(x-c_{i})\right)\right)^{i}},
\end{equation}
which seems to be a very powerful model, and will guarantee the smoothness
required of $f$, which needs to be differentiated twice. At the same
time we expand the energy histogram $n(E)$ in a set of cubic B-spline
polynomials to reach a smooth result (this is less critical, since
we only need first derivatives). 

The method is typically driven to ``convergence'' by increasing
the number of Monte Carlo steps. The potential weakness, especially
when the current of walkers is small, as it will be in our simulations,
is that the update (\ref{eq:TTupd}) does not converge, but is dominated
by the numerical noise in the low-sample regions. To that end we have
chosen to add a limiting function to the update,
\begin{equation}
\ln w_{\text{opt}}(E)=\ln w(E)+\frac{1}{2}\ln\left[\text{sm}\left(\frac{df}{dE}/n_{w}(E)\right)\right],\label{eq:TTupd-1}
\end{equation}
where we have used 
\begin{equation}
\text{sm}(x)=\frac{1+2x^{3}}{2+x^{3}},
\end{equation}
(any similar function as a ratio of polynomials with retrograde coefficients
could have been used, since we have $\mbox{\text{sm}}(1/x)=1/\text{sm}(x)$,
which links in with the logarithmic update, but the one chosen seems
quite sensible.)

\section{Density of states in a crystal\label{app:cryst}}

The density of states for the crystalline state can be evaluated rather straightforwardly;
for each of the $N$ crystal states, we can use the harmonic approximation
to the potential
\begin{equation}
E\approx E_{0}+\frac{1}{2}\sum_{i=1}^{2N}\omega_{i}^{2}x_{i}^{2}.
\end{equation}
The number of states below a certain energy $E$ can be found by simple
integration and multiplication with the degeneracy factor $N$
\begin{equation}
n(E)=N\int\prod_{i}dx_{i}\theta\left(E-E_{0}-\frac{1}{2}\sum_{i=1}^{2N}\omega_{i}^{2}x_{i}^{2}\right),
\end{equation}
which is simply the volume of an ellipsoid,
\begin{equation}
n(E)=N\prod_{i=1}^{2N}\frac{\sqrt{2(E-E_{0})}}{\omega_{i}}=N\frac{\left[2(E-E_{0})\right]^{N}}{\sqrt{\det E^{(2)}}},
\end{equation}
where $E^{(2)}$ denotes the matrix of second derivatives at one of
the crystal minima. If we express this in crystal units, 
\begin{equation}
\mathcal{E}=\frac{2\beta_{A}}{N}E,
\end{equation}
 we find
\begin{equation}
n(\mathcal{E})=N\frac{\left[2(\mathcal{E}+1)\right]^{N}}{\sqrt{\det\mathcal{E}^{(2)}}}.
\end{equation}
The density of states thus takes the form 
\begin{equation}
g(\mathcal{E})=\frac{d}{d\mathcal{E}}n(\mathcal{E})=2N^{2}\frac{\left[2(\mathcal{E}+1)\right]^{N-1}}{\sqrt{\det\mathcal{E}^{(2)}}},
\end{equation}
and
\begin{equation}
\ln g(\mathcal{E})=c+(N-1)\ln(\mathcal{E}+1),
\end{equation}
with 
\begin{equation}
c=\ln2^{N}N^{2}-\frac{1}{2}\ln\det\mathcal{E}^{(2)}.
\end{equation}
Thus finally,
\begin{equation}
S_{\text{mc}}=(N-1)/(\mathcal{E}+1).
\end{equation}

\bibliographystyle{apsrev4-1}
\bibliography{Paper}

\end{document}